\documentclass[floatfix,showpacs,preprintnumbers,amsmath,amssymb,pra,superscriptaddress,longbibliography]{revtex4-2}
\usepackage{color}
\usepackage[usenames,dvipsnames,svgnames,table]{xcolor}
\usepackage[colorlinks=true,linkcolor=blue,urlcolor=blue,citecolor=blue]{hyperref}
\usepackage{mathtools}
\usepackage{graphicx}
\usepackage{dcolumn}
\usepackage{array}
\usepackage{lipsum}
\usepackage{bm}
\usepackage{subfigure}
\usepackage{amssymb}
\usepackage{pifont} 
\usepackage{multirow}
\usepackage{tabularx}
\usepackage{amsmath}
\usepackage{braket}
\usepackage{csquotes}
\usepackage{enumitem}
\graphicspath{{plots/}}
 \usepackage{lipsum}
\usepackage{mathrsfs}
\usepackage{MnSymbol}

\newcommand{\beq}{\begin{equation}}
\newcommand{\eeq}{\end{equation}}
\newcommand{\bea}{\begin{eqnarray}}
\newcommand{\eea}{\end{eqnarray}}

\usepackage{booktabs}
\usepackage{tikz}
\usepackage{pgfplots}
\pgfplotsset{compat=1.18} 
\usepackage{float}
\makeatletter
\let\newfloat\newfloat@ltx
\makeatother

\usepackage{algorithm}
\usepackage{algpseudocode}
\usepackage{todonotes}

\begin{document}
\title{Estimates of the dynamic structure factor for the finite temperature electron liquid via analytic continuation of path integral Monte Carlo data}

\author{Thomas Chuna}
\email{t.chuna@hzdr.de}
\affiliation{Center for Advanced Systems Understanding (CASUS), D-02826 G\"orlitz, Germany}
\affiliation{Helmholtz-Zentrum Dresden-Rossendorf (HZDR), D-01328 Dresden, Germany}

\author{Nicholas Barnfield}%
\affiliation{Department of Statistics, Harvard University, Cambridge, Massachusetts 02138, USA}

\author{Jan Vorberger}
\affiliation{Helmholtz-Zentrum Dresden-Rossendorf (HZDR), D-01328 Dresden, Germany}

\author{Michael P. Friedlander}
\affiliation{Department of Computer Science and Mathematics, University of British Columbia, Vancouver, BC V6T 1Z4, Canada}

\author{Tim Hoheisel}
\affiliation{Department of Mathematics and Statistics, McGill University, Montreal, Quebec H3A 0G4, Canada}

\author{Tobias Dornheim}
\affiliation{Center for Advanced Systems Understanding (CASUS), D-02826 G\"orlitz, Germany}
\affiliation{Helmholtz-Zentrum Dresden-Rossendorf (HZDR), D-01328 Dresden, Germany}



\begin{abstract}
Understanding the dynamic properties of the uniform electron gas (UEG) is important for numerous applications ranging from semiconductor physics to exotic warm dense matter.
In this work, we apply the maximum entropy method (MEM), as implemented in Chuna \emph{et al.}~[arXiv:2501.01869], to \emph{ab initio} path integral Monte Carlo (PIMC) results for the imaginary-time correlation function $F(q,\tau)$ to estimate the dynamic structure factor $S(q,\omega)$ over an unprecedented range of densities at the electronic Fermi temperature. To conduct the MEM, we propose to construct the Bayesian prior $\mu$ from the PIMC data. Constructing the static approximation leads to a drastic improvement in $S(q,\omega)$ estimate over using the more simple random phase approximation (RPA) as the Bayesian prior.
We find good agreement with existing results by Dornheim \emph{et al.}~[\textit{Phys.~Rev.~Lett.}~\textbf{121}, 255001 (2018)], where they are available.  In addition, we present new results for the strongly coupled electron liquid regime with $r_s=50,...,200$, which reveal a pronounced roton-type feature and an incipient double peak structure in $S(q,\omega)$ at intermediate wavenumbers. We also find that our dynamic structure factors satisfy known sum rules, even though these sum rules are not enforced explicitly.
An advantage of our set-up is that it is not specific to the UEG, thereby opening up new avenues to study the dynamics of real warm dense matter systems based on cutting-edge PIMC simulations in future works.

\end{abstract}

\maketitle

\section{Introduction}

The uniform electron gas (UEG) is an important and well-studied physical system ~\cite{GiulianiVignale2008quantumtheory, mahan1990many, hansen2013theory3rdEd, Dornheim2018UEGreview, loos}. Investigations on this topic have continued for decades because there is no closed analytic model which can describe the spatial and temporal structure of the plasma at all experimentally accessible number densities and temperatures. A current frontier of the UEG is the finite temperature electron liquid, where the plasma is both quantum degenerate $\Theta = k_\text{B} T /E_F \approx 1 $ and strongly coupled $\Gamma = \frac{1}{r_s} \left( (k_\text{B} T)^2 + E_F^2 \right)^{-1/2} \geq 1$. We have adopted Hartree atomic units in this work with $r_s$ being the Wigner-Seitz radius that is defined by $n_e = (4 \pi r_s^3 / 3)^{-1}$ (in units of Bohr radius $a_B$), $E_F$ is the Fermi energy in Hartree, and $\beta=(k_\textnormal{B} T)^{-1}$ is the inverse temperature. When the plasma is both quantum degenerate and strongly coupled, path integral Monte Carlo (PIMC) simulations are one of the few methods that can describe the structure of the UEG~\cite{bonitz2020WDMUEG}. In this work, we study PIMC simulations of the electron liquid at fixed $\Theta=1$ and vary $\Gamma$ between approximately $1$ and $150$; this gives us access to a broad range of physical conditions.

In principle, PIMC simulations give one access to the complete equilibrium dynamics of the simulated system in the form of various imaginary-time correlation functions (ITCF)~\cite{Rabani_PNAS_2002,Filinov_PRA_2012,Dornheim_JCP_ITCF_2021}. In particular, the imaginary-time density--density correlation function $F(q,\tau)=\braket{\hat{n}(q,\tau)\hat{n}(-q,0)}$, which corresponds to the usual intermediate scattering function evaluated at the imaginary time argument $t=-i\hbar\tau$ with $\tau\in[0,\beta]$ and $\beta=1/k_\textnormal{B}T$ the inverse temperature in energy units, is directly related to the dynamic structure factor (DSF) $S(q,\omega)$ (in units of inverse energy) via a two-sided Laplace transform
\begin{align}\label{eq:periodicLaplace}
    F(q,\tau) = \int_{-\infty}^\infty d\omega\ e^{-\tau\omega} S(q,\omega) =  \int_0^{\infty} d\omega \left( e^{-\tau \omega} + e^{-(\beta - \tau) \omega} \right) S(q,\omega)\ ,
\end{align}
where the second equality follows from the detailed balance condition $S(q,-\omega)=S(q,\omega)e^{-\beta\omega}$.
Matching $\beta$, $\omega$ is the frequency in energy units. Since the LHS is a function of imaginary-time and the RHS is a function of real-time frequency, estimating the DSF is an analytic continuation problem. The DSF is of key importance e.g.~for the interpretation of x-ray Thomson scattering experiments~\cite{Gregori_PRE_2003,siegfried_review,Dornheim_review}, and directly gives one access to a number of density response and dielectric properties~\cite{Hamann_CPP_2020,Hamann_PRB_2020}.
Unfortunately, inverting \eqref{eq:periodicLaplace} to reconstruct $S(q,\omega)$ from $F(q,\tau)$ is a well-known exponentially ill-posed problem~\cite{epstein2008badtruth, istratov1999} and plagues many fields of science~\cite{Fournier_PRL_2020,Yoon2018ACviaML,Goulko_PRB_2017,otsuki2017sparse,huang2023acflow,beach2004identifying,fuchs2010analytic,JARRELL1996MEM,BurnierRothkopf2013bayesianreconstruction,Fischer2018SmoothedBRM}. For the UEG, Dornheim \textit{et al.}~\cite{dornheim2018dynamiclocalfieldcorrection,dynamic_folgepaper,Dornheim_PRE_2020} have partially solved this problem by evoking the fluctuation--dissipation theorem, which has allowed to exactly fulfill a number of exact constrains that are known for the UEG. However, this \emph{stochastic sampling} approach breaks down with increasing coupling strength, and cannot be easily generalized to real systems beyond the UEG model. Consequently, a more general approach is needed.

The recent work of Chuna et al.~\cite{chuna2025dual} has demonstrated that the all-purpose maximum entropy method (MEM) can reliably estimate the DSF from PIMC data in some situations. This method is not specifically tailored for the UEG and can be applied directly to two-component systems that are comprised of both electrons and nuclei. In the present work, we apply the MEM with a dual Newton optimizer to estimate the UEG DSF, at $\Theta=1$ over a broad range of densities identified by the corresponding Wigner-Seitz radius. In particular, we investigate densities spanning the warm dense matter regime~\cite{wdm_book} ($r_s=2$) to the strongly coupled electron liquid~\cite{dornheim_electron_liquid,Tolias_JCP_2021,Tolias_JCP_2023} ($r_s=50,...,200$). We compare our estimates with the more limited stochastic inversion where it is available, i.e., for $r_s\leq10$. As we apply the MEM, we have given special attention to our uncertainty procedure. The MEM requires ITCF data and a best guess of the solution (\textit{i.e.} Bayesian prior). In our application, we construct the Bayesian prior directly from the ITCF data based on the \emph{static approximation} to the DSF that approximates the full dynamic local field correction (LFC) by its exact static limit, i.e., $G(q,\omega)\approx G(q,0)$. The static approximation has been shown, in many finite temperature electron liquid applications, to yield a good approximation of the DSF~\cite{dornheim2018dynamiclocalfieldcorrection, groth2019stochasticsampling, dornheim2019MLstatic}. We consider the impact of using such a Bayesian prior that depends on the data, and compare the static approximation to the more simple random phase approximation (RPA) that neglects the LFC entirely. In addition to this, we quantify the uncertainty in our DSF estimates via leave-one-out binning~\cite{berg2004markov, berg2004introduction, james2013statisticallearning}, which estimates the variance arising from uncertainty in the MEM regularization weight and from sampling error in the two-particle correlation function.

Thus, this work establishes how to use the general purpose tool, the MEM, to estimate the UEG DSF from PIMC data for the ITCF. In the weak coupling limit $\Gamma \rightarrow 0$, our estimates resemble the RPA, as expected.  In addition, our estimates match results obtained by stochastically sampling the dynamic local field corrections~\cite{dornheim2018dynamiclocalfieldcorrection} for $r_s\leq10$, and our set-up works well even at large $r_s$ where the latter approach fails. We find a pronounced roton-type feature in our new estimates for $S(q,\omega)$ at intermediate wavenumbers $q$ that emerges with increasing coupling strength. This feature has been observed in a number of recent theoretical studies of strongly correlated quantum Coulomb systems~\cite{Hamann2023rotonfeature, Dornheim_Nature_2022, koskelo2025short, Filinov_PRB_2023,Takada_PRB_2016} and phenomenologically resembles the roton feature reported in both fermionic He-3~\cite{Godfrin2012, Dornheim_SciRep_2022}, bosonic He-4 quantum liquids~\cite{Trigger, Ferre_PRB_2016}, and dusty plasmas \cite{murillo1998SLFC, murillo2000DLFC}. Our estimates also show an incipient double peak structure, which emerges at strong coupling and matches strongly coupled molecular dynamics simulations~\cite{Korolov_CPP_2015, Choi_PRE_2019} and investigations of the UEG~\cite{Takada_PRL_2002, Takada_PRB_2016, Filinov_PRB_2023}. Finally, our estimates recover the appropriate sum rule behaviors, even though the sum rules are not explicitly enforced on the solution as a constraint.

There are many future applications of the DSFs estimated in this work; we name only a few here. First, determining the full frequency dependence of the dynamic LFC $G(q,\omega)$ provides an estimate for a nonadiabatic dynamic exchange-correlation kernel for linear response time-dependent density functional theory (TD-DFT)~\cite{ullrich2011time,Panholzer_PRL_2018,Ruzsinszky_PRB_2020}. Such kernels would inform simulations of a host of dynamic and transport properties of real multi-component systems, and we expect that an improved treatment of dynamic exchange--correlation effects might be particularly important to study electron--electron effects such as the roton mode that has recently been predicted to occur in warm dense hydrogen~\cite{Hamann2023rotonfeature}. Second, the estimated DSF is related to various dielectric and transport properties via Kramer-Kronig relations or sum rules~\cite{boon1991molecular,ichimaru2018plasmavol1, hansen2013theory3rdEd, Hamann_CPP_2020, Hamann_PRB_2020}. Third, the DSF of the UEG is important for the interpretation of x-ray Thomson scattering experiments within the widely used Chihara decomposition~\cite{siegfried_review,Gregori_PRE_2003,boehme2023evidence}. Finally, these estimated DSFs establish a benchmark for new dynamic theories in the strongly coupled regime, such as (non)equilibrium Green's functions~\cite{kwong_prl-00, Kas_PRL_2017, Schluenzen_PRL_2020} and dynamic dielectric theories~\cite{Holas_PRB_1987, tolias2024density,Tolias_JCP_2023}.


The remainder of this paper is organized as follows: In Section \ref{sec:methods}, we discuss Methods. We give a short review of the MEM, discuss the static LFC and the RPA, contemplate the implications of using the data twice (\textit{i.e.}, first to estimate the Bayesian prior and second to conduct the MEM), and finally we describe the leave-one-out uncertainty quantification procedure. In Section \ref{sec:DSFplots}, we present heatmaps of the DSF, heatmaps of the DSF error, comparative plots of Bayesian priors, plots of the estimated dispersion relation, plots of stacked $q$ cross sections from the DSF heat maps. In section~\ref{sec:stochasticcomparison}, we compare our DSF estimates to the stochastic analysis conducted on similar finite temperature electron liquid data. In Section~\ref{sec:DSFsumrules}, demonstrate that certain sum rules are satisfied even though they are not enforced by the MEM. Finally in Section \ref{sec:conclusions}, we draw conclusions from our results.

\section{Methods} \label{sec:methods}
In this work, we use the all-purpose MEM to estimate the DSF by inverting Eq.~\eqref{eq:periodicLaplace}. We supply two practical considerations that support this choice. The MEM has a large literature base to make practitioners familiar with the approach and many investigations establish the systematics of the MEM. In particular, entropic regularizers tend to contain ringing in the solution~\cite{KimPetreczkyRothkopf2015ComparisonBRMandMEM} and among entropic regularizers the MEM tends to be the smoothest~\cite{fuchs2010analytic, Fischer2018SmoothedBRM}. The second reason is that it allows for a rigorous error analysis. The MEM quantifies the uncertainty arising from the regularization weight~\cite{Asakawa2001MEM} and the MEM is computationally cheaper than other methods~\cite{fuchs2010analytic}, so it can be used in a resampling procedure to estimate the uncertainty arising from the data.

Many thorough reviews of the MEM exist~\cite{JARRELL1996MEM, Asakawa2001MEM}. The MEM relies on a Bayesian framework created by Gull~\cite{gull1989MEMBayesianWeighting}, which shows how to average over a collection of solutions produced by solving the optimization problem \eqref{eq:GLSEntropy} at different regularization weights $\alpha$,
\begin{align} \label{eq:GLSEntropy}
    \max_x \quad  -\frac{1}{2} \Vert A x - b \Vert_C + \alpha S_{SJ}[ x \mid \mu].
\end{align}
Here $\Vert A x - b \Vert_C = (Ax-b)^T C^{-1} (Ax-b)$ is the generalized least squares criterion and $S_{SJ}[ x| \mu] = - \sum_i x_j \ln x_j /\mu_i$ is the Shannon-Jaynes relative entropy regularizer. For this problem, we have the matrix $A_{i,j} = \exp(- \tau_i \, \omega_j ) + \exp(- (\beta - \tau_i) \omega_j)$, the data vector $b_i = F(q,\tau_i)$, the error of the data 
\begin{align}
    C^{\text{bin}\, 0}_{ij} = \frac{\sigma^2_{F^{\text{bin}\, 0}}}{\sqrt{N-1}} \delta_{ij},
\end{align}
the Bayesian prior $\mu_j$, and the desired solution vector $x_j = S(q,\omega_j)$ all at fixed wavenumber $q$. From the collection of solutions obtained varying $\alpha$, the MEM yields an estimate of the solution and the uncertainty arising from the choice of $\alpha$. In this work, we estimate $50$ $\alpha$ solutions, sampling logarithmically from $10^{-3} - 10^7$,

Typically, the MEM relies on Bryan's algorithm to solve Eq.~\eqref{eq:GLSEntropy} at various values of $\alpha$~\cite{bryan1990algorithm}. However, Bryan's algorithm only approximates the solution to \eqref{eq:GLSEntropy}~\cite{bryan1990algorithm, rothkopf2020bryan, chuna2025dual}. Many applications of MEM with Bryan's algorithm can be found~\cite{JARRELL1996MEM, Asakawa2001MEM, Boninsegni1996MEM}. Chuna et al. propose an alternative dual Newton algorithm which avoids the issues concerning Bryan's algorithm~\cite{chuna2025dual}. Furthermore, in the same work they establish rigorous limits on the error in $x$ arising from the data $b$, and their numerical investigations indicate that the MEM substantively outperforms the analytic bound.

\subsection{Choice of Bayesian prior $\mu$}
\subsubsection{Models of the finite temperature electron liquid dynamic structure factor}\label{sec:DSFmodels}
Early models of the dynamic response of one-component plasmas were developed for either classical or quantum weakly coupled systems~\cite{kugler1975LFCs, VashishtaSingwi1972VS-LFC, Singwi1968STLS-LFC, Sjostrom2013STLS2-LFC},  $\Gamma \ll 1$, meaning that only the self-interaction energy was accounted for~\cite{atwal2002fullyconserving, Morawetz2000quantumliquids}. Later models were developed for the dynamic response in classical strongly coupled systems~\cite{murillo2004strongly, ichimaru1982stronglycoupledplasma}, $\Gamma \geq 1$, meaning that $N$-body correlation effects were also being taken into account. However, the finite temperature electron liquid is both strongly coupled and quantum degenerate, $\Theta \approx 1$. Thus, $N$-body exchange effects are important as well.
In selecting a suitable default model for the Bayesian prior, we consider the random phase approximation, which is accurate in the high-density limit where the UEG is weakly coupled. In addition, we consider a second, vastly improved default model for the Bayesian prior that takes into account PIMC results for exchange--correlation effects on the static level.

The DSF is related to the system's dynamic linear density response function $\chi(q,\omega)$, where $q$ and $\omega$ are the wavenumber and frequency of a weak external harmonic perturbation~\cite{Dornheim_review}, via the fluctuation dissipation theorem~\cite{GiulianiVignale2008quantumtheory},
\begin{align} \label{eq:FDT}
    S(q,\omega) =  - \frac{1}{ \pi n} \frac{\text{Im} \chi(q,\omega)}{1-e^{- \beta \hbar \omega} }.
\end{align}
Without loss of generality, the density response is often expressed as~\cite{ichimaru2018plasmavol1, ichimaru1982stronglycoupledplasma, kugler1975LFCs} 
\begin{align}\label{eq:susceptibility}
    \chi(q,\omega) = \frac{\chi^0(q,\omega)}{1 - v(q) \left[1 - G(q,\omega)\right] \chi^0(q,\omega)}.
\end{align}
Here $v(q)$ is the Coulomb interaction and $\chi^0(q, \omega)$ describes the density response of the non-interacting Fermi gas at finite temperatures, which is given by
\begin{align} \label{eq:idealgas}
    \chi^{0}(q,\omega) = 2 \int \frac{d^3k}{(2\pi)^3} \frac{f_0 (k + q) - f_0(k)}{ (\epsilon_{k+q} - \epsilon_k) - \hbar \omega},
\end{align}
where $\epsilon_k = \hbar^2 k^2 / 2 m_e$ is the Fermi energy. We relied on Giuliani and Vignale's textbook~\cite{GiulianiVignale2008quantumtheory} as well as the work by Tolias et al.~\cite{tolias2024density} to implement this function. Kugler indicates a few $q \rightarrow 0$ sanity checks that we verified as well~\cite{kugler1970bounds}.

The dynamic local field correction $G(q,\omega)$ in \eqref{eq:susceptibility} contains the full wavenumber and frequency resolved information about exchange--correlation effects. It is formally related to the dynamic exchange correlation kernel from linear response time dependent density functional theory as~\cite{olsen2019beyond,Moldabekov_PRR_2023}
\begin{align}
    K_{XC}(q, \omega) = - v(q) G(q, \omega)\ .
\end{align}
In other words, $G(q,\omega)$ modifies the interaction potential such that the exchange and correlation effects are included, in which case this seemingly one-body description includes all the $N$-body physics~\cite{RG84}. 

The random phase approximation corresponds to $G(q,\omega)\equiv0$ and, thus, describes the dynamic density response on the mean-field level, neglecting exchange-correlation effects. In contrast, the static approximation only neglects the dynamic effects by setting $G(q,\omega) \approx G(q,0)$. As we show in Sec.~\ref{sec:results}, including the static LFC into the Bayesian prior leads to substantially improved performance and reduced uncertainty for $r_s\gtrsim10$.

In the context of the present work, the foremost advantage of the static approximation is that $G(q, \omega=0)$ is computed directly from the PIMC data; avoiding the problematic and arbitrary selection of a model that is only valid in a certain density and temperature limit. The procedure for constructing the static LFC is given e.g.~in Ref.~\cite{dornheim2019MLstatic}; we do not use the neural network representation reported therein, as it has only been trained for $r_s\leq20$~\cite{dornheim2019MLstatic}. Instead, before conducting the analytic continuation, we compute the static LFC $G(q,\omega=0)$, and construct the Bayesian prior. The formula to calculate $G(q, \omega=0)$ from the PIMC data is derived as follows. Start by inverting \eqref{eq:susceptibility} to isolate $G(q,\omega)$. 
Then let $\omega=0$ and simplify to arrive at 
\begin{align}\label{eq:Gq}
    G(q,\omega=0) &= 1 + \frac{1}{v(q)} \left(\frac{1}{\chi(q,\omega=0)} - \frac{1}{\chi^0(q,\omega=0)}\right).
\end{align}
Here $\chi(q,\omega=0)$ can be estimated from the data via the imaginary time version of the fluctuation dissipation theorem
\begin{align}\label{eq:chi_q}
    \frac{\chi(q, \omega=0)}{n \beta} = - \frac{1}{\beta} \int^{\beta}_0 d \tau F(q,\tau),
\end{align}
derived e.g.~in the appendix of Ref.~\cite{Dornheim_MRE_2023}.

Many investigations have been conducted comparing the RPA and the static approximation for the UEG~\cite{dornheim2018dynamiclocalfieldcorrection,groth2019stochasticsampling,Dornheim_PRB_ESA_2021}. Both RPA and static approximation exhibit a single sharp peak at the plasma frequency at small $q$, where the effect of $G(q,0)\sim q^2$ vanishes. Conversely, $v(q) G(q,\omega=0)$ becomes constant at large $q$~\cite{Hou_PRB_2022} (i.e., small length scales $\lambda=2\pi/q$), and its effects on the dispersion relation vanish. The discrepancies between the RPA and static approximation occur where the wavelength $\lambda$ is roughly the interparticle spacing $2 \pi / q \approx r_s$ and are explained by the electron pair alignment model \cite{Dornheim_Nature_2022}. The authors argue that density fluctuations caused by an external perturbation of wavelength $2 \pi / q \approx r_s$ need less energy compared to other density fluctuations because this wavenumber coincides with a spatial pattern that minimizes the potential energy of the system. Since the potential energy of the system is dependent on the exchange-correlation interaction energy and RPA entirely neglects these effects, the static approximation will greatly differ from the RPA in this region. Further, since exchange-correlation effects become more important with increasing coupling, this disagreement becomes more pronounced with increasing $r_s$.

Investigations have found that the static approximation exhibits a negative dispersion relation (\textit{i.e.}, roton feature)~\cite{dornheim2018dynamiclocalfieldcorrection, Hamann2023rotonfeature, Dornheim_Nature_2022, koskelo2025short, Filinov_PRB_2023,Takada_PRB_2016}, satisfies the small $q$ limit for $S(q)$~\cite{dornheim2024QuantumDelocalization}, violates the large $q$ asymptotic convergence for $S(q)$~\cite{Dornheim_PRL_2020_ESA, Dornheim_PRB_2021,Dornheim_PRB_2024}, empirically satisfies the frequency sum rule (\textit{i.e.}, F-sum rule) and higher sum rules.~\cite{Dornheim_moments_2023, Dornheim_PTR_2023, Dornheim_MRE_2023}. Further, investigation has also seen that the inverse moment sum rule is satisfied~\cite{dornheim2024QuantumDelocalization}; we provide an analytical proof for the latter in Appendix~\ref{app:inversemoment}. Lastly, going beyond the static approximation, investigations have shown that including the dynamic dependence in the local field correction corrects the large $q$ limit for $S(q)$~\cite{Dornheim_PRB_2024}, enhances the roton feature \cite{dornheim2018dynamiclocalfieldcorrection}, and creates an incipient double peak structure \cite{Dornheim_Nature_2022}. In summary, the static LFC has a record of including much of the important physics, but by definition, it does not include the full dynamics dependencies desired in a DSF model.

\subsubsection{Discussion of Extracting the Bayesian Prior from the Data}
Using the data to extract both the Bayesian prior and the final DSF estimate is less \textit{ad hoc} than more traditional Bayesian priors; this is expected to mitigate the bias introduced by the choice of the Bayesian prior. However, complications may arise, and so in the following text we contemplate our approach.

We argue that our method of extracting the Bayesian prior is comparable to common data de-trending approaches seen in data science~\cite{wu2007trend}. Examples of detrending can be found in introductory textbooks~\cite{brockwell2002TSA, adhikari2013TSA, james2013statisticallearning}; SciPy Signal library's even has a built-in detrend function~\cite{SciPy2020NMeth}. Detrending identifies and removes the easily spotted trends in the data before beginning data analysis procedures; this focuses the algorithm's inference capability on the parts of the data that are poorly understood. A simple example of this is known to all scientific fields. When analyzing probabilistic samples, we first center the samples about the sample mean and then scale the samples by the sample standard deviation. In the case of this analytic continuation algorithm, we are doing something only slightly more sophisticated than centering the data and standardizing the data. Equation \eqref{eq:chi_q}, has the form of the mean value theorem from introductory calculus. Thus, $\frac{\chi(q, \omega=0)}{n \beta}$ is the average value of the ITCF. In this sense, the static LFC is informing our default model of the data's mean value, so that the inference power is not wasted on resolving the mean. To relate this discussion to the physics presented in Section~\ref{sec:DSFmodels}, we are doing some initial work so that the MEM is only extracting the dynamic corrections to the static approximations.

From a statistics perspective, we argue that extracting the Bayesian prior from the data reduces the bias introduced in entropic regularization. In statistical inference, regularization is introduced to produce a better estimator. Regularization introduces a bias to reduce the estimators sensitivity to the data. This is known as the bias-variance trade-off~\cite{james2013statisticallearning}. Both the choice of regularizer and, if the regularizer includes a Bayesian prior, the choice of Bayesian prior introduce bias. Axiomatic arguments conclude that, unlike other regularizers, entropic regularization does not introduce spurious correlations in $S(q,\omega)$ across $\omega$ \cite{JARRELL1996MEM}. Thus, our primary concern is the choice of Bayesian prior. In the discussion of possible Bayesian priors (see section~\ref{sec:DSFmodels}), we established that $G(q,\omega)$ determines the prior. In fact, determining $G(q,\omega)$ is equivalent to solving the analytic continuation problem \cite{groth2019stochasticsampling}. Clearly, choosing $G(q,\omega)=0$ introduces a bias. By comparison, the static approximation will recover $G(q,\omega) = 0$ where it is reasonable, but can deviate from this value. This suggests that the bias arising from using the static approximation can be no worse than the RPA. Maybe a $G(q,\omega)$ that does not neglect the $\omega$-dependence can be computed to further reduce the bias of the static approximation, but as stated, determining the dynamic dependencies is equivalent to solving the problem. So this becomes a \textit{chicken-and-egg} problem. This suggests that the static approximation is an optimal balance of reducing bias without \textit{a priori} knowing the true solution. To relate this discussion to the physics presented in Section~\ref{sec:DSFmodels}, we are extracting the static exchange-correlation effects to create a Bayesian prior that is more reflective of the true DSF, hence focusing the MEM on extracting the dynamic exchange-correlation effects. 


Next we consider Stein's paradox, a statistical concern~\cite{samworth2012stein}. Stein's paradox examines how to estimate quantities from data. The general idea is communicated by a simple example. Consider a set of $M$ Gaussian distributed quantities ($a$,$b$,$c$,.., $m$), that you sample N times (\textit{i.e.} $a_0, \dots, a_{N-1}$). If you want to estimate the center of the Gaussian distribution ($<a>$, $<b>$, $<c>$... ) then taking the average of all the $a$'s, $b$'s, $c$'s individually is NOT the best way to do this. Instead, Stein's estimator dominates this naive approach when $M \geq 3$. In general, Stein's paradox and later proofs are a warning, that if the error of every parameter is important then a combined fit is the best way to get the lowest total error. Assuming that Stein's paradox applies, then it informs us that if we estimated our Bayesian prior and MEM solution simultaneously, we may be able to concoct an estimator that improves the total error of both. But this again becomes a \textit{chicken-and-egg} problem because if we knew the best prior we would not need to solve the problem. Thus, we are only focused on producing the best MEM estimate, the Bayesian prior is only a concern in that it effects this estimate. To relate this discussion to the physics presented in Section~\ref{sec:DSFmodels}, we are not concerned with the accuracy of the static approximation because we know it is wrong. Hence we do not make efforts to improve the estimate. 

The only remaining concern given the discussions above is the general concern of how does error in the default model propagate into the MEM estimate, which would apply to any Bayesian prior we select. we implement the leave-one-out procedure to quantify this error.


\subsection{Leave-one-out binning}
To estimate both the DSF and its related statistical uncertainty from the PIMC data for $F(q,\tau)$, we use the leave-one-out binning over independent PIMC simulations. This procedure is also known as jackknife binning~\cite{berg2004markov, berg2004introduction, james2013statisticallearning}. We are motivated to use leave-one-out-binning procedure for two reasons. First, we need an uncertainty quantification tool that can include uncertainty arising from estimating the Bayesian prior from the data. Second, we need a tool that can handle biased non-linear error estimates (note that the entropic regularization makes the MEM, itself, a biased non-linear function acting on the data to produce a DSF estimate). Based on Berg~\cite{berg2004markov}, leave-one-out binning is the simplest tool we can expect to handle both issues.

Leave-one-out-binning is a resampling algorithm that produces $N$ estimates from $N$ independent samples by creating $N$ subsets of the data, known as bins. As the name indicates, the zeroth subset of data, known as bin 0, leaves out data sample $0$ (\textit{i.e.}, contains samples $1, \dots, N-1$). For this bin, we estimate the ITCF $F^{\text{bin}\, 0}$ and the error $C^{\text{bin}\, 0}_{ij}$ using Hatano's error analysis~\cite{hatano1994data}. With this bin's ITCF, we construct a Bayesian prior $\mu^{\text{bin}\, 0}$ by computing $G(q, \omega=0)$ via  \eqref{eq:Gq} and \eqref{eq:chi_q}. Then we use both the Bayesian prior and ITCF together to solve \eqref{eq:GLSEntropy} at a collection of $\alpha$'s. In total, this procedure yields, for each bin, an unparameterized estimate of the DSF $S(q,\omega)_{\text{bin}}$, the uncertainty with respect to $\alpha$ $\text{Var}_\alpha \, S(q,\omega)_{\text{bin}}$, and the MEM posterior weighting function. 

Using the outputs of the leave-one-out-binning procedure, we can estimate the variance in the solution with respect to perturbations in the data $\delta F$ as, 
\begin{align}
    \text{Var}_{\delta F} \left[ S(q,\omega) \right] = \sum_{\text{bins}} (\langle S(q,\omega)\rangle - S(q,\omega)_{\text{bin}})^2.
\end{align}
We can also estimate the variance in the solution with respect to regularization weight $\alpha$ as,
\begin{align}
    \text{Var}_\alpha \left[ S(q,\omega) \right] = \frac{1}{N} \sum_{\text{bins}} \text{Var}_\alpha \left[ S(q,\omega)_{\text{bin}} \right].
\end{align}
We find that the variance in the estimate across bins is of the same order of magnitude as the variance in the estimate across $\alpha$. So both statistical and regularization uncertainty are important. To investigate the correlations between statistical uncertainty and $\alpha$ uncertainty, we looked at the collection of MEM posterior weighting functions, finding that they varied little across bins. We concluded that the $\alpha$ uncertainty was not sensitive to perturbations in $b$. Due to this insensitivity, we added the uncertainties in quadrature.
\begin{align}\label{eq:totalMEMerror}
    \delta S(q,\omega) = \sqrt{\text{Var}_\alpha \left[ S(q,\omega) \right] + \text{Var}_{\delta F} \left[ S(q,\omega) \right]}
\end{align}

\section{Results} \label{sec:results}
In this section, we present the estimated DSFs obtained using the leave-one-out binning procedure discussed in Section~\ref{sec:methods} on the PIMC simulation data described in Table~\ref{tab:simulations}. Our PIMC data contains $M=1000$ independent runs, generated via the open-source ISHTAR code~\cite{ISHTAR}, which is a canonical implementation~\cite{Dornheim_PRB_nk_2021} of the worm algorithm by Boninsegni et al.~\cite{boninsegni1, boninsegni2}. The error analysis on this data must account for the fermion cancellation problem (\textit{i.e.} sign problem); we use Hatano's procedure~\cite{hatano1994data}, which efficiently deals with the underlying ratio estimation. From Table~\ref{tab:simulations}, the typical uncertainty in the ITCF values decreases with $r_s$, where the UEG behavior is well-understood. Both the PIMC raw data for $F(q,\tau)$ and the reconstructed results for $S(q,\omega)$ are freely available in an online repository.

\begin{table}
\caption{Tabulation of the physical parameters describing the path integral Monte Carlo simulations used in this work. We list the quantum degeneracy parameter $\Theta = k_\textnormal{B} T / E_F$, the coupling parameter $\Gamma = \frac{1}{r_s} \left( (k_\text{B} T)^2 + E_F^2 \right)^{-1/2}$, the Fermi wavenumber $k_F$, the plasma frequency $\omega_{p,e}$, the inverse temperature $\beta$ in units of energy, the number of particles in the simulation, the volume of the simulation (\textit{i.e.} length of the periodic boundary cubed), the independent seeds (\textit{i.e.} the number of independent MCMC simulations), and the typical error in the ITCF values $\mathcal{O}[\delta F]$ obtained using Hatano's error analysis~\cite{hatano1994data}. Note that ITCF values are of the order 1, so $\mathcal{O}[\delta F]$ is also indicative of the relative error.}
\vspace{.1cm}
\begin{ruledtabular} \label{tab:simulations}
    \begin{tabular}{|c|c|c|c|c|c|c|c|}
         $r_s$ & 2 & 5 & 10 & 20 & 50 & 100 & 200
         \\ \hline $\Theta$ & 1 & 1 & 1 & 1 & 1 & 1 & 1
         \\ $\Gamma$ & 1.54 & 3.84 & 7.68 & 15.4 & 38.4 & 76.8 & 154
         \\ $k_F$ (1/Bohr) & $9.596 \times 10^{-1}$  & $3.838 \times 10^{-1}$ & $1.919 \times 10^{-1}$ & $9.596 \times 10^{-2}$  & $3.838 \times 10^{-2}$ & $1.919 \times 10^{-2}$ & $9.596 \times 10^{-3}$
         \\ $\omega_{p,e}$ (Hartree) & $6.124 \times 10^{-1}$  & $1.549 \times 10^{-1}$ & $5.477 \times 10^{-2}$ & $1.936 \times 10^{-2}$  & $4.899 \times 10^{-3}$ & $1.732 \times 10^{-3}$ & $6.124 \times 10^{-4}$
         \\ $\beta$ (Hartree) & $2.172 \times 10^{0}$  & $1.358 \times 10^{1}$ & $5.430 \times 10^{1}$ & $2.172 \times 10^{2}$  & $1.357 \times 10^{3}$ & $5.430 \times 10^{3}$ & $2.172 \times 10^{4}$
         \\ particles & 34 & 34 & 34 & 34 & 34 & 34 & 34
         \\ volume ($\text{cm}^3$) & $1.69 \times 10^{-22}$  & $2.64 \times 10^{-21}$ & $2.11 \times 10^{-20}$ & $1.69 \times 10^{-19}$  & $2.64 \times 10^{-18}$ & $2.11 \times 10^{-17}$ & $1.69 \times 10^{-16}$ 
         \\ ind. seeds & 1000 & 1000 & 1000 & 1000 & 1000 & 1000 & 1000
         \\ $\mathcal{O} [\delta F]$ & $10^{-2}$ & $10^{-3}$ & $10^{-4}$ & $10^{-4}$ & $10^{-4}$ & $10^{-4}$ & $10^{-4}$
    \end{tabular}
\end{ruledtabular}
\end{table}

\subsection{Presentation of Dynamics Structure Factor Estimates with Static Approximation}\label{sec:DSFplots}

\begin{figure}[h]
    \centering
    \includegraphics[width=\linewidth]{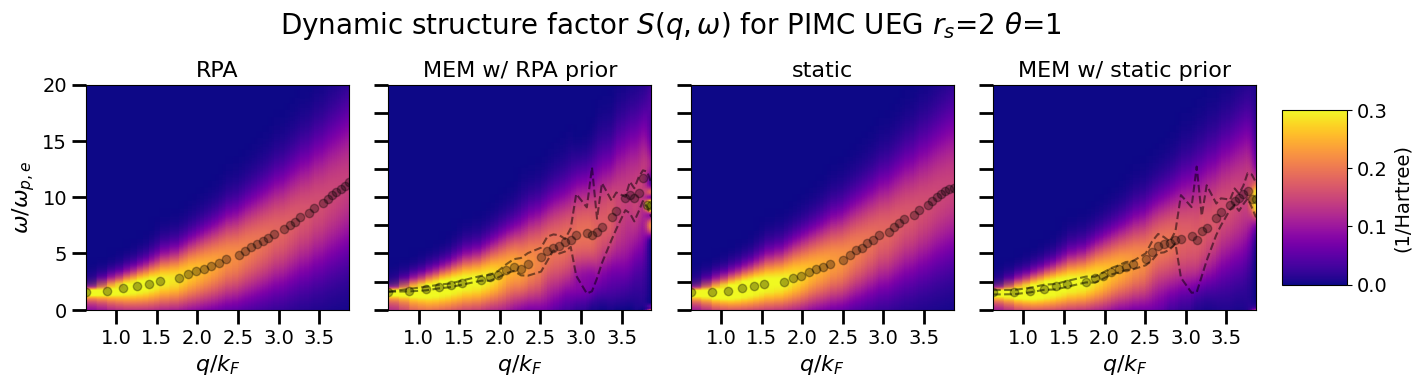}
    \includegraphics[width=\linewidth]{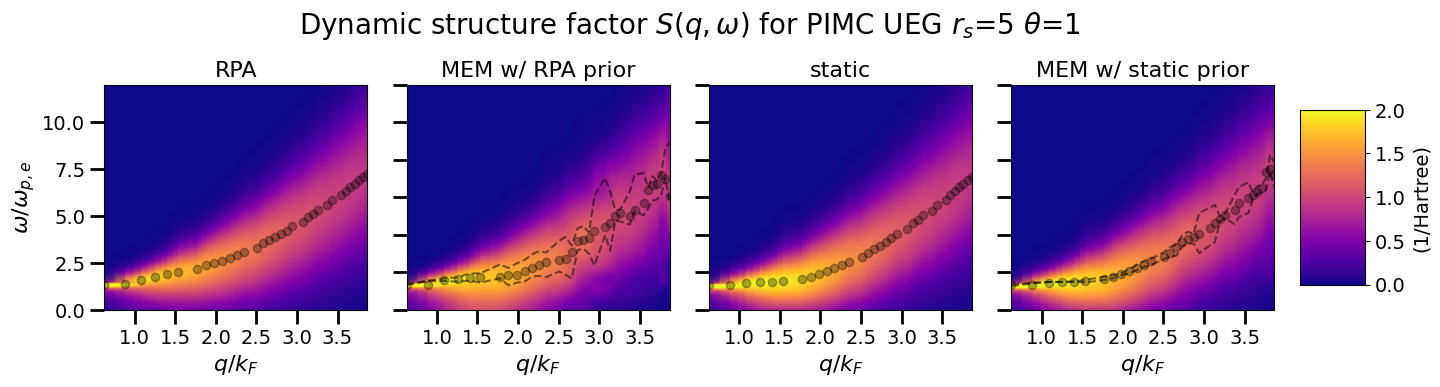}
    \includegraphics[width=\linewidth]{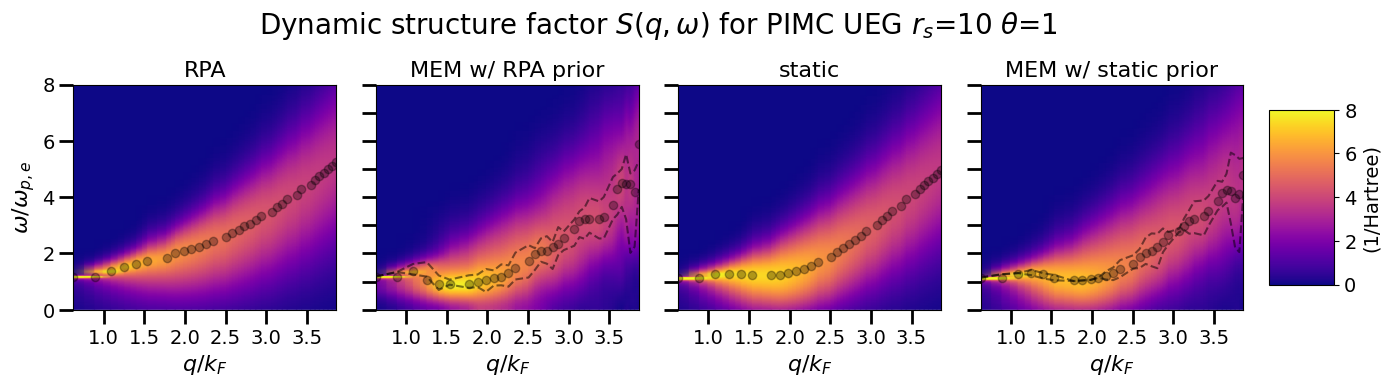}
    \includegraphics[width=\linewidth]{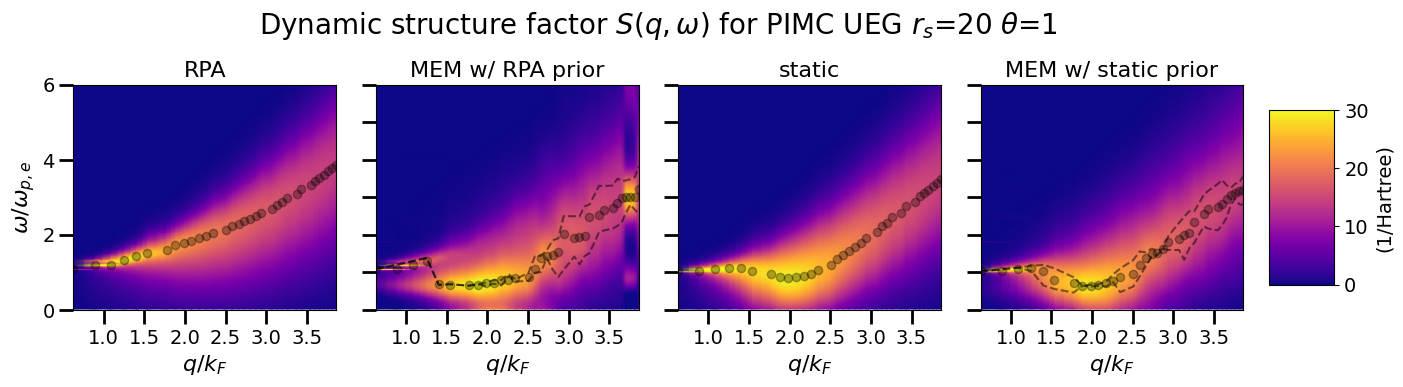}
    \caption{Heat maps of $S(q,\omega)$ at quantum degeneracy parameter $\Theta=1$ and Wigner-Seitz radius $r_s = 2, 5, 10, 20$. The first column from the left contains the RPA DSF given by \eqref{eq:FDT} and \eqref{eq:susceptibility} with $G(q,\omega)=0$. The second column contains the MEM estimate of the DSF using the RPA DSF as the Bayesian prior. The third column contains static LFC DSF, given by \eqref{eq:FDT} and \eqref{eq:susceptibility} with $G(q,\omega)=G(q, \omega=0)$ from \eqref{eq:Gq}. The fourth column contains the MEM estimate of the DSF using the static DSF as the Bayesian prior. All analysis was conducted using leave-one-out binning on the data described in Table~\ref{tab:simulations}.}
    \label{fig:Skw_heatmaps1}
\end{figure}

\begin{figure}[h]
    \includegraphics[width=\linewidth]{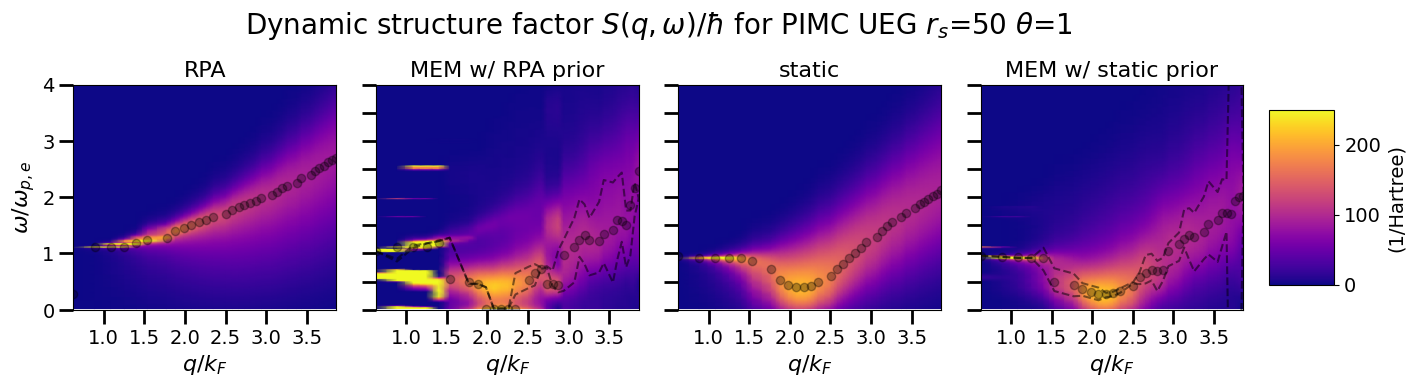}
    \includegraphics[width=\linewidth]{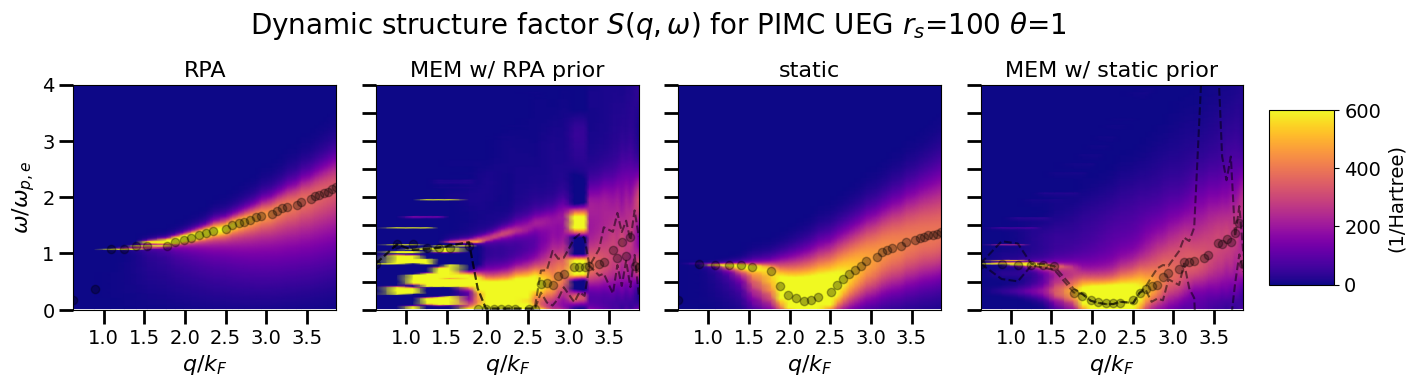}
    \includegraphics[width=\linewidth]{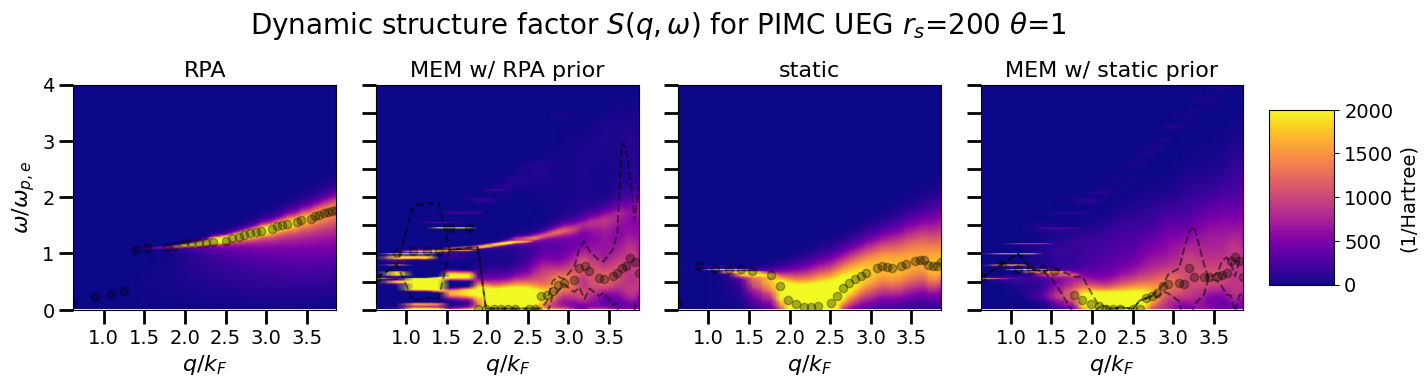}
    \caption{Heat maps of $S(q,\omega)$ at quantum degeneracy parameter $\Theta=1$ and Wigner-Seitz radius $r_s = 50,100, 200$. The first column from the left contains the RPA DSF given by \eqref{eq:FDT} and \eqref{eq:susceptibility} with $G(q,\omega)=0$. The second column contains the MEM estimate of the DSF using the RPA DSF as the Bayesian prior. The third column contains static LFC DSF, given by \eqref{eq:FDT} and \eqref{eq:susceptibility} with $G(q,\omega)=G(q, \omega=0)$ from \eqref{eq:Gq}. The fourth column contains the MEM estimate of the DSF using the static DSF as the Bayesian prior. All analysis was conducted using leave-one-out binning on the data described in Table~\ref{tab:simulations}.}
    \label{fig:Skw_heatmaps2}
\end{figure}

First, we compare our RPA and static LFC Bayesian priors. Heatmaps of the Bayesian priors and the associated MEM results are given in Figure~\ref{fig:Skw_heatmaps1} and Figure~\ref{fig:Skw_heatmaps2}.
Let us start with a qualitative discussion of the underlying physics. In the limit of $q\to0$, the system is probed on large length scales that feature the collective plasmon excitation, which manifests as a nascent delta peak around $\omega_{p,e}$ in both priors and the MEM. 
For $q\gg k_\textnormal{F}$, the DSF attains the single particle limit with the usual parabolic dispersion relation; this too is captured by all the models and the MEM.
The physically most interesting regime is given by $q\sim k_\textnormal{F}$, where $\lambda\sim r_s$. In this regime, the DSF is dominated by electronic exchange--correlation effects, which are neglected in the RPA and approximated within the static approximation.
From Figure~\ref{fig:Skw_heatmaps1} and Figure~\ref{fig:Skw_heatmaps2}, we see that using the RPA as a Bayesian prior produces DSF estimates that are less smooth than the static LFC, though at $r_s=2$ the difference between the Bayesian priors is marginal. We observe the deterioration of the RPA's results at large $r_s$. this is to be expected since strong coupling $\Gamma \gg 1$ violates the assumptions of the RPA. Note that $r_s$ values from $2$ to $20$ have corresponding coupling parameters $\Gamma = 1.54$ to $15.4$. At $r_s = 50, 100,200$, corresponding to a coupling parameter $\Gamma = 38.4$, $76.8$, $154$, using the RPA as the Bayesian prior produces extremely low quality results. 

Examining the behavior of the DSFs in Figure~\ref{fig:Skw_heatmaps1} and Figure~\ref{fig:Skw_heatmaps2}, we find that the dispersion relation of the MEM estimates differ from the RPA. MEM reconstructions using either Bayesian prior exhibit a roton feature, which indicates that this is a genuine feature of the data. These results can be understood via the pair alignment argument discussed in Section~\ref{sec:DSFmodels} and originally proposed in Ref.~\cite{Dornheim_Nature_2022}. 
In the case of the RPA as the Bayesian prior, which neglects exchange-correlation effects, the roton feature is created completely by minimizing the difference from the PIMC input data for $F(q,\tau)$. In the case of the static LFC as the Bayesian prior, the existing roton feature in the prior is enhanced by the PIMC input. This is evidence that the MEM fidelity term is including the dynamic exchange-correlation effects, contained in the ITCF, into the DSF estimate. We note that any significant deviations between the static approximation and the full MEM result further attest to the importance of dynamic exchange--correlation effects in this regime~\cite{Dornheim_PRB_2024}.

Again examining the behavior of the DSFs in Figure~\ref{fig:Skw_heatmaps1} and Figure~\ref{fig:Skw_heatmaps2}, we also find that outside the $q \rightarrow 0$ limit, the MEM DSF estimates have their largest values at intermediate wavenumbers $2 < q/ k_F < 2.5$. This is a qualitative disagreement with the RPA model, which has a peak height that only reduces with $q$ and a peak width that only broadens. This same qualitative disagreement persists in the $\omega \rightarrow 0$ limit. As we will see in Section~\ref{sec:DSFsumrules} Figure~\ref{fig:wneg1}, the system's response to a static perturbation $\chi(q,\omega=0)$ has a sharper peak than the RPA in this region; this qualitative disagreement grows with coupling, indicating that the exclusion of exchange correlation effects is the root cause.

\begin{figure}[h]
    \includegraphics[width=0.5\linewidth]{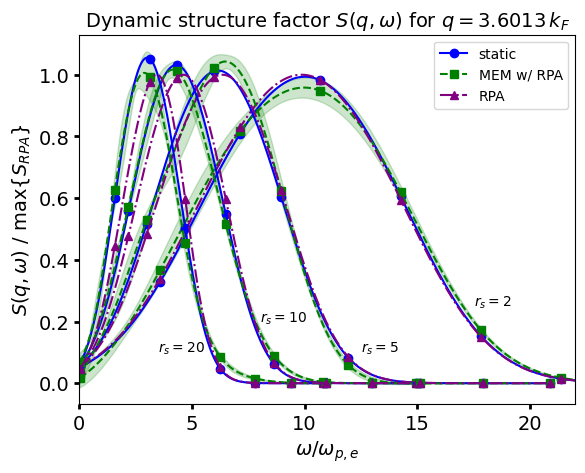}%
    \includegraphics[width=0.5\linewidth]{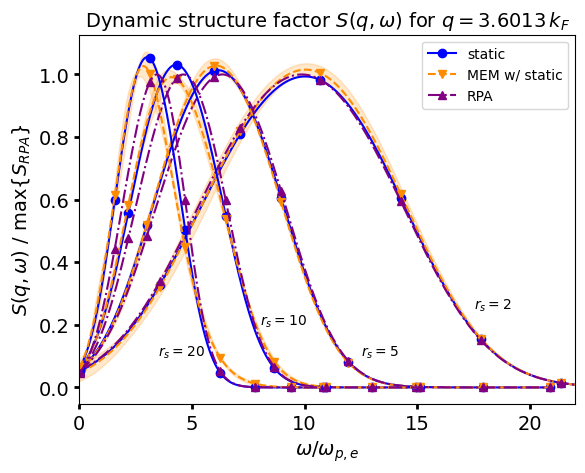}
    \includegraphics[width=0.5\linewidth]{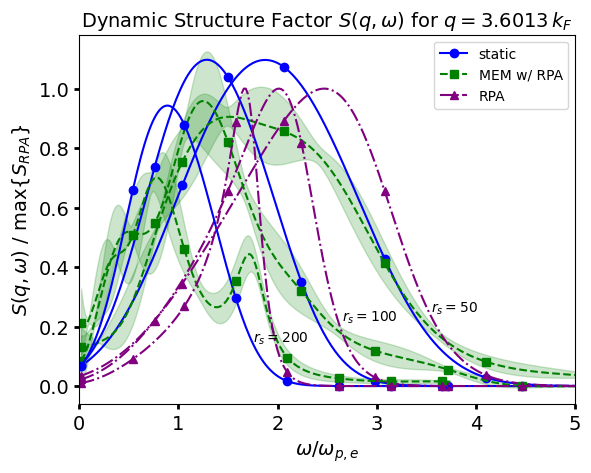}%
    \includegraphics[width=0.5\linewidth]{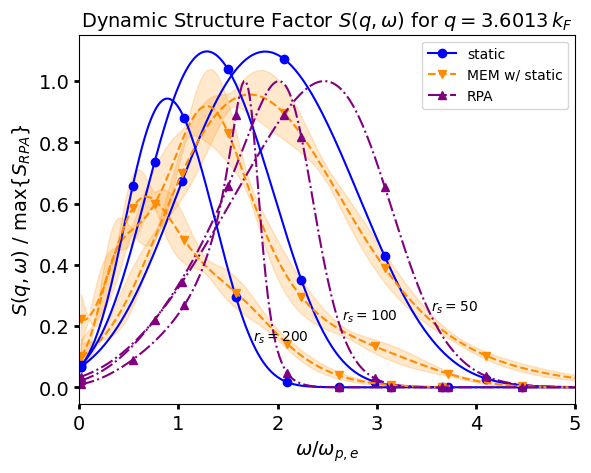}
    \caption{Plots of $S(q,\omega)$ for fixed $q= 3.6013\, k_F$, where $k_D$ is the Debye wavenumber. The legend indicates which colors/symbols correspond to the RPA DSF (purple, \ding{115}), the static LFC DSF (blue, \ding{108}), the MEM fit with RPA Bayesian prior (green, \ding{110}), and the MEM fit with static LFC Bayesian prior (darkorange, \ding{116}). In the left column, MEM fits are conducted using the RPA as the the Bayesian prior. In the right column, MEM fits are conducted using the static LFC as the Bayesian prior. The top row contains fits for $r_s = 2,5,10,20$ and the bottom row contains fits for $r_s = 50,100,200$. In each graph, text $r_s = ...$ have been placed to indicate which curve corresponds to which $r_s$. All $S(q,\omega)$ curves are normalized by the RPA DSF's maximum value. For $r_s = 2,5,10,20$ all curves and error are similar. For $r_s = 50,100,200$ the static LFC and the RPA DSFs differ, leading to different MEM estimates. However, both estimates exhibit a double peak structure.}
    \label{fig:DSF-kslice}
\end{figure}

\begin{figure}[h!]
    \centering
    \includegraphics[width=.5\linewidth]{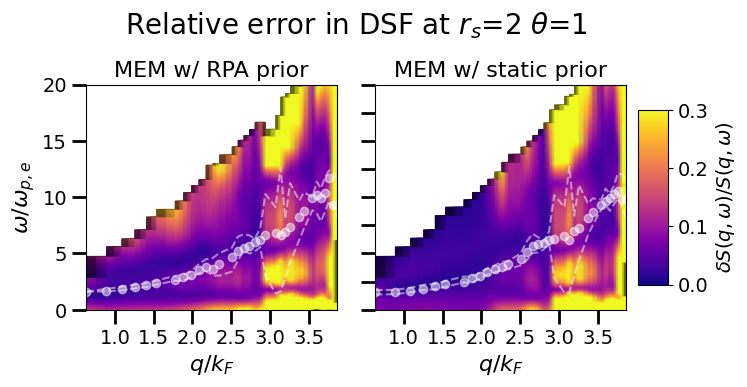}%
    \includegraphics[width=.5\linewidth]{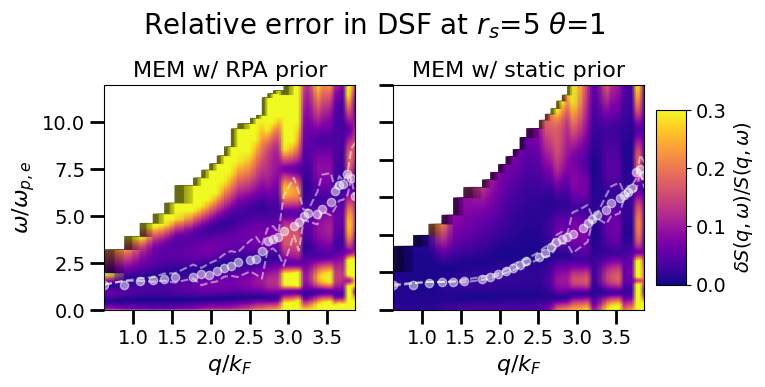}
    \includegraphics[width=.5\linewidth]{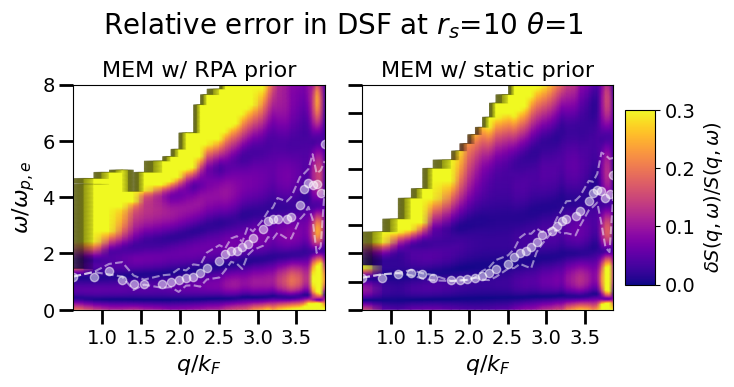}%
    \includegraphics[width=.5\linewidth]{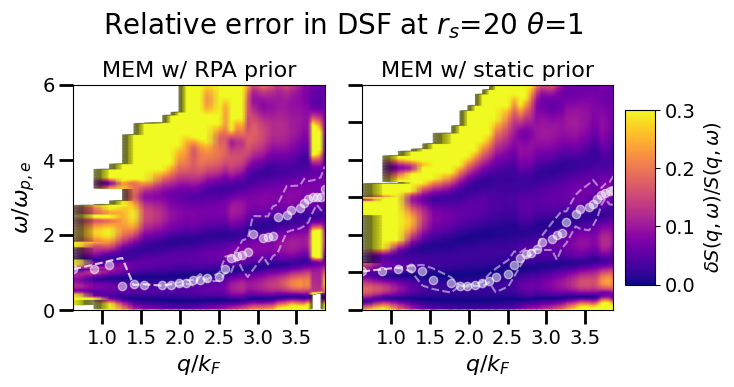}
    \caption{Heat maps of the relative error in the MEM estimate $\delta S(q,\omega) / S(q,\omega)$, plot titles indicate the corresponding $r_s$ and $\Theta$ values, as well as which Bayesian prior was used. Error $\delta S(q,\omega)$ is quantified using \eqref{eq:totalMEMerror}. As $S(q,\omega) \rightarrow 0$ goes to zero, $\delta S(q,\omega) / S(q,\omega)$ becomes unstable. Thus, for visual clarity we do not plot results where $S(q,\omega)<10^{-3}$; this manifests as a white region. Using the RPA model results in more pronounced ringing in the uncertainty than using the static LFC.} 
    \label{fig:Err_heatmaps1}
\end{figure}

\begin{figure}[h!]
    \includegraphics[width=0.5\linewidth]{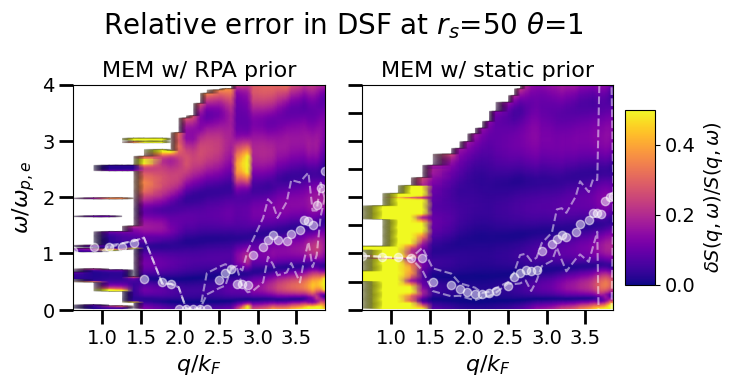}%
    \includegraphics[width=0.5\linewidth]{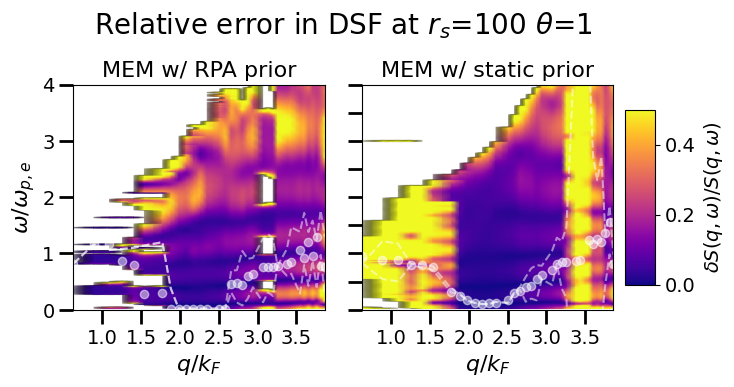}
    \includegraphics[width=0.5\linewidth]{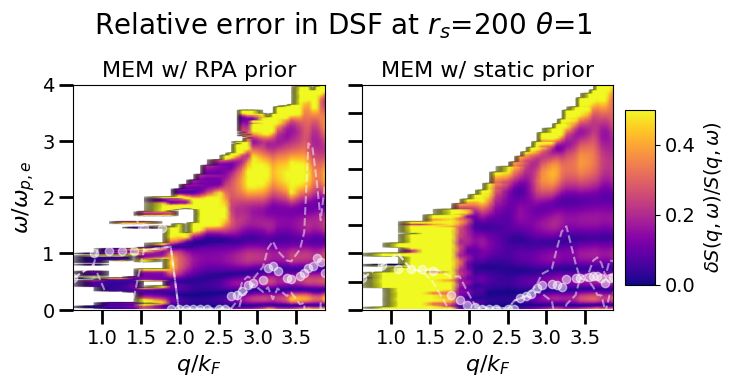}
    \caption{Heat maps of the relative error in the MEM estimate $\delta S(q,\omega) / S(q,\omega)$, plot titles indicate the corresponding $r_s$ and $\Theta$ values, as well as which Bayesian prior was used. Error $\delta S(q,\omega)$ is quantified using \eqref{eq:totalMEMerror}. As $S(q,\omega) \rightarrow 0$ goes to zero $\delta S(q,\omega) / S(q,\omega)$ becomes unstable. Thus, for visual clarity we do not plot results where $S(q,\omega)<10^{-3}$; this manifests as a white region. Using the RPA model results in a break down of the MEM estimate at small $q$.}
    \label{fig:Err_heatmaps2}
\end{figure}

Next, we examine in more detail a $q$ cross section from the $r_s=100$ heatmap in Figure~\ref{fig:Skw_heatmaps1} and Figure~\ref{fig:Skw_heatmaps2}. We plot the $q=3.546 k_F$ cross section in Figure~\ref{fig:DSF-kslice}. For $r_s = 2,5,10,20$, the static LFC and RPA DSFs are similar; unsurprisingly the resulting estimates are therefore insensitive to the prior. For $r_s = 50,100,200$ the static LFC and RPA DSFs differ substantially; this leads to MEM estimates, which also differ, although significantly less than the underlying priors. Taken as an individual cross section at large $q$ it is difficult to assess which Bayesian prior is best. However, later plots, such as the uncertainty heatmaps (Figures~\ref{fig:Err_heatmaps1} and Figure~\ref{fig:Err_heatmaps2}), will show that using the static LFC as the Bayesian prior yields the most stable behavior across $q$, particularly at small $q$. From these $q$ cross sections, we can see interesting physics. First, at large $r_s$ (\textit{i.e.}, large coupling) both reconstructions observe an incipient double peak, which has also been observed in classical one-component strongly coupled Coulomb systems (SCCS)~\cite{Korolov_CPP_2015, Choi_PRE_2019} as well as in the UEG~\cite{Takada_PRL_2002, Takada_PRB_2016, Filinov_PRB_2023}. 

\begin{figure}[h]
    \includegraphics[width=0.5\linewidth]{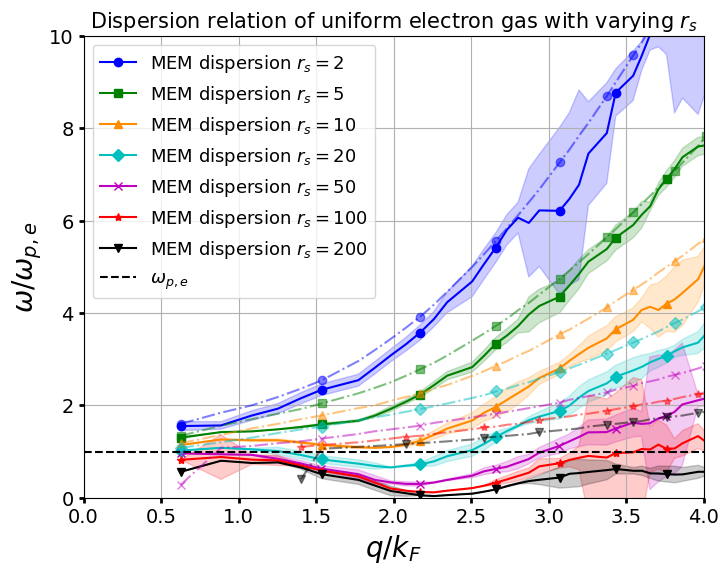}
    \caption{Plot of the dispersion relation for $S(q,\omega)$ with varying $r_s$. The solid lines indicate our estimate using MEM with the static LFC as the Bayesian prior. For comparison we include the RPA's dispersion relation as dashed lines. We set each dispersion curve's color and symbol based on the corresponding $r_s$ value. For small $r_s$ (\textit{i.e.}, weak coupling), our MEM reconstructions agree with the RPA dispersion curve within the error \eqref{eq:totalMEMerror}. For large $r_s$ a clear roton-type feature emerges, which is substantiated by the narrow error bands.}
    \label{fig:DSF-dispersion}
\end{figure}

We combine the dispersion relations and their uncertainty across $r_s$ from the heatmaps in Figure~\ref{fig:Skw_heatmaps1} and Figure~\ref{fig:Skw_heatmaps2} into Figure~\ref{fig:DSF-dispersion}, limiting ourselves to the MEM results using the static LFC prior; as a reference, RPA dispersions (dash-dotted curves) are also shown. Examining the behavior, at large $r_s$, the MEM dispersion relation differs substantially from the RPA. In particular, the peak $\omega(q)$ of the dynamic structure factor $S(q,\omega)$ exhibits a nonmonotonic wave number dependence with an inflection point $q_I$, such that $\partial_q \omega(q) < 0$ for $q<q_I$ and $\partial_q \omega(q) > 0$ for $q > q_I$. 
These curves establish that exchange-correlation effects grow with $r_s$, as expected. 

\begin{figure}[h]
    \includegraphics[width=0.492\linewidth]{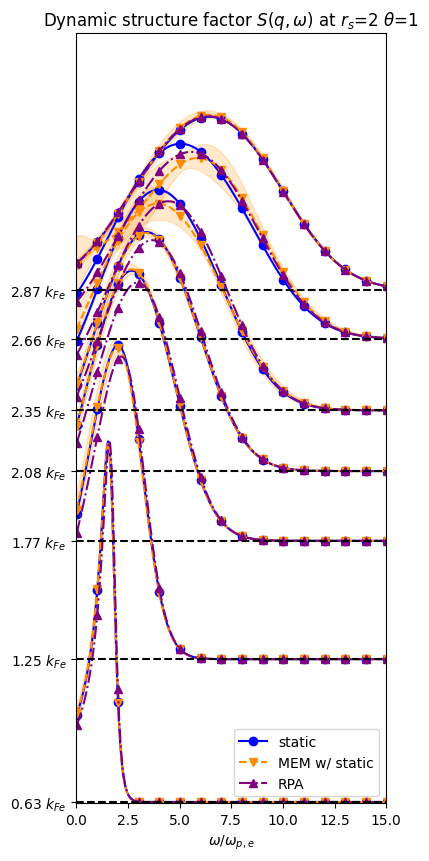}%
    \includegraphics[width=0.5\linewidth]{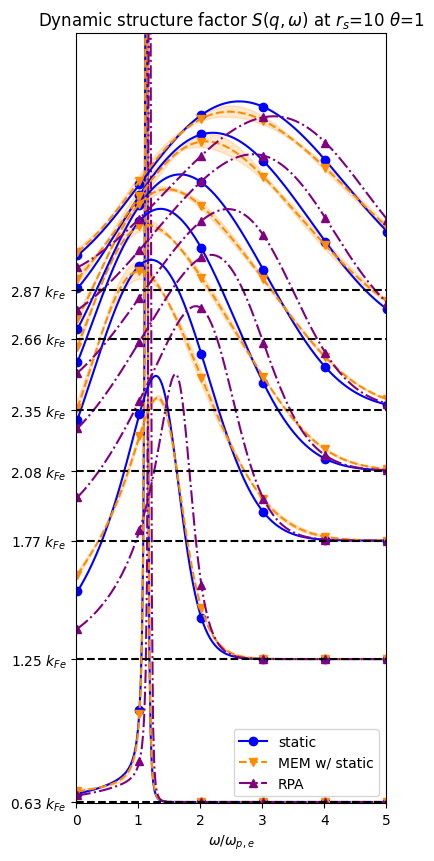}    
    \caption{Stacked $q$ cross sections for $S(q,\omega)$, $\omega$ is varied along the x-axis and dashed black lines indicate the zero position for each corresponding $q$ cross section. The plot titles indicate the corresponding $r_s$ and $\Theta$ values. The shaded dark orange regions indicate the error of the MEM estimate computed via \eqref{eq:totalMEMerror}. Relative to the RPA DSF (RPA) and static local field correction DSF (static), the MEM's estimate of the DSF (MEM w/ static) tends to reduce the dispersion peak's width and shift the peak to smaller frequencies $\omega$.}
    \label{fig:DSF-stacked-kslices1}
\end{figure}

\begin{figure}[h]
    \includegraphics[width=0.495\linewidth]{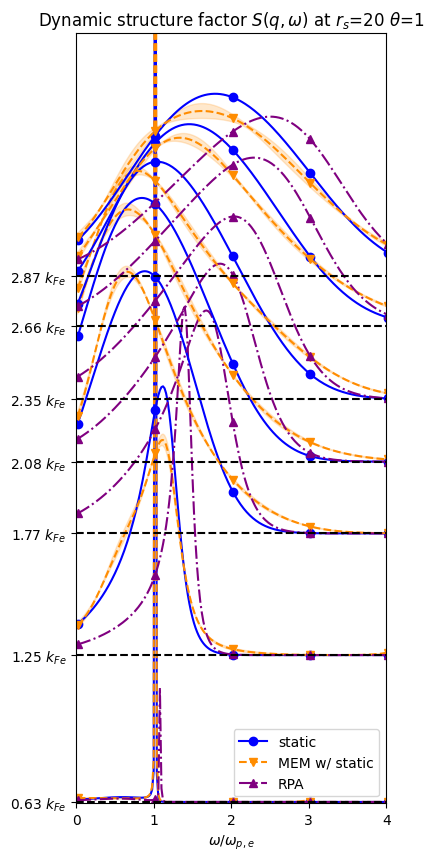}%
    \includegraphics[width=0.5\linewidth]{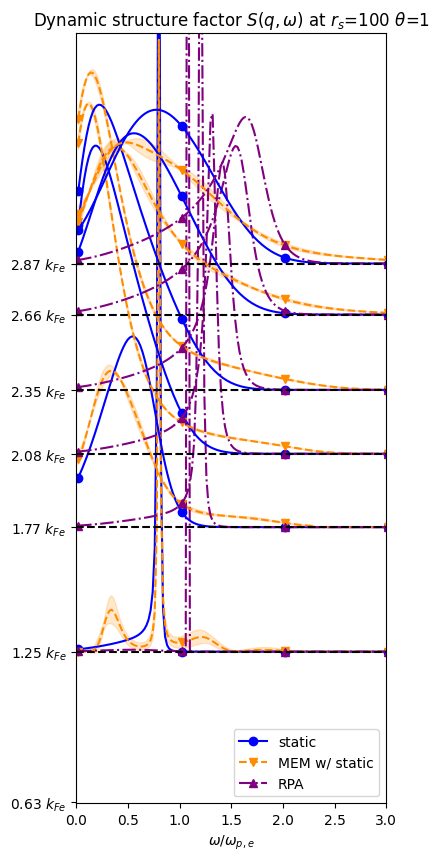}
    \caption{Stacked $q$ cross sections for $S(q,\omega)$, $\omega$ is varied along the x-axis and dashed black lines indicate the zero position for each corresponding $q$ cross section. The plot titles indicate the corresponding $r_s$ and $\Theta$ values. The shaded dark orange regions indicate the error of the MEM estimate computed via \eqref{eq:totalMEMerror}. We neglect the $q = 0.63 k_F$ curves in the right plot ($r_s=100, \Theta = 1$) because the plasmon peak is too narrow to be resolved at our $\omega$ discretization. Relative to the RPA DSF (RPA) and static local field correction DSF (static), the MEM's estimate of the DSF (MEM w/ static) tends to reduce the dispersion peak's width and shift the peak to smaller frequencies $\omega$.}
    \label{fig:DSF-stacked-kslices2}
\end{figure}

Next, we plot DSF $q$ cross sections, which vary the $q$ value rather than the $r_s$ value in Figure~\ref{fig:DSF-stacked-kslices1} and Figure~\ref{fig:DSF-stacked-kslices2} and examine how the MEM estimates differ from the RPA and static approximation across $q$. As $q$ decreases, the dispersion peak becomes narrow and the MEM estimate converges to the Bayesian prior. In particular, for $r_s=100$ (Figure~\ref{fig:DSF-stacked-kslices2} right), the dispersion peak is too narrow for our $\omega$ discretization for the smallest wavenumbers. This is a natural occurrence, in the $q \rightarrow 0$ the DSF must converge to a Dirac delta located at $\omega = \omega_{p,e}$ $\delta(\omega - \omega_{p,e})$ to satisfy the continuity equation~\cite{chuna2024conservative, atwal2002fullyconserving}. The MEM DSF reduces the dispersion peak's width and shift the peak to smaller frequencies. These alterations match what dynamic local field corrections produce~\cite{Dornheim_Nature_2022}, thus Figure~\ref{fig:DSF-stacked-kslices1} and Figure~\ref{fig:DSF-stacked-kslices2} provide evidence that using the static approximation as the Bayesian prior focuses the MEM on extracting the dynamic physics we know is missing from the static model. Additionally, we observe that these dynamic effects become more prominent for $r_s > 10$. We take this as evidence that the dynamic dependence of the exchange-correlation energy becomes more important with increasing coupling.

Finally, we examine our uncertainty quantification. Heatmaps of the relative error associated to these plots are given in Figure~\ref{fig:Err_heatmaps1} and Figure~\ref{fig:Err_heatmaps2}. From these heatmaps, we see that using the RPA as a Bayesian prior produces DSF estimates that are less certain. Across all $r_s$, the uncertainty in the dispersion relation grows with $q$ after the dispersion relation's inflection point. As the dispersion peaks move to large $\omega$ the fewer $\tau$ points inform the fit, resulting in greater uncertainty. We also observe a systematic ringing as the MEM solutions flatten this matches previous investigations of entropic regularizers~\cite{Fischer2018SmoothedBRM}. By inspection, the solution does not contain this ringing. We take this as evidence that the Bayesian posterior is weighting the solutions appropriately. Another feature of these heat maps is that, as $r_s$ grows, the uncertainty in small $q$ grows. This occurs because, in this limit, the DSF is very narrow Lorentz curve, converging to a Dirac delta function. Meaning that many DSF values are close to numerical zero and any finite error yields a large relative error. For our application, convergence to a Dirac delta peak is desired and we further investigate the convergence in Section~\ref{sec:DSFsumrules}.

\subsection{Comparison to stochastic approach} \label{sec:stochasticcomparison}
As mentioned in the introduction section, a stochastic sampling algorithm has already been employed for $r_s=10, \Theta=1$ UEG PIMC data in Refs.~\cite{dornheim2018dynamiclocalfieldcorrection,Dornheim_PRE_2020}. This approach is specific to the UEG, for which it automatically fulfills a number of exact constraints and sum rules. By comparison, the MEM has no guarantee of satisfying sum rules. We compare our DSF estimates to the stochastic estimates and expect that the general shape of the stochastic estimate is physical and provides information through comparison even if the stochastic estimate is not the exact inversion either. We find good qualitative agreement.
\begin{figure}[h!]
    \includegraphics[width=0.8\linewidth]{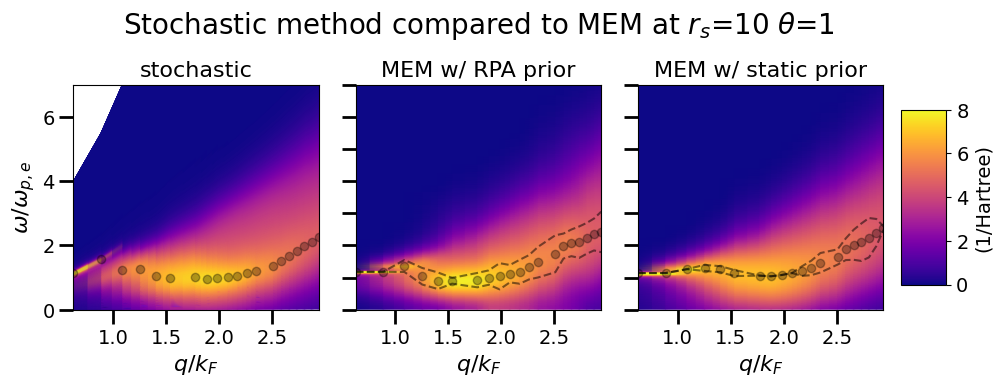}
    \caption{Heat maps of the DSF $S(q,\omega)$, which compare  our MEM inversions to the results of Dornheim et al.~\cite{dornheim2018dynamiclocalfieldcorrection}, which were computed with a stochastic algorithm~\cite{groth2019stochasticsampling}.}
    \label{fig:stochasticcomparison}
\end{figure}

\subsection{Sum rules of the DSF} \label{sec:DSFsumrules}
A crucial assessment for a DSF is the fulfillment of relevant sum rules~\cite{selchow1999dielectric, Choi_PRE_2019}. Each sum rule establishes whether certain physics is present in the DSF model. We consider the following sum rules 
\begin{align}
    \langle \omega^{-1} \rangle &= - \frac{\chi(q,0)}{2n} = \frac{1}{2} \int_0^\beta d\tau \, F(q,\tau) \label{eq:inverse}\,,
    \\ \langle \omega^{0} \rangle &= S(q) = F(q,\tau=0) \label{eq:Sq}\,,
    \\ \langle \omega \rangle &= E_F \left(\frac{q}{k_F}\right)^2 = - \left. \frac{\partial}{\partial \tau} F(q,\tau) \right\rvert_{\tau=0} \label{eq:fsum}\,,
\end{align}
where the average is defined $\langle \omega^{k} \rangle  \equiv \int d \omega \, \omega^k S(q,\omega)$. Consider \eqref{eq:inverse}, the first equality, presented in~\cite{Vitali_PRB_2010, dornheim2024QuantumDelocalization}, tells us that the static density response function is sensitive to when the dispersion peak is located at small $\omega$. So for systems with a roton feature we expect larger density response to intermediate wavenumbers $q$. The second equality, presented e.g.~in~\cite{dornheim2024QuantumDelocalization}, tells us that if the integral of the ITCF is small then there is little density response. This integral is strongly dependent on the decay rate of the ITCF which by definition expresses the quantum delocalization in the system. Hence, quantum delocalization reduces density response. Equation \eqref{eq:Sq} defines the static structure factor $S(q)$, which represents the average spatial correlations in the system~\cite{hansen2013theory3rdEd, zhang2016Sq}. Finally, Equation \eqref{eq:fsum} defines the F-sum rule, which tells us whether the continuity equation $\textbf{k} \cdot (n_0 \delta\textbf{u}) = \omega \delta n$ is satisfied~\cite{chuna2024conservative, boon1991molecular, hansen2013theory3rdEd}.

As stated, the MEM is not guaranteed to satisfy any of the sum rules \eqref{eq:inverse}, \eqref{eq:Sq}, or \eqref{eq:fsum}. However, it is known that moments of the DSF can be related to the derivatives of the ITCF at $\tau=0$~\cite{sandvik1998numerical, Dornheim_moments_2023}
\begin{align}\label{eq:moments}
    (-1)^n \left. \left( \frac{\partial}{\partial \tau}\right)^n F(q,\tau) \right\rvert_{\tau=0} = \int^{\infty}_{-\infty} d \omega \, \omega^n S(q,\omega).
\end{align}
Consider \eqref{eq:moments} as a finite difference expression, since there is no data for $\tau<0$ estimating the $n^\text{th}$ derivative (and hence the $n^\text{th}$ moment) requires $n+1$ points from $\tau=0$. Further, consider a Taylor expansion of $F(q,\tau)$ 
\begin{align}\label{eq:Ftau-taylorexpansion}
    F(q,\tau) =  \sum_n \frac{1}{n!} \left( \left. \left( \frac{\partial}{\partial \tau}\right)^n F(q,\tau) \right\rvert_{\tau=0}\right) \tau^n,
\end{align}
which shows that when $\tau$ is small the higher moments are exponentially suppressed. From \eqref{eq:moments} and \eqref{eq:Ftau-taylorexpansion}, we infer that information for lower moments is contained in the ITCF behavior near $\tau=0$ and information for higher moments is contained at larger $\tau$. The main takeaway is that, since the PIMC ITCF data satisfies the sum rules, this information is contained in the structure of the data. Thus, minimizing the generalized least squares criteria \eqref{eq:GLSEntropy} (\textit{i.e.} selecting a DSF that matches the data) implicitly enforces the correct sum rule behavior. From this perspective, satisfying known sum rules is a non-trivial test of our analytic continuation algorithm's ability to extract information from the data. Furthermore, higher power moments are suppressed by a factor $1/n!$, so each successively higher moment is a more challenging test than the previous. We note that for large $r_s$, the dispersion peak narrows in the $q \rightarrow 0$ limit and numerical integration fails, leading to false zeros for the static and RPA $S(q)$. This is an unavoidable occurrence, since the DSF is converging to a Dirac delta. 

\begin{figure}[h!]
    \centering
    \includegraphics[width=0.62\linewidth]{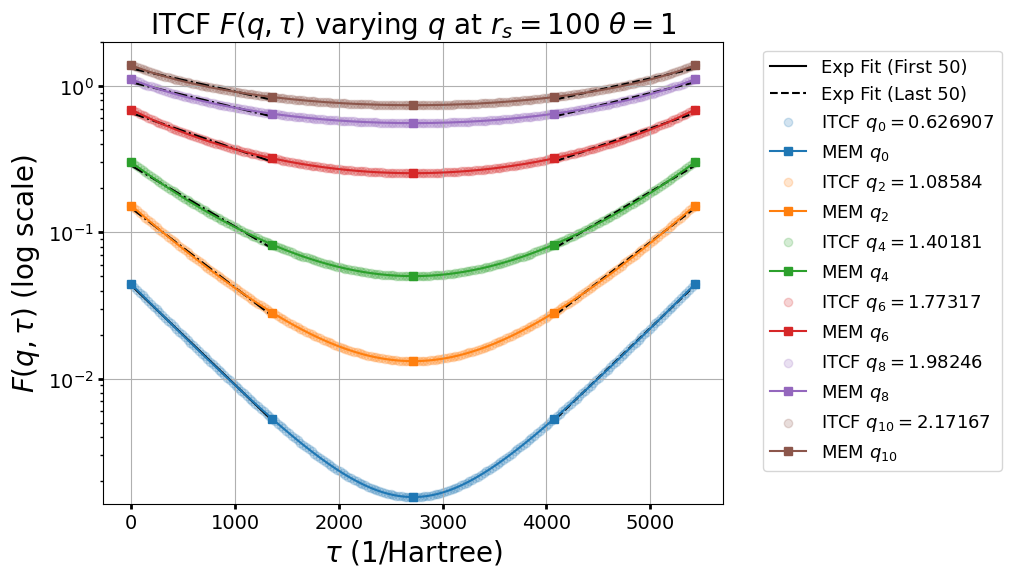}%
    \includegraphics[width=0.38\linewidth]{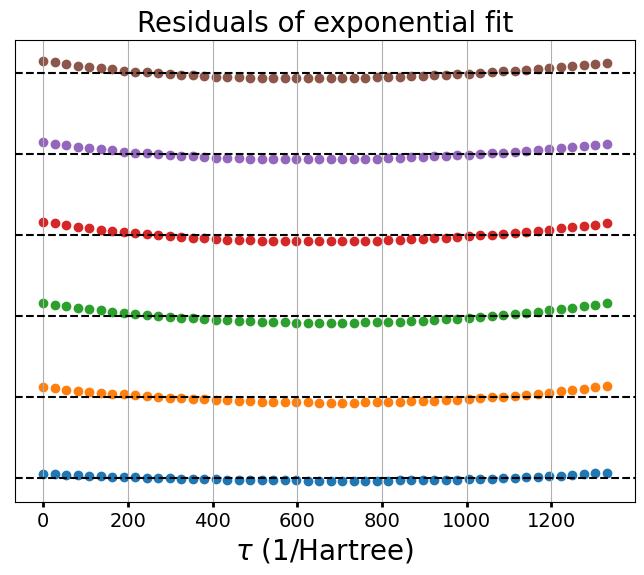}
    \caption{The left plot is of the Laplace transform of the MEM DSF estimate compared with the ITCF data for $r_s=100$ $\Theta=1$; these two curves are indistinguishable. We also include exponential fits on the first and last $50$ data points of the ITCF and show in the right plot the residuals of the single exponential fit on the first $50$ points. As $q \rightarrow 0$ the systematic errors in the residuals vanish. This implies that the ITCF is best explained by a single Dirac delta function DSF and explains the integration convergence issues for large $r_s$ at small $q$ in Figure~\ref{fig:Sk}, Figure~\ref{fig:fsum}, and Figure~\ref{fig:wneg1}.}
    \label{fig:ITCFcomparison}
\end{figure}

For $r_s=100$ and $\Theta=1$, we visualize the agreement between the ITCF data and the Laplace transform of the DSF estimate; the Laplace transformed DSF is indistinguishable from the ITCF (see Figure~\ref{fig:ITCFcomparison} left). Based on \eqref{eq:moments} and \eqref{eq:Ftau-taylorexpansion} as well as the discussion around these equations, the numerical agreement between the Laplace transform of the DSF and the ITCF indicates we should expect that the reconstruction will satisfy sum rules, even though the MEM criteria does not enforce them explicitly. We also include in Figure~\ref{fig:ITCFcomparison} an exponential fit and residuals of the exponential fit. These residuals show that, as $q/k_F \rightarrow 0$, a single exponential fit does not have systematic bias. This implies that the DSF is indeed converging to a Dirac delta.  

\begin{figure}[h!]
    \centering
    \includegraphics[width=0.33\linewidth]{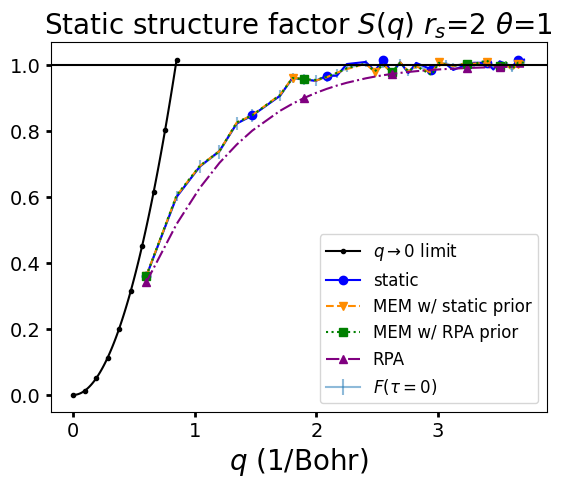}%
    \includegraphics[width=0.33\linewidth]{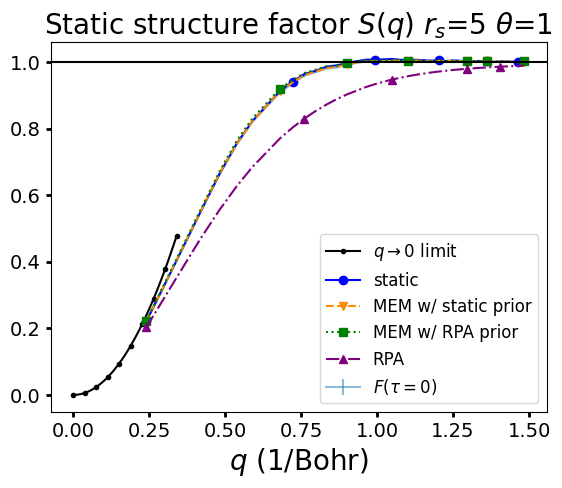}%
    \includegraphics[width=0.33\linewidth]{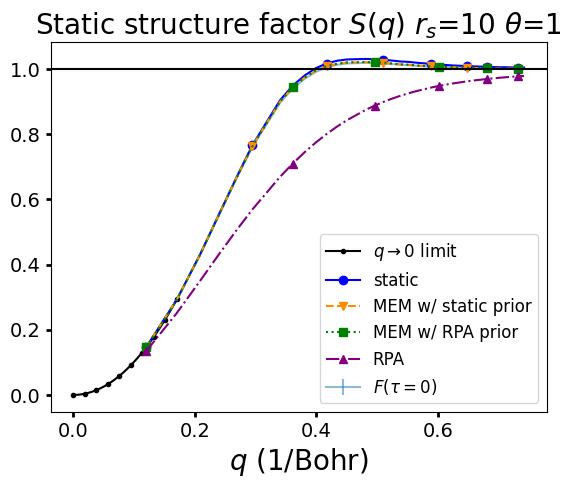}
    \includegraphics[width=0.33\linewidth]{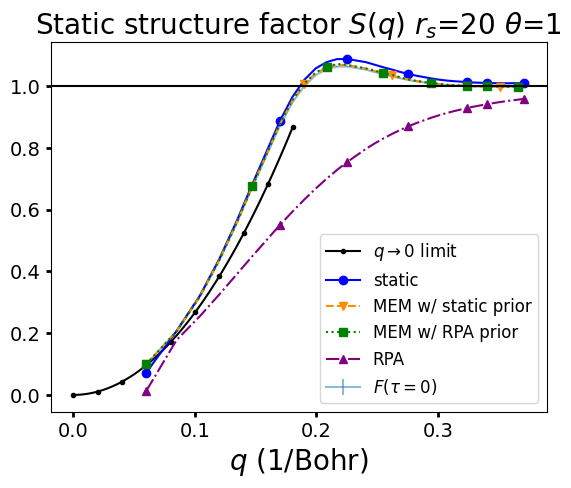}%
    \includegraphics[width=0.33\linewidth]{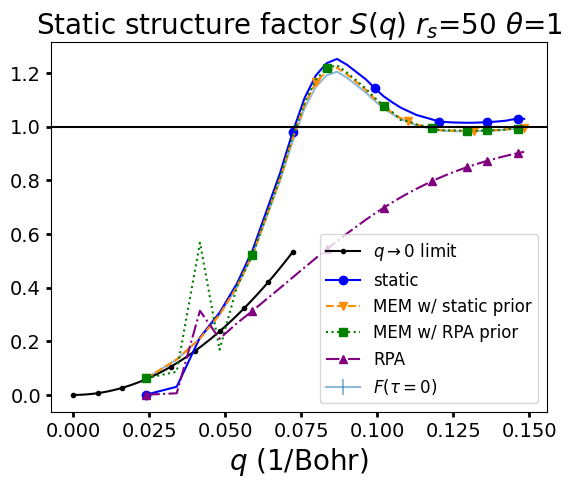}%
    \includegraphics[width=0.33\linewidth]{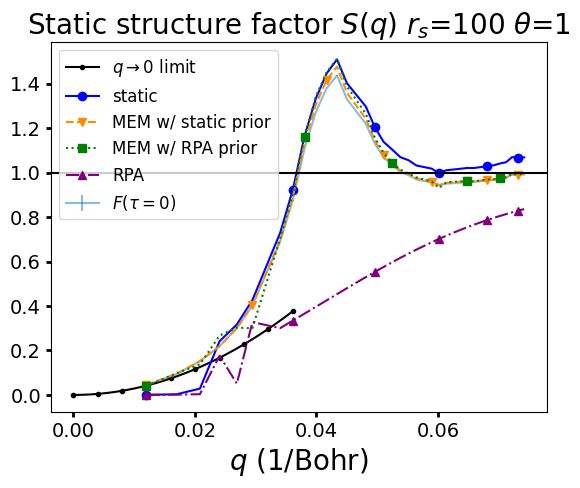}
    \caption{We plot the static structure factor \eqref{eq:Sq} for the various computed DSF functions. The lack of smoothness in the $r_s=2$ arises from the error in ITCF data, this is also the only plot where the error in $F(\tau=0)$ is visible. For large $r_s$ ($r_s > 20$), the dispersion peak is so narrow in the $q \rightarrow 0$ limit that numerical integration fails. This is an unavoidable occurrence, since the DSF is converging to a Dirac delta.}
    \label{fig:Sk}
\end{figure}

Let us next investigate the performance of the MEM solutions wrt.~the sum rules, Eqs.~(\ref{eq:inverse}-\ref{eq:fsum}).
First, we investigate the static structure factor \eqref{eq:Sq}. Plots are given in Figure~\ref{fig:Sk} which demonstrate that our MEM results closely follow the static approximation, but match the PIMC data $F(\tau=0)$ better for large $r_s$. The MEM with static LFC as the Bayesian prior matches better than the MEM with the RPA as the Bayesian prior. Overall the MEM estimates have a smoother convergence to $S(q)=1$ in the $q \rightarrow \infty$ limit and to $S(q)= q^2/k_{D,e}^2$ in the $q \rightarrow 0$ limit than either default model. 

\begin{figure}[h!]
    \centering
    \includegraphics[width=0.33\linewidth]{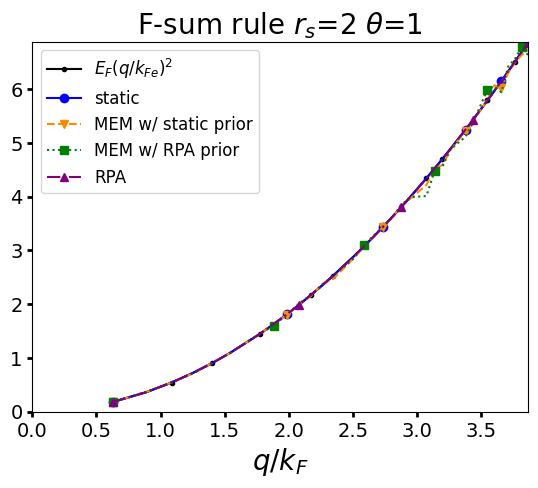}%
    \includegraphics[width=0.33\linewidth]{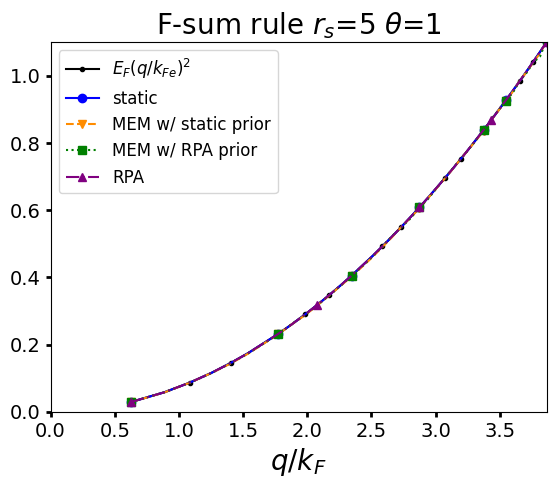}%
    \includegraphics[width=0.33\linewidth]{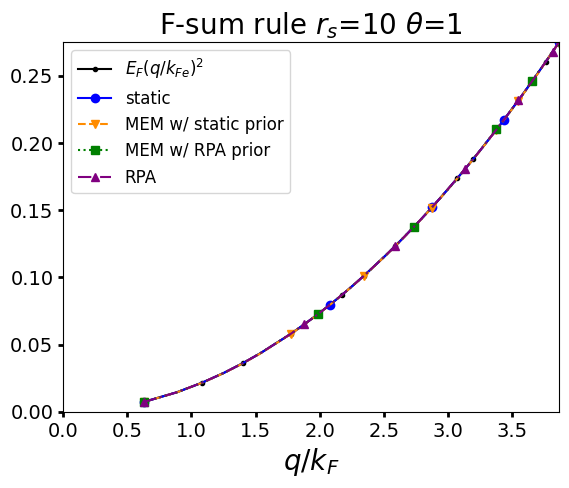}
    \includegraphics[width=0.33\linewidth]{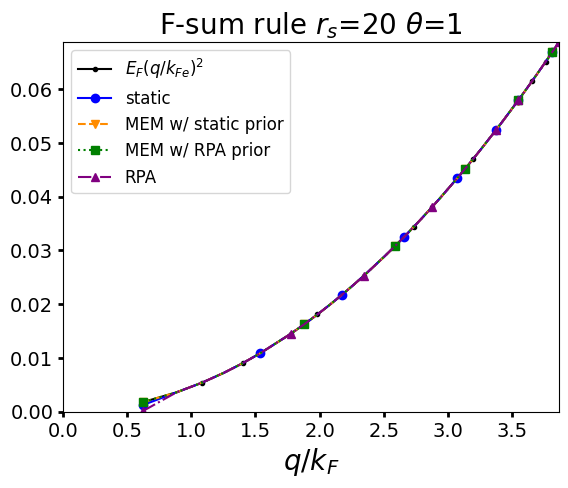}%
    \includegraphics[width=0.33\linewidth]{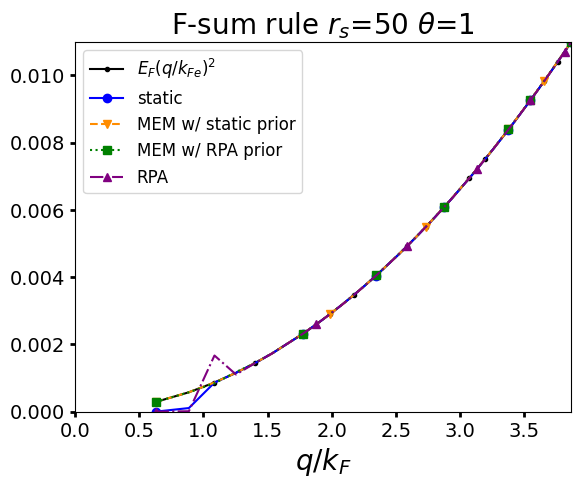}%
    \includegraphics[width=0.33\linewidth]{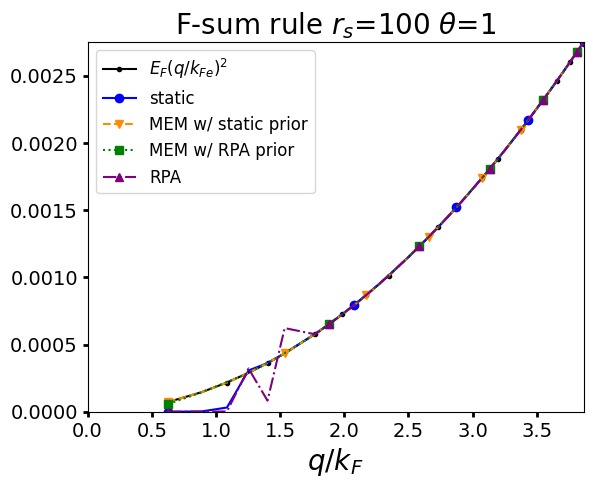}
    \caption{We plot the F-sum rule \eqref{eq:fsum} for the various computed DSF functions. All plots show numerical agreement. For large $r_s$, the dispersion peak narrows in the $q \rightarrow 0$ limit and numerical integration begins to fail, since the DSF is converging to a Dirac delta this is an unavoidable occurrence.}
    \label{fig:fsum}
\end{figure}

Next, we investigate the frequency sum rule (F-sum rule) \eqref{eq:fsum}. Plots are given in Figure~\ref{fig:fsum}, which demonstrate that our MEM results numerically satisfy the F-sum rule for all $q$ values. Based on \eqref{eq:moments}, this tells us that the Laplace transform of our DSF estimates have a first derivative that closely match the first derivative of the ITCF data near $\tau=0$ and that the DSF is converging to a Dirac delta in the small $q$ limit. 

Next, we investigate the MEM estimate's inverse moment sum rule \eqref{eq:inverse}, which is sensitive to when the dispersion peak is located at small $\omega$. We have numerical agreement for all $r_s$. Comparatively, the static approximation has a slight deviation from this sum rule for $r_s =50, 100$ between $q = 2 k_F$ and $q=2.5 k_F$. Although this slight deviation is hard to notice, it is still worth mentioning because the static LFC is guaranteed to satisfy the inverse moment sum rule, see appendix~\ref{app:inversemoment}. This feature is likely a small inconsistency of combining a static LFC that has been extracted from PIMC results for $F(q,\tau)$ for a finite number of particles with $\chi^0(q,\omega)$ computed for the thermodynamic limit, i.e., the limit of $N\to\infty$ with $r_s$ being kept constant. While finite-size effects in PIMC results are generally negligible for $q$-resolved properties~\cite{Chiesa_PRL_2006,dornheim_prl}, this breaks down for large $r_s$ as an incipient long-range order is forming; this, in turn, leads to commensurability effects, which might explain the small observed discrepancy. In appendix~\ref{app:inversemoment}, we also discuss why the MEM estimates, which are not guaranteed to satisfy the inverse moment sum rule, perform better than the static approximation.
\begin{figure}[h!]
    \centering
    \includegraphics[width=0.33\linewidth]{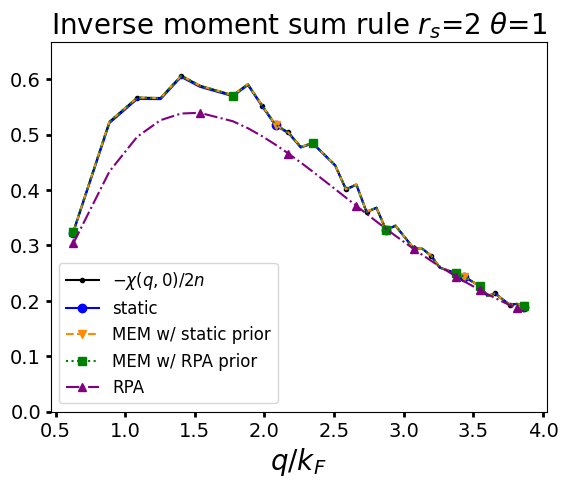}%
    \includegraphics[width=0.33\linewidth]{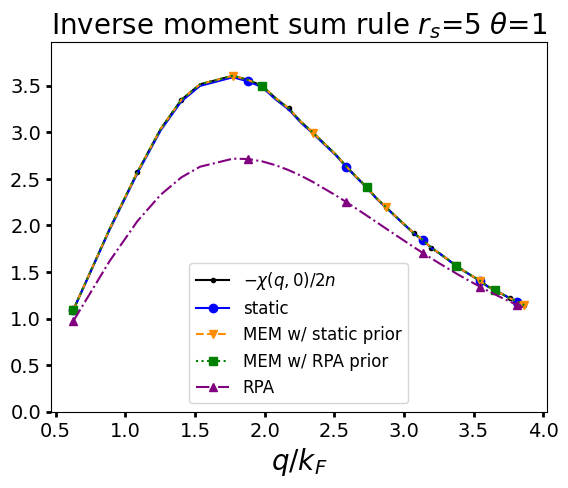}%
    \includegraphics[width=0.33\linewidth]{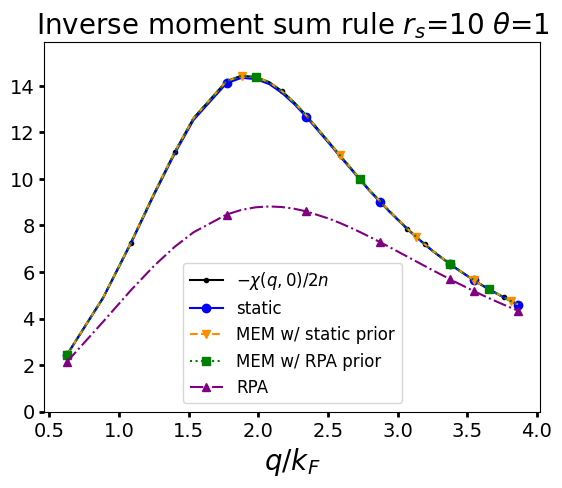}
    \includegraphics[width=0.33\linewidth]{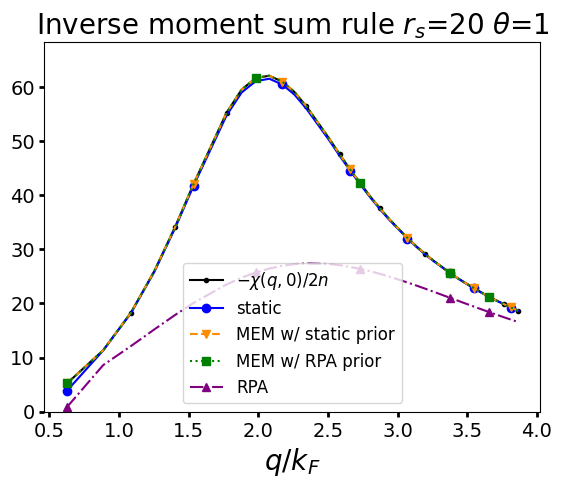}%
    \includegraphics[width=0.33\linewidth]{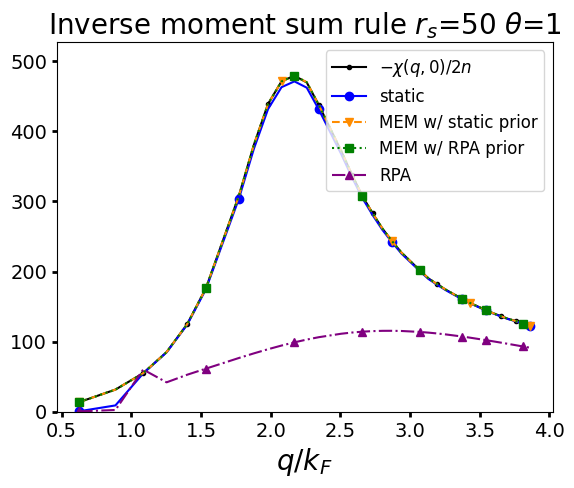}%
    \includegraphics[width=0.33\linewidth]{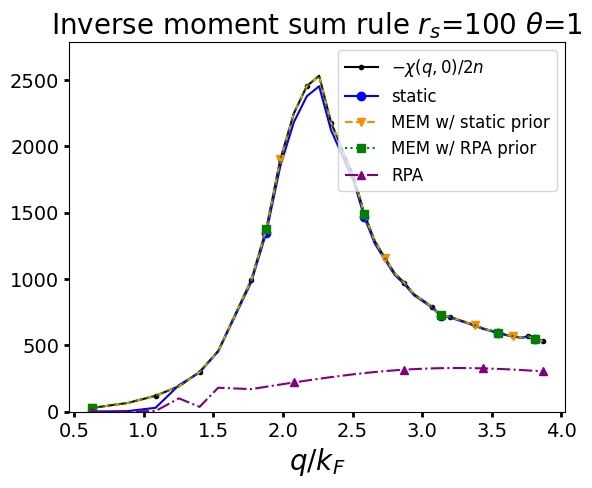}
    \caption{We plot the inverse moment sum rule \eqref{eq:inverse} for the various computed DSF functions. Numerical agreement exists for the MEM estimates for all $r_s$ values. For large $r_s$, slight deviations exist between the sum rule $-\chi(q,0)/2n$ and the static LFC DSF and are discussed in Appendix~\ref{app:inversemoment}. Further, at large $r_s$, the dispersion peak narrows in the $q \rightarrow 0$ limit and numerical integration begins to fail, since the DSF is converging to a Dirac delta this is an unavoidable occurrence.}
    \label{fig:wneg1}
\end{figure}

\section{Summary and Conclusions} \label{sec:conclusions}

%

In this work, we have estimated DSFs of the finite temperature electron liquid from PIMC ITCF data (see Table~\ref{tab:simulations}) by direct inversion of the two-sided Laplace transform using the MEM. We constructed the MEM's Bayesian prior via \eqref{eq:FDT} and \eqref{eq:susceptibility} after computing the static LFC $G(q, \omega=0)$ from our data. We conducted the MEM within a leave-one-out resampling routine, which provided appropriate error quantification \eqref{eq:totalMEMerror}, see Figure~\ref{fig:DSF-dispersion}, Figure~\ref{fig:Err_heatmaps1}, and Figure~\ref{fig:Err_heatmaps2}.

We have considered the physics and data science of constructing the static LFC from the data to be used as the MEM's Bayesian prior. Investigations have shown the static approximation gives a reasonable description of the dispersion relation, but fails to capture all of the collisional damping in a plasma. As such, using the static LFC \eqref{eq:Gq} as the Bayesian prior means we are focusing the MEM on extracting the correct frequency dependence of $G(q,\omega)$. In particular, at large $r_s$ values, the RPA does not even qualitatively describe the dispersion peak and using the RPA as the Bayesian prior produces low quality results. By comparison, the static LFC Bayesian prior remains valid and the MEM results are stable (see Figure~\ref{fig:Skw_heatmaps2}). We discuss from the statistics perspective, that using an entropic regularization with a Bayesian prior estimated from the data mitigates the bias introduced by regularization. Further, from the data science perspective, computing $G(q, \omega=0)$ from the data is informing the MEM of the ITCF's average value \eqref{eq:chi_q}. This resembles typical detrending approaches in data science, focusing the MEM's inference abilities on deviations from the mean. 

We have shown that our DSFs recover certain known limits and properties. In particular, our DSF estimates recover the RPA form at small $r_s$ and $\Theta=1$, see Figure~\ref{fig:Skw_heatmaps1}. Our DSFs also resemble the pre-existing stochastic calculations of the DSF at moderate $r_s$ values, see Figure~\ref{fig:stochasticcomparison}. In particular, when the MEM uses the static approximation as a Bayesian prior, it produces a DSF estimate that differs from the static approximation in ways that are similar to how dynamic local field correction approaches differ from the static approximation. We take this as evidence that the MEM is using the data to incorporate the dynamic dependency of the local field correction, which was neglected by the static approximation. Additionally, we have checked that our DSFs satisfy the large and small $q$ limits for the static structure factor (see Figure~\ref{fig:Sk}), the F-sum rule (see Figure~\ref{fig:fsum}), and the $\omega^{-1}$ sum rule (see Figure~\ref{fig:wneg1}). Satisfying these sum rules is significant because the MEM does not enforce them as solution constraints. Since the sum rules are contained in the derivatives of the ITCF data, whether or not they manifest in the MEM estimates is determined solely by the goodness of fit (\textit{i.e.} least squares criteria). 

Our investigations have established interesting new results. Using the static LFC as our Bayesian prior, the MEM produces DSFs which contain a roton feature for $r_s\gtrsim10$ (see Figure~\ref{fig:DSF-dispersion}). Our DSFs show a more pronounced roton feature than the static LFC (see Figure~\ref{fig:DSF-kslice}, Figure~\ref{fig:DSF-stacked-kslices1} and Figure~\ref{fig:DSF-stacked-kslices2}), as it is expected. Moreover, the depth of the roton minimum substantially increases with $r_s$, as it is predicted by the recent semi-empirical pair alignment model~\cite{Dornheim_Nature_2022}. We note that the present study covers the regime of substantially lower density and, hence, stronger coupling compared to previous investigations~\cite{dornheim2018dynamiclocalfieldcorrection,Filinov_PRB_2023}.
Finally, we find at large $r_s$ values that the MEM estimates exhibit an incipient double peak structure (see Figure~\ref{fig:DSF-kslice}).

Future work can apply these DSF estimates to the applications discussed in the introduction: establishing frequency dependent exchange-correlation kernels for TD-DFT, computing dielectric functions and transport properties for kinetic or hydrodynamic simulations, using UEG DSFs as input for the interpretation of x-ray Thomson scattering experiments within the Chihara decomposition, and benchmarking new dynamic theories in the strongly coupled regime. Future work on the UEG can explore different default models and simulations parameters; the latter would be particularly helpful to inform new analytical or neural network representations of the full dynamic local field correction in the real-frequency domain. Finally, we stress that the present MEM set-up is not limited to the UEG and can be used to compute the electronic DSF of real warm dense matter systems such as hydrogen and beryllium, for which accurate PIMC calculation of the ITCF have already been demonstrated~\cite{dornheim_MRE_2024,Dornheim_JCP_2024,Dornheim_NatComm_2024}. This will allow for the PIMC informed prediction of x-ray Thomson scattering measurements e.g.~at the National Ignition Facility~\cite{Tilo_Nature_2023} or at the European XFEL~\cite{Gawne_PRB_2024}, for the development of improved model descriptions for the electronic dynamics of warm dense matter e.g.~within the Chihara decomposition~\cite{bellenbaum2025estimatingionizationstatescontinuum}, and to rigorously assess the accuracy of different time-dependent DFT methodologies~\cite{Moldabekov_PRR_2023,Baczewski_PRL_2016,Schoerner_PRE_2023,moldabekov2025applyingliouvillelanczosmethodtimedependent,White_2025}.


\begin{acknowledgements}

This work was partially supported by the Center for Advanced Systems Understanding (CASUS), financed by Germany’s Federal Ministry of Education and Research (BMBF) and the Saxon state government out of the State budget approved by the Saxon State Parliament.
This work has received funding from the European Union's Just Transition Fund (JTF) within the project \emph{R\"ontgenlaser-Optimierung der Laserfusion} (ROLF), contract number 5086999001, co-financed by the Saxon state government out of the State budget approved by the Saxon State Parliament.
This work has received funding from the European Research Council (ERC) under the European Union’s Horizon 2022 research and innovation programme
(Grant agreement No. 101076233, "PREXTREME"). 
Views and opinions expressed are however those of the authors only and do not necessarily reflect those of the European Union or the European Research Council Executive Agency. Neither the European Union nor the granting authority can be held responsible for them. Computations were performed on a Bull Cluster at the Center for Information Services and High-Performance Computing (ZIH) at Technische Universit\"at Dresden and at the Norddeutscher Verbund f\"ur Hoch- und H\"ochstleistungsrechnen (HLRN) under grant mvp00024.
\end{acknowledgements}

\appendix
\section{Inverse moment sum rule} \label{app:inversemoment}

In this appendix, we establish that the static local field correction satisfies the inverse moment sum rule. For the reader's convenience we repeat the inverse moment sum rule \eqref{eq:inverse}
\begin{align}\label{eq:app-inversesumrule}
    \langle \omega^{-1} \rangle &= \int_{-\infty}^{+\infty} \frac{1}{\omega} S(q,\omega) = \frac{1}{2} \int_0^\beta d\tau \, F(q,\tau)\ .
\end{align}
Inserting the definition of the DSF~\eqref{eq:susceptibility} and simplifying, we find
\begin{align}
    \langle \omega^{-1} \rangle &=  - \frac{1}{2 \pi n} \int_{-\infty}^{+\infty} d\omega \,\frac{1}{\omega} \text{Im} \chi(q,\omega)\ .
\end{align}
The Kramer's Kronig relations inform us that
\begin{align}
    \frac{1}{2 \pi n}  \int_{-\infty}^{+\infty} \frac{1}{\omega} \text{Im} \chi(q,\omega) &= \frac{1}{2n} \text{Re} \chi(q, 0)\ ,
\end{align}
which has two consequences: first, it holds
\begin{align}
    \langle \omega^{-1} \rangle = -\frac{1}{2n} \text{Re} \chi(q, 0)\ ,
\end{align}
which establishes the inverse moment sum rule; second, we have
\begin{align}
    \int_{-\infty}^{+\infty} \frac{1}{\omega} S^\text{SLFC}(q,\omega)&= \frac{1}{2n} \text{Re} \chi^\text{SLFC}(q, 0),
\end{align}
but we also have by definition $\text{Re} \chi^\text{SLFC}(q, 0) = \text{Re} \chi(q, 0)$. Thus, $S^\text{SLFC}$ satisfies the inverse sum rule. Notice that this derivation works for the inverse moment because we have Kramer-Kronig relations and does not work for positive moments $\omega^1,\omega^2,...$, where Kramers-Kronig relations cannot be invoked.

\begin{figure}
    \centering
    \includegraphics[width=0.515\linewidth]{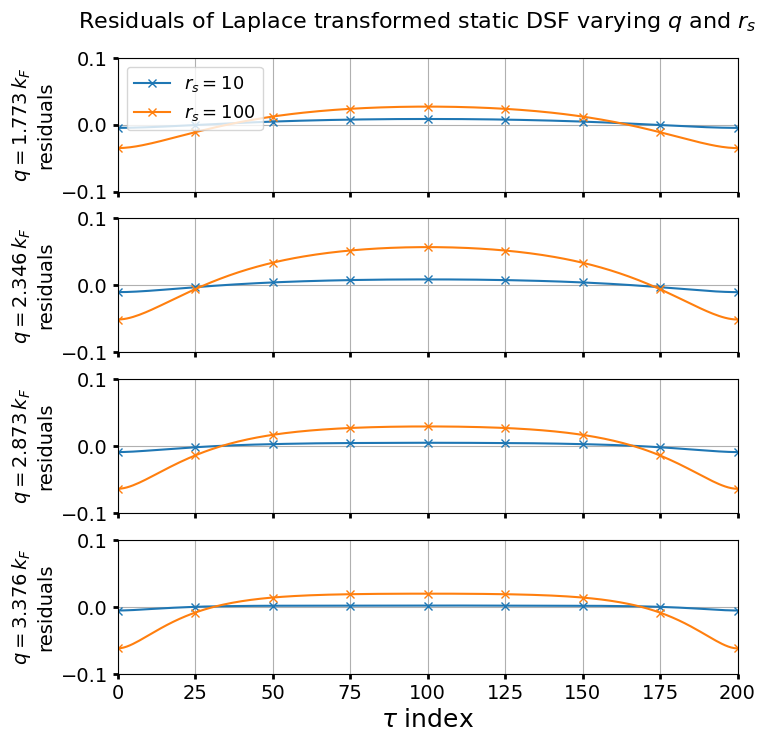}%
    \includegraphics[width=0.485\linewidth]{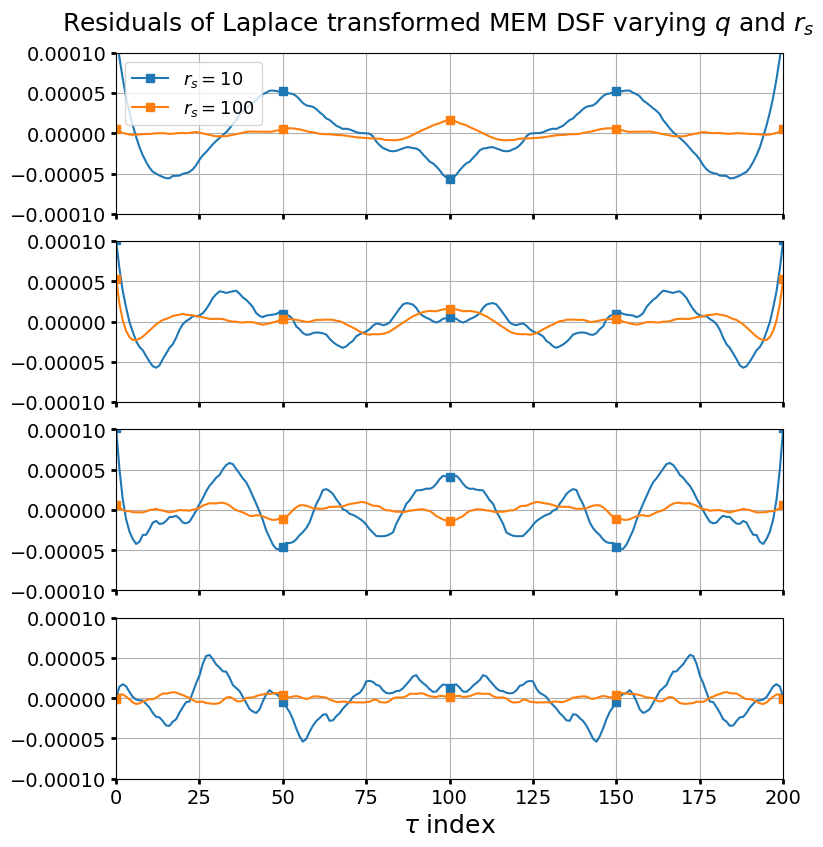}
    \caption{Plot of the residuals from the ITCF data for the static approximation DSF (left) and the MEM estimated DSF (right). Notice, the y-axes are not the same for the left and right plots. The residuals associated to the static approximation have a unique systematic trend, where the magnitude of this systematic trend varies with both $r_s$ and $q$. Comparatively, the residuals associated to the MEM are orders of magnitude smaller and appear as oscillations with no definitive $q$ and $r_s$ dependence.}
    \label{fig:Ftaudiscrepancies}
\end{figure}

It is important to consider that the MEM, which is not guaranteed to satisfy this sum rule, does a better job than the static model for which the sum rule is, in theory, guaranteed. To understand this, consider \eqref{eq:app-inversesumrule}, which informs us that the inverse moment sum rule is essentially a check of whether the Laplace transformed DSF model $F^\text{Model}(\tau) = \int d \omega  \, e^{-\tau \omega} S^\text{Model}(\omega,q)$  and PIMC ITCF $F(\tau)$ integrate to the same value. Hence, a breakdown in the sum rule is a statement about the discrepancies between $F^\text{Model}(\tau)$ and $F(\tau)$. We investigate these discrepancies in Figure~\ref{fig:Ftaudiscrepancies}. First, we see that such discrepancies exist in all the models plotted in Figure~~\ref{fig:wneg1}. However, these discrepancies are much smaller for the MEM estimates because the fidelity term in the MEM regularization minimizes the $F^\text{Model}(\tau)$ deviation from $F(\tau)$. Thus, the fidelity term is automatically enforcing the sum rules, but only up to some numerical degree and not exactly. Also note that while the differences in the ITCF between the static approximation and the PIMC reference data for $r_s=100$ are substantial, they are largely averaged out by the integration of $\tau$. It is only the small residual of the $\tau$-integral that violates the inverse moment sum rule.

As it has been stated in the main text, a possible explanation for the latter is given by the potential inconsistency between the static LFC that has been extracted from PIMC results for a finite number of electrons $N$ and the reference function $\chi^{(0)}(q,\omega)$ that is considered in the thermodynamic limit.
This might lead to a violation of the sum rule if finite-size effects depended on $\omega$, which is to be expected.


\bibliography{bibliography}

\begin{thebibliography}{118}%
\makeatletter
\providecommand \@ifxundefined [1]{%
 \@ifx{#1\undefined}
}%
\providecommand \@ifnum [1]{%
 \ifnum #1\expandafter \@firstoftwo
 \else \expandafter \@secondoftwo
 \fi
}%
\providecommand \@ifx [1]{%
 \ifx #1\expandafter \@firstoftwo
 \else \expandafter \@secondoftwo
 \fi
}%
\providecommand \natexlab [1]{#1}%
\providecommand \enquote  [1]{``#1''}%
\providecommand \bibnamefont  [1]{#1}%
\providecommand \bibfnamefont [1]{#1}%
\providecommand \citenamefont [1]{#1}%
\providecommand \href@noop [0]{\@secondoftwo}%
\providecommand \href [0]{\begingroup \@sanitize@url \@href}%
\providecommand \@href[1]{\@@startlink{#1}\@@href}%
\providecommand \@@href[1]{\endgroup#1\@@endlink}%
\providecommand \@sanitize@url [0]{\catcode `\\12\catcode `\$12\catcode `\&12\catcode `\#12\catcode `\^12\catcode `\_12\catcode `\%12\relax}%
\providecommand \@@startlink[1]{}%
\providecommand \@@endlink[0]{}%
\providecommand \url  [0]{\begingroup\@sanitize@url \@url }%
\providecommand \@url [1]{\endgroup\@href {#1}{\urlprefix }}%
\providecommand \urlprefix  [0]{URL }%
\providecommand \Eprint [0]{\href }%
\providecommand \doibase [0]{https://doi.org/}%
\providecommand \selectlanguage [0]{\@gobble}%
\providecommand \bibinfo  [0]{\@secondoftwo}%
\providecommand \bibfield  [0]{\@secondoftwo}%
\providecommand \translation [1]{[#1]}%
\providecommand \BibitemOpen [0]{}%
\providecommand \bibitemStop [0]{}%
\providecommand \bibitemNoStop [0]{.\EOS\space}%
\providecommand \EOS [0]{\spacefactor3000\relax}%
\providecommand \BibitemShut  [1]{\csname bibitem#1\endcsname}%
\let\auto@bib@innerbib\@empty
\bibitem [{\citenamefont {Giuliani}\ and\ \citenamefont {Vignale}(2008)}]{GiulianiVignale2008quantumtheory}%
  \BibitemOpen
  \bibfield  {author} {\bibinfo {author} {\bibfnamefont {G.}~\bibnamefont {Giuliani}}\ and\ \bibinfo {author} {\bibfnamefont {G.}~\bibnamefont {Vignale}},\ }\href@noop {} {\emph {\bibinfo {title} {Quantum Theory of the Electron Liquid}}}\ (\bibinfo  {publisher} {Cambridge University Press},\ \bibinfo {address} {Cambridge},\ \bibinfo {year} {2008})\BibitemShut {NoStop}%
\bibitem [{\citenamefont {Mahan}(1990)}]{mahan1990many}%
  \BibitemOpen
  \bibfield  {author} {\bibinfo {author} {\bibfnamefont {G.}~\bibnamefont {Mahan}},\ }\href {https://books.google.de/books?id=v8du6cp0vUAC} {\emph {\bibinfo {title} {Many-Particle Physics}}},\ Physics of Solids and Liquids\ (\bibinfo  {publisher} {Springer US},\ \bibinfo {year} {1990})\BibitemShut {NoStop}%
\bibitem [{\citenamefont {Hansen}\ and\ \citenamefont {McDonald}(2013)}]{hansen2013theory3rdEd}%
  \BibitemOpen
  \bibfield  {author} {\bibinfo {author} {\bibfnamefont {J.}~\bibnamefont {Hansen}}\ and\ \bibinfo {author} {\bibfnamefont {I.}~\bibnamefont {McDonald}},\ }\href {https://books.google.de/books?id=agbEswEACAAJ} {\emph {\bibinfo {title} {Theory of simple liquids: with applications to soft matter}}}\ (\bibinfo  {publisher} {Academic Press},\ \bibinfo {year} {2013})\BibitemShut {NoStop}%
\bibitem [{\citenamefont {Dornheim}\ \emph {et~al.}(2018{\natexlab{a}})\citenamefont {Dornheim}, \citenamefont {Groth},\ and\ \citenamefont {Bonitz}}]{Dornheim2018UEGreview}%
  \BibitemOpen
  \bibfield  {author} {\bibinfo {author} {\bibfnamefont {T.}~\bibnamefont {Dornheim}}, \bibinfo {author} {\bibfnamefont {S.}~\bibnamefont {Groth}},\ and\ \bibinfo {author} {\bibfnamefont {M.}~\bibnamefont {Bonitz}},\ }\bibfield  {title} {\bibinfo {title} {The uniform electron gas at warm dense matter conditions},\ }\href {https://www.sciencedirect.com/science/article/abs/pii/S0370157318300516} {\bibfield  {journal} {\bibinfo  {journal} {Phys. Reports}\ }\textbf {\bibinfo {volume} {744}},\ \bibinfo {pages} {1} (\bibinfo {year} {2018}{\natexlab{a}})}\BibitemShut {NoStop}%
\bibitem [{\citenamefont {Loos}\ and\ \citenamefont {Gill}(2016)}]{loos}%
  \BibitemOpen
  \bibfield  {author} {\bibinfo {author} {\bibfnamefont {P.-F.}\ \bibnamefont {Loos}}\ and\ \bibinfo {author} {\bibfnamefont {P.~M.~W.}\ \bibnamefont {Gill}},\ }\bibfield  {title} {\bibinfo {title} {The uniform electron gas},\ }\href {http://onlinelibrary.wiley.com/doi/10.1002/wcms.1257/abstract} {\bibfield  {journal} {\bibinfo  {journal} {Comput. Mol. Sci}\ }\textbf {\bibinfo {volume} {6}},\ \bibinfo {pages} {410} (\bibinfo {year} {2016})}\BibitemShut {NoStop}%
\bibitem [{\citenamefont {Bonitz}\ \emph {et~al.}(2020)\citenamefont {Bonitz}, \citenamefont {Dornheim}, \citenamefont {Moldabekov}, \citenamefont {Zhang}, \citenamefont {Hamann}, \citenamefont {K{\"a}hlert}, \citenamefont {Filinov}, \citenamefont {Ramakrishna},\ and\ \citenamefont {Vorberger}}]{bonitz2020WDMUEG}%
  \BibitemOpen
  \bibfield  {author} {\bibinfo {author} {\bibfnamefont {M.}~\bibnamefont {Bonitz}}, \bibinfo {author} {\bibfnamefont {T.}~\bibnamefont {Dornheim}}, \bibinfo {author} {\bibfnamefont {Z.~A.}\ \bibnamefont {Moldabekov}}, \bibinfo {author} {\bibfnamefont {S.}~\bibnamefont {Zhang}}, \bibinfo {author} {\bibfnamefont {P.}~\bibnamefont {Hamann}}, \bibinfo {author} {\bibfnamefont {H.}~\bibnamefont {K{\"a}hlert}}, \bibinfo {author} {\bibfnamefont {A.}~\bibnamefont {Filinov}}, \bibinfo {author} {\bibfnamefont {K.}~\bibnamefont {Ramakrishna}},\ and\ \bibinfo {author} {\bibfnamefont {J.}~\bibnamefont {Vorberger}},\ }\bibfield  {title} {\bibinfo {title} {Ab initio simulation of warm dense matter},\ }\href@noop {} {\bibfield  {journal} {\bibinfo  {journal} {Physics of Plasmas}\ }\textbf {\bibinfo {volume} {27}} (\bibinfo {year} {2020})}\BibitemShut {NoStop}%
\bibitem [{\citenamefont {Rabani}\ \emph {et~al.}(2002)\citenamefont {Rabani}, \citenamefont {Reichman}, \citenamefont {Krilov},\ and\ \citenamefont {Berne}}]{Rabani_PNAS_2002}%
  \BibitemOpen
  \bibfield  {author} {\bibinfo {author} {\bibfnamefont {E.}~\bibnamefont {Rabani}}, \bibinfo {author} {\bibfnamefont {D.~R.}\ \bibnamefont {Reichman}}, \bibinfo {author} {\bibfnamefont {G.}~\bibnamefont {Krilov}},\ and\ \bibinfo {author} {\bibfnamefont {B.~J.}\ \bibnamefont {Berne}},\ }\bibfield  {title} {\bibinfo {title} {The calculation of transport properties in quantum liquids using the maximum entropy numerical analytic continuation method: Application to liquid <i>para</i>-hydrogen},\ }\href {https://doi.org/10.1073/pnas.261540698} {\bibfield  {journal} {\bibinfo  {journal} {Proceedings of the National Academy of Sciences}\ }\textbf {\bibinfo {volume} {99}},\ \bibinfo {pages} {1129} (\bibinfo {year} {2002})},\ \Eprint {https://arxiv.org/abs/https://www.pnas.org/doi/pdf/10.1073/pnas.261540698} {https://www.pnas.org/doi/pdf/10.1073/pnas.261540698} \BibitemShut {NoStop}%
\bibitem [{\citenamefont {Filinov}\ and\ \citenamefont {Bonitz}(2012)}]{Filinov_PRA_2012}%
  \BibitemOpen
  \bibfield  {author} {\bibinfo {author} {\bibfnamefont {A.}~\bibnamefont {Filinov}}\ and\ \bibinfo {author} {\bibfnamefont {M.}~\bibnamefont {Bonitz}},\ }\bibfield  {title} {\bibinfo {title} {Collective and single-particle excitations in two-dimensional dipolar bose gases},\ }\href {https://doi.org/10.1103/PhysRevA.86.043628} {\bibfield  {journal} {\bibinfo  {journal} {Phys. Rev. A}\ }\textbf {\bibinfo {volume} {86}},\ \bibinfo {pages} {043628} (\bibinfo {year} {2012})}\BibitemShut {NoStop}%
\bibitem [{\citenamefont {Dornheim}\ \emph {et~al.}(2021{\natexlab{a}})\citenamefont {Dornheim}, \citenamefont {Moldabekov},\ and\ \citenamefont {Vorberger}}]{Dornheim_JCP_ITCF_2021}%
  \BibitemOpen
  \bibfield  {author} {\bibinfo {author} {\bibfnamefont {T.}~\bibnamefont {Dornheim}}, \bibinfo {author} {\bibfnamefont {Z.~A.}\ \bibnamefont {Moldabekov}},\ and\ \bibinfo {author} {\bibfnamefont {J.}~\bibnamefont {Vorberger}},\ }\bibfield  {title} {\bibinfo {title} {Nonlinear density response from imaginary-time correlation functions: Ab initio path integral monte carlo simulations of the warm dense electron gas},\ }\href {https://doi.org/10.1063/5.0058988} {\bibfield  {journal} {\bibinfo  {journal} {The Journal of Chemical Physics}\ }\textbf {\bibinfo {volume} {155}},\ \bibinfo {pages} {054110} (\bibinfo {year} {2021}{\natexlab{a}})}\BibitemShut {NoStop}%
\bibitem [{\citenamefont {Gregori}\ \emph {et~al.}(2003)\citenamefont {Gregori}, \citenamefont {Glenzer}, \citenamefont {Rozmus}, \citenamefont {Lee},\ and\ \citenamefont {Landen}}]{Gregori_PRE_2003}%
  \BibitemOpen
  \bibfield  {author} {\bibinfo {author} {\bibfnamefont {G.}~\bibnamefont {Gregori}}, \bibinfo {author} {\bibfnamefont {S.~H.}\ \bibnamefont {Glenzer}}, \bibinfo {author} {\bibfnamefont {W.}~\bibnamefont {Rozmus}}, \bibinfo {author} {\bibfnamefont {R.~W.}\ \bibnamefont {Lee}},\ and\ \bibinfo {author} {\bibfnamefont {O.~L.}\ \bibnamefont {Landen}},\ }\bibfield  {title} {\bibinfo {title} {Theoretical model of x-ray scattering as a dense matter probe},\ }\href {https://doi.org/10.1103/PhysRevE.67.026412} {\bibfield  {journal} {\bibinfo  {journal} {Phys. Rev. E}\ }\textbf {\bibinfo {volume} {67}},\ \bibinfo {pages} {026412} (\bibinfo {year} {2003})}\BibitemShut {NoStop}%
\bibitem [{\citenamefont {Glenzer}\ and\ \citenamefont {Redmer}(2009)}]{siegfried_review}%
  \BibitemOpen
  \bibfield  {author} {\bibinfo {author} {\bibfnamefont {S.~H.}\ \bibnamefont {Glenzer}}\ and\ \bibinfo {author} {\bibfnamefont {R.}~\bibnamefont {Redmer}},\ }\bibfield  {title} {\bibinfo {title} {X-ray thomson scattering in high energy density plasmas},\ }\href {https://journals.aps.org/rmp/abstract/10.1103/RevModPhys.81.1625} {\bibfield  {journal} {\bibinfo  {journal} {Rev. Mod. Phys}\ }\textbf {\bibinfo {volume} {81}},\ \bibinfo {pages} {1625} (\bibinfo {year} {2009})}\BibitemShut {NoStop}%
\bibitem [{\citenamefont {Dornheim}\ \emph {et~al.}(2023{\natexlab{a}})\citenamefont {Dornheim}, \citenamefont {Moldabekov}, \citenamefont {Ramakrishna}, \citenamefont {Tolias}, \citenamefont {Baczewski}, \citenamefont {Kraus}, \citenamefont {Preston}, \citenamefont {Chapman}, \citenamefont {Böhme}, \citenamefont {Döppner}, \citenamefont {Graziani}, \citenamefont {Bonitz}, \citenamefont {Cangi},\ and\ \citenamefont {Vorberger}}]{Dornheim_review}%
  \BibitemOpen
  \bibfield  {author} {\bibinfo {author} {\bibfnamefont {T.}~\bibnamefont {Dornheim}}, \bibinfo {author} {\bibfnamefont {Z.~A.}\ \bibnamefont {Moldabekov}}, \bibinfo {author} {\bibfnamefont {K.}~\bibnamefont {Ramakrishna}}, \bibinfo {author} {\bibfnamefont {P.}~\bibnamefont {Tolias}}, \bibinfo {author} {\bibfnamefont {A.~D.}\ \bibnamefont {Baczewski}}, \bibinfo {author} {\bibfnamefont {D.}~\bibnamefont {Kraus}}, \bibinfo {author} {\bibfnamefont {T.~R.}\ \bibnamefont {Preston}}, \bibinfo {author} {\bibfnamefont {D.~A.}\ \bibnamefont {Chapman}}, \bibinfo {author} {\bibfnamefont {M.~P.}\ \bibnamefont {Böhme}}, \bibinfo {author} {\bibfnamefont {T.}~\bibnamefont {Döppner}}, \bibinfo {author} {\bibfnamefont {F.}~\bibnamefont {Graziani}}, \bibinfo {author} {\bibfnamefont {M.}~\bibnamefont {Bonitz}}, \bibinfo {author} {\bibfnamefont {A.}~\bibnamefont {Cangi}},\ and\ \bibinfo {author} {\bibfnamefont {J.}~\bibnamefont {Vorberger}},\ }\bibfield  {title} {\bibinfo {title} {Electronic density response of warm dense
  matter},\ }\href {https://doi.org/10.1063/5.0138955} {\bibfield  {journal} {\bibinfo  {journal} {Physics of Plasmas}\ }\textbf {\bibinfo {volume} {30}},\ \bibinfo {pages} {032705} (\bibinfo {year} {2023}{\natexlab{a}})}\BibitemShut {NoStop}%
\bibitem [{\citenamefont {Hamann}\ \emph {et~al.}(2020{\natexlab{a}})\citenamefont {Hamann}, \citenamefont {Vorberger}, \citenamefont {Dornheim}, \citenamefont {Moldabekov},\ and\ \citenamefont {Bonitz}}]{Hamann_CPP_2020}%
  \BibitemOpen
  \bibfield  {author} {\bibinfo {author} {\bibfnamefont {P.}~\bibnamefont {Hamann}}, \bibinfo {author} {\bibfnamefont {J.}~\bibnamefont {Vorberger}}, \bibinfo {author} {\bibfnamefont {T.}~\bibnamefont {Dornheim}}, \bibinfo {author} {\bibfnamefont {Z.~A.}\ \bibnamefont {Moldabekov}},\ and\ \bibinfo {author} {\bibfnamefont {M.}~\bibnamefont {Bonitz}},\ }\bibfield  {title} {\bibinfo {title} {Ab initio results for the plasmon dispersion and damping of the warm dense electron gas},\ }\href {https://doi.org/https://doi.org/10.1002/ctpp.202000147} {\bibfield  {journal} {\bibinfo  {journal} {Contributions to Plasma Physics}\ }\textbf {\bibinfo {volume} {60}},\ \bibinfo {pages} {e202000147} (\bibinfo {year} {2020}{\natexlab{a}})}\BibitemShut {NoStop}%
\bibitem [{\citenamefont {Hamann}\ \emph {et~al.}(2020{\natexlab{b}})\citenamefont {Hamann}, \citenamefont {Dornheim}, \citenamefont {Vorberger}, \citenamefont {Moldabekov},\ and\ \citenamefont {Bonitz}}]{Hamann_PRB_2020}%
  \BibitemOpen
  \bibfield  {author} {\bibinfo {author} {\bibfnamefont {P.}~\bibnamefont {Hamann}}, \bibinfo {author} {\bibfnamefont {T.}~\bibnamefont {Dornheim}}, \bibinfo {author} {\bibfnamefont {J.}~\bibnamefont {Vorberger}}, \bibinfo {author} {\bibfnamefont {Z.~A.}\ \bibnamefont {Moldabekov}},\ and\ \bibinfo {author} {\bibfnamefont {M.}~\bibnamefont {Bonitz}},\ }\bibfield  {title} {\bibinfo {title} {Dynamic properties of the warm dense electron gas based on $ab initio$ path integral monte carlo simulations},\ }\href {https://doi.org/10.1103/PhysRevB.102.125150} {\bibfield  {journal} {\bibinfo  {journal} {Phys. Rev. B}\ }\textbf {\bibinfo {volume} {102}},\ \bibinfo {pages} {125150} (\bibinfo {year} {2020}{\natexlab{b}})}\BibitemShut {NoStop}%
\bibitem [{\citenamefont {Epstein}\ and\ \citenamefont {Schotland}(2008)}]{epstein2008badtruth}%
  \BibitemOpen
  \bibfield  {author} {\bibinfo {author} {\bibfnamefont {C.~L.}\ \bibnamefont {Epstein}}\ and\ \bibinfo {author} {\bibfnamefont {J.}~\bibnamefont {Schotland}},\ }\bibfield  {title} {\bibinfo {title} {The bad truth about laplace's transform},\ }\href@noop {} {\bibfield  {journal} {\bibinfo  {journal} {SIAM review}\ }\textbf {\bibinfo {volume} {50}},\ \bibinfo {pages} {504} (\bibinfo {year} {2008})}\BibitemShut {NoStop}%
\bibitem [{\citenamefont {Istratov}\ and\ \citenamefont {Vyvenko}(1999)}]{istratov1999}%
  \BibitemOpen
  \bibfield  {author} {\bibinfo {author} {\bibfnamefont {A.~A.}\ \bibnamefont {Istratov}}\ and\ \bibinfo {author} {\bibfnamefont {O.~F.}\ \bibnamefont {Vyvenko}},\ }\bibfield  {title} {\bibinfo {title} {Exponential analysis in physical phenomena},\ }\href@noop {} {\bibfield  {journal} {\bibinfo  {journal} {Review of Scientific Instruments}\ }\textbf {\bibinfo {volume} {70}},\ \bibinfo {pages} {1233} (\bibinfo {year} {1999})}\BibitemShut {NoStop}%
\bibitem [{\citenamefont {Fournier}\ \emph {et~al.}(2020)\citenamefont {Fournier}, \citenamefont {Wang}, \citenamefont {Yazyev},\ and\ \citenamefont {Wu}}]{Fournier_PRL_2020}%
  \BibitemOpen
  \bibfield  {author} {\bibinfo {author} {\bibfnamefont {R.}~\bibnamefont {Fournier}}, \bibinfo {author} {\bibfnamefont {L.}~\bibnamefont {Wang}}, \bibinfo {author} {\bibfnamefont {O.~V.}\ \bibnamefont {Yazyev}},\ and\ \bibinfo {author} {\bibfnamefont {Q.}~\bibnamefont {Wu}},\ }\bibfield  {title} {\bibinfo {title} {Artificial neural network approach to the analytic continuation problem},\ }\href {https://doi.org/10.1103/PhysRevLett.124.056401} {\bibfield  {journal} {\bibinfo  {journal} {Phys. Rev. Lett.}\ }\textbf {\bibinfo {volume} {124}},\ \bibinfo {pages} {056401} (\bibinfo {year} {2020})}\BibitemShut {NoStop}%
\bibitem [{\citenamefont {Yoon}\ \emph {et~al.}(2018)\citenamefont {Yoon}, \citenamefont {Sim},\ and\ \citenamefont {Han}}]{Yoon2018ACviaML}%
  \BibitemOpen
  \bibfield  {author} {\bibinfo {author} {\bibfnamefont {H.}~\bibnamefont {Yoon}}, \bibinfo {author} {\bibfnamefont {J.-H.}\ \bibnamefont {Sim}},\ and\ \bibinfo {author} {\bibfnamefont {M.~J.}\ \bibnamefont {Han}},\ }\bibfield  {title} {\bibinfo {title} {Analytic continuation via domain knowledge free machine learning},\ }\href {https://doi.org/10.1103/PhysRevB.98.245101} {\bibfield  {journal} {\bibinfo  {journal} {Phys. Rev. B}\ }\textbf {\bibinfo {volume} {98}},\ \bibinfo {pages} {245101} (\bibinfo {year} {2018})}\BibitemShut {NoStop}%
\bibitem [{\citenamefont {Goulko}\ \emph {et~al.}(2017)\citenamefont {Goulko}, \citenamefont {Mishchenko}, \citenamefont {Pollet}, \citenamefont {Prokof'ev},\ and\ \citenamefont {Svistunov}}]{Goulko_PRB_2017}%
  \BibitemOpen
  \bibfield  {author} {\bibinfo {author} {\bibfnamefont {O.}~\bibnamefont {Goulko}}, \bibinfo {author} {\bibfnamefont {A.~S.}\ \bibnamefont {Mishchenko}}, \bibinfo {author} {\bibfnamefont {L.}~\bibnamefont {Pollet}}, \bibinfo {author} {\bibfnamefont {N.}~\bibnamefont {Prokof'ev}},\ and\ \bibinfo {author} {\bibfnamefont {B.}~\bibnamefont {Svistunov}},\ }\bibfield  {title} {\bibinfo {title} {Numerical analytic continuation: Answers to well-posed questions},\ }\href {https://doi.org/10.1103/PhysRevB.95.014102} {\bibfield  {journal} {\bibinfo  {journal} {Phys. Rev. B}\ }\textbf {\bibinfo {volume} {95}},\ \bibinfo {pages} {014102} (\bibinfo {year} {2017})}\BibitemShut {NoStop}%
\bibitem [{\citenamefont {Otsuki}\ \emph {et~al.}(2017)\citenamefont {Otsuki}, \citenamefont {Ohzeki}, \citenamefont {Shinaoka},\ and\ \citenamefont {Yoshimi}}]{otsuki2017sparse}%
  \BibitemOpen
  \bibfield  {author} {\bibinfo {author} {\bibfnamefont {J.}~\bibnamefont {Otsuki}}, \bibinfo {author} {\bibfnamefont {M.}~\bibnamefont {Ohzeki}}, \bibinfo {author} {\bibfnamefont {H.}~\bibnamefont {Shinaoka}},\ and\ \bibinfo {author} {\bibfnamefont {K.}~\bibnamefont {Yoshimi}},\ }\bibfield  {title} {\bibinfo {title} {Sparse modeling approach to analytical continuation of imaginary-time quantum monte carlo data},\ }\href@noop {} {\bibfield  {journal} {\bibinfo  {journal} {Physical Review E}\ }\textbf {\bibinfo {volume} {95}},\ \bibinfo {pages} {061302} (\bibinfo {year} {2017})}\BibitemShut {NoStop}%
\bibitem [{\citenamefont {Huang}(2023)}]{huang2023acflow}%
  \BibitemOpen
  \bibfield  {author} {\bibinfo {author} {\bibfnamefont {L.}~\bibnamefont {Huang}},\ }\bibfield  {title} {\bibinfo {title} {Acflow: An open source toolkit for analytic continuation of quantum monte carlo data},\ }\href@noop {} {\bibfield  {journal} {\bibinfo  {journal} {Computer Physics Communications}\ }\textbf {\bibinfo {volume} {292}},\ \bibinfo {pages} {108863} (\bibinfo {year} {2023})}\BibitemShut {NoStop}%
\bibitem [{\citenamefont {Beach}(2004)}]{beach2004identifying}%
  \BibitemOpen
  \bibfield  {author} {\bibinfo {author} {\bibfnamefont {K.}~\bibnamefont {Beach}},\ }\bibfield  {title} {\bibinfo {title} {Identifying the maximum entropy method as a special limit of stochastic analytic continuation},\ }\href@noop {} {\bibfield  {journal} {\bibinfo  {journal} {arXiv preprint cond-mat/0403055}\ } (\bibinfo {year} {2004})}\BibitemShut {NoStop}%
\bibitem [{\citenamefont {Fuchs}\ \emph {et~al.}(2010)\citenamefont {Fuchs}, \citenamefont {Pruschke},\ and\ \citenamefont {Jarrell}}]{fuchs2010analytic}%
  \BibitemOpen
  \bibfield  {author} {\bibinfo {author} {\bibfnamefont {S.}~\bibnamefont {Fuchs}}, \bibinfo {author} {\bibfnamefont {T.}~\bibnamefont {Pruschke}},\ and\ \bibinfo {author} {\bibfnamefont {M.}~\bibnamefont {Jarrell}},\ }\bibfield  {title} {\bibinfo {title} {Analytic continuation of quantum monte carlo data by stochastic analytical inference},\ }\href@noop {} {\bibfield  {journal} {\bibinfo  {journal} {Physical Review E—Statistical, Nonlinear, and Soft Matter Physics}\ }\textbf {\bibinfo {volume} {81}},\ \bibinfo {pages} {056701} (\bibinfo {year} {2010})}\BibitemShut {NoStop}%
\bibitem [{\citenamefont {Jarrell}\ and\ \citenamefont {Gubernatis}(1996)}]{JARRELL1996MEM}%
  \BibitemOpen
  \bibfield  {author} {\bibinfo {author} {\bibfnamefont {M.}~\bibnamefont {Jarrell}}\ and\ \bibinfo {author} {\bibfnamefont {J.}~\bibnamefont {Gubernatis}},\ }\bibfield  {title} {\bibinfo {title} {Bayesian inference and the analytic continuation of imaginary-time quantum monte carlo data},\ }\href {https://doi.org/https://doi.org/10.1016/0370-1573(95)00074-7} {\bibfield  {journal} {\bibinfo  {journal} {Physics Reports}\ }\textbf {\bibinfo {volume} {269}},\ \bibinfo {pages} {133} (\bibinfo {year} {1996})}\BibitemShut {NoStop}%
\bibitem [{\citenamefont {Burnier}\ and\ \citenamefont {Rothkopf}(2013)}]{BurnierRothkopf2013bayesianreconstruction}%
  \BibitemOpen
  \bibfield  {author} {\bibinfo {author} {\bibfnamefont {Y.}~\bibnamefont {Burnier}}\ and\ \bibinfo {author} {\bibfnamefont {A.}~\bibnamefont {Rothkopf}},\ }\bibfield  {title} {\bibinfo {title} {Bayesian approach to spectral function reconstruction<? format?> for euclidean quantum field theories},\ }\href@noop {} {\bibfield  {journal} {\bibinfo  {journal} {Physical review letters}\ }\textbf {\bibinfo {volume} {111}},\ \bibinfo {pages} {182003} (\bibinfo {year} {2013})}\BibitemShut {NoStop}%
\bibitem [{\citenamefont {Fischer}\ \emph {et~al.}(2018)\citenamefont {Fischer}, \citenamefont {Pawlowski}, \citenamefont {Rothkopf},\ and\ \citenamefont {Welzbacher}}]{Fischer2018SmoothedBRM}%
  \BibitemOpen
  \bibfield  {author} {\bibinfo {author} {\bibfnamefont {C.~S.}\ \bibnamefont {Fischer}}, \bibinfo {author} {\bibfnamefont {J.~M.}\ \bibnamefont {Pawlowski}}, \bibinfo {author} {\bibfnamefont {A.}~\bibnamefont {Rothkopf}},\ and\ \bibinfo {author} {\bibfnamefont {C.~A.}\ \bibnamefont {Welzbacher}},\ }\bibfield  {title} {\bibinfo {title} {Bayesian analysis of quark spectral properties from the dyson-schwinger equation},\ }\href@noop {} {\bibfield  {journal} {\bibinfo  {journal} {Physical Review D}\ }\textbf {\bibinfo {volume} {98}},\ \bibinfo {pages} {014009} (\bibinfo {year} {2018})}\BibitemShut {NoStop}%
\bibitem [{\citenamefont {Dornheim}\ \emph {et~al.}(2018{\natexlab{b}})\citenamefont {Dornheim}, \citenamefont {Groth}, \citenamefont {Vorberger},\ and\ \citenamefont {Bonitz}}]{dornheim2018dynamiclocalfieldcorrection}%
  \BibitemOpen
  \bibfield  {author} {\bibinfo {author} {\bibfnamefont {T.}~\bibnamefont {Dornheim}}, \bibinfo {author} {\bibfnamefont {S.}~\bibnamefont {Groth}}, \bibinfo {author} {\bibfnamefont {J.}~\bibnamefont {Vorberger}},\ and\ \bibinfo {author} {\bibfnamefont {M.}~\bibnamefont {Bonitz}},\ }\bibfield  {title} {\bibinfo {title} {Ab initio path integral monte carlo results for the dynamic structure factor of correlated electrons: From the electron liquid to warm dense matter},\ }\href@noop {} {\bibfield  {journal} {\bibinfo  {journal} {Physical review letters}\ }\textbf {\bibinfo {volume} {121}},\ \bibinfo {pages} {255001} (\bibinfo {year} {2018}{\natexlab{b}})}\BibitemShut {NoStop}%
\bibitem [{\citenamefont {Groth}\ \emph {et~al.}(2019{\natexlab{a}})\citenamefont {Groth}, \citenamefont {Dornheim},\ and\ \citenamefont {Vorberger}}]{dynamic_folgepaper}%
  \BibitemOpen
  \bibfield  {author} {\bibinfo {author} {\bibfnamefont {S.}~\bibnamefont {Groth}}, \bibinfo {author} {\bibfnamefont {T.}~\bibnamefont {Dornheim}},\ and\ \bibinfo {author} {\bibfnamefont {J.}~\bibnamefont {Vorberger}},\ }\bibfield  {title} {\bibinfo {title} {Ab initio path integral {M}onte {C}arlo approach to the static and dynamic density response of the uniform electron gas},\ }\href {https://link.aps.org/doi/10.1103/PhysRevB.99.235122} {\bibfield  {journal} {\bibinfo  {journal} {Phys. Rev. B}\ }\textbf {\bibinfo {volume} {99}},\ \bibinfo {pages} {235122} (\bibinfo {year} {2019}{\natexlab{a}})}\BibitemShut {NoStop}%
\bibitem [{\citenamefont {Dornheim}\ and\ \citenamefont {Vorberger}(2020)}]{Dornheim_PRE_2020}%
  \BibitemOpen
  \bibfield  {author} {\bibinfo {author} {\bibfnamefont {T.}~\bibnamefont {Dornheim}}\ and\ \bibinfo {author} {\bibfnamefont {J.}~\bibnamefont {Vorberger}},\ }\bibfield  {title} {\bibinfo {title} {Finite-size effects in the reconstruction of dynamic properties from ab initio path integral monte carlo simulations},\ }\href {https://doi.org/10.1103/PhysRevE.102.063301} {\bibfield  {journal} {\bibinfo  {journal} {Phys. Rev. E}\ }\textbf {\bibinfo {volume} {102}},\ \bibinfo {pages} {063301} (\bibinfo {year} {2020})}\BibitemShut {NoStop}%
\bibitem [{\citenamefont {Chuna}\ \emph {et~al.}(2025)\citenamefont {Chuna}, \citenamefont {Barnfield}, \citenamefont {Dornheim}, \citenamefont {Friedlander},\ and\ \citenamefont {Hoheisel}}]{chuna2025dual}%
  \BibitemOpen
  \bibfield  {author} {\bibinfo {author} {\bibfnamefont {T.}~\bibnamefont {Chuna}}, \bibinfo {author} {\bibfnamefont {N.}~\bibnamefont {Barnfield}}, \bibinfo {author} {\bibfnamefont {T.}~\bibnamefont {Dornheim}}, \bibinfo {author} {\bibfnamefont {M.~P.}\ \bibnamefont {Friedlander}},\ and\ \bibinfo {author} {\bibfnamefont {T.}~\bibnamefont {Hoheisel}},\ }\bibfield  {title} {\bibinfo {title} {Dual formulation of the maximum entropy method applied to analytic continuation of quantum monte carlo data},\ }\href@noop {} {\bibfield  {journal} {\bibinfo  {journal} {arXiv preprint arXiv:2501.01869}\ } (\bibinfo {year} {2025})}\BibitemShut {NoStop}%
\bibitem [{\citenamefont {Graziani}\ \emph {et~al.}(2014)\citenamefont {Graziani}, \citenamefont {Desjarlais}, \citenamefont {Redmer},\ and\ \citenamefont {Trickey}}]{wdm_book}%
  \BibitemOpen
  \bibinfo {editor} {\bibfnamefont {F.}~\bibnamefont {Graziani}}, \bibinfo {editor} {\bibfnamefont {M.~P.}\ \bibnamefont {Desjarlais}}, \bibinfo {editor} {\bibfnamefont {R.}~\bibnamefont {Redmer}},\ and\ \bibinfo {editor} {\bibfnamefont {S.~B.}\ \bibnamefont {Trickey}},\ eds.,\ \href@noop {} {\emph {\bibinfo {title} {Frontiers and Challenges in Warm Dense Matter}}}\ (\bibinfo  {publisher} {Springer},\ \bibinfo {address} {International Publishing},\ \bibinfo {year} {2014})\BibitemShut {NoStop}%
\bibitem [{\citenamefont {Dornheim}\ \emph {et~al.}(2020{\natexlab{a}})\citenamefont {Dornheim}, \citenamefont {Sjostrom}, \citenamefont {Tanaka},\ and\ \citenamefont {Vorberger}}]{dornheim_electron_liquid}%
  \BibitemOpen
  \bibfield  {author} {\bibinfo {author} {\bibfnamefont {T.}~\bibnamefont {Dornheim}}, \bibinfo {author} {\bibfnamefont {T.}~\bibnamefont {Sjostrom}}, \bibinfo {author} {\bibfnamefont {S.}~\bibnamefont {Tanaka}},\ and\ \bibinfo {author} {\bibfnamefont {J.}~\bibnamefont {Vorberger}},\ }\bibfield  {title} {\bibinfo {title} {Strongly coupled electron liquid: Ab initio path integral monte carlo simulations and dielectric theories},\ }\href {https://doi.org/10.1103/PhysRevB.101.045129} {\bibfield  {journal} {\bibinfo  {journal} {Phys. Rev. B}\ }\textbf {\bibinfo {volume} {101}},\ \bibinfo {pages} {045129} (\bibinfo {year} {2020}{\natexlab{a}})}\BibitemShut {NoStop}%
\bibitem [{\citenamefont {Tolias}\ \emph {et~al.}(2021)\citenamefont {Tolias}, \citenamefont {Lucco~Castello},\ and\ \citenamefont {Dornheim}}]{Tolias_JCP_2021}%
  \BibitemOpen
  \bibfield  {author} {\bibinfo {author} {\bibfnamefont {P.}~\bibnamefont {Tolias}}, \bibinfo {author} {\bibfnamefont {F.}~\bibnamefont {Lucco~Castello}},\ and\ \bibinfo {author} {\bibfnamefont {T.}~\bibnamefont {Dornheim}},\ }\bibfield  {title} {\bibinfo {title} {Integral equation theory based dielectric scheme for strongly coupled electron liquids},\ }\href {https://doi.org/10.1063/5.0065988} {\bibfield  {journal} {\bibinfo  {journal} {The Journal of Chemical Physics}\ }\textbf {\bibinfo {volume} {155}},\ \bibinfo {pages} {134115} (\bibinfo {year} {2021})}\BibitemShut {NoStop}%
\bibitem [{\citenamefont {Tolias}\ \emph {et~al.}(2023)\citenamefont {Tolias}, \citenamefont {Lucco~Castello},\ and\ \citenamefont {Dornheim}}]{Tolias_JCP_2023}%
  \BibitemOpen
  \bibfield  {author} {\bibinfo {author} {\bibfnamefont {P.}~\bibnamefont {Tolias}}, \bibinfo {author} {\bibfnamefont {F.}~\bibnamefont {Lucco~Castello}},\ and\ \bibinfo {author} {\bibfnamefont {T.}~\bibnamefont {Dornheim}},\ }\bibfield  {title} {\bibinfo {title} {{Quantum version of the integral equation theory-based dielectric scheme for strongly coupled electron liquids}},\ }\href {https://doi.org/10.1063/5.0145687} {\bibfield  {journal} {\bibinfo  {journal} {The Journal of Chemical Physics}\ }\textbf {\bibinfo {volume} {158}},\ \bibinfo {pages} {141102} (\bibinfo {year} {2023})}\BibitemShut {NoStop}%
\bibitem [{\citenamefont {Groth}\ \emph {et~al.}(2019{\natexlab{b}})\citenamefont {Groth}, \citenamefont {Dornheim},\ and\ \citenamefont {Vorberger}}]{groth2019stochasticsampling}%
  \BibitemOpen
  \bibfield  {author} {\bibinfo {author} {\bibfnamefont {S.}~\bibnamefont {Groth}}, \bibinfo {author} {\bibfnamefont {T.}~\bibnamefont {Dornheim}},\ and\ \bibinfo {author} {\bibfnamefont {J.}~\bibnamefont {Vorberger}},\ }\bibfield  {title} {\bibinfo {title} {Ab initio path integral monte carlo approach to the static and dynamic density response of the uniform electron gas},\ }\href@noop {} {\bibfield  {journal} {\bibinfo  {journal} {Physical Review B}\ }\textbf {\bibinfo {volume} {99}},\ \bibinfo {pages} {235122} (\bibinfo {year} {2019}{\natexlab{b}})}\BibitemShut {NoStop}%
\bibitem [{\citenamefont {Dornheim}\ \emph {et~al.}(2019)\citenamefont {Dornheim}, \citenamefont {Vorberger}, \citenamefont {Groth}, \citenamefont {Hoffmann}, \citenamefont {Moldabekov},\ and\ \citenamefont {Bonitz}}]{dornheim2019MLstatic}%
  \BibitemOpen
  \bibfield  {author} {\bibinfo {author} {\bibfnamefont {T.}~\bibnamefont {Dornheim}}, \bibinfo {author} {\bibfnamefont {J.}~\bibnamefont {Vorberger}}, \bibinfo {author} {\bibfnamefont {S.}~\bibnamefont {Groth}}, \bibinfo {author} {\bibfnamefont {N.}~\bibnamefont {Hoffmann}}, \bibinfo {author} {\bibfnamefont {Z.}~\bibnamefont {Moldabekov}},\ and\ \bibinfo {author} {\bibfnamefont {M.}~\bibnamefont {Bonitz}},\ }\bibfield  {title} {\bibinfo {title} {The static local field correction of the warm dense electron gas: An ab initio path integral {M}onte {C}arlo study and machine learning representation},\ }\href {https://aip.scitation.org/doi/full/10.1063/1.5123013} {\bibfield  {journal} {\bibinfo  {journal} {J. Chem. Phys}\ }\textbf {\bibinfo {volume} {151}},\ \bibinfo {pages} {194104} (\bibinfo {year} {2019})}\BibitemShut {NoStop}%
\bibitem [{\citenamefont {Berg}(2004{\natexlab{a}})}]{berg2004markov}%
  \BibitemOpen
  \bibfield  {author} {\bibinfo {author} {\bibfnamefont {B.~A.}\ \bibnamefont {Berg}},\ }\href@noop {} {\emph {\bibinfo {title} {Markov chain Monte Carlo simulations and their statistical analysis: with web-based Fortran code}}}\ (\bibinfo  {publisher} {World Scientific Publishing Company},\ \bibinfo {year} {2004})\BibitemShut {NoStop}%
\bibitem [{\citenamefont {Berg}(2004{\natexlab{b}})}]{berg2004introduction}%
  \BibitemOpen
  \bibfield  {author} {\bibinfo {author} {\bibfnamefont {B.~A.}\ \bibnamefont {Berg}},\ }\bibfield  {title} {\bibinfo {title} {Introduction to markov chain monte carlo simulations and their statistical analysis},\ }\href@noop {} {\bibfield  {journal} {\bibinfo  {journal} {arXiv preprint cond-mat/0410490}\ } (\bibinfo {year} {2004}{\natexlab{b}})}\BibitemShut {NoStop}%
\bibitem [{\citenamefont {James}\ \emph {et~al.}(2013)\citenamefont {James}, \citenamefont {Witten}, \citenamefont {Hastie}, \citenamefont {Tibshirani} \emph {et~al.}}]{james2013statisticallearning}%
  \BibitemOpen
  \bibfield  {author} {\bibinfo {author} {\bibfnamefont {G.}~\bibnamefont {James}}, \bibinfo {author} {\bibfnamefont {D.}~\bibnamefont {Witten}}, \bibinfo {author} {\bibfnamefont {T.}~\bibnamefont {Hastie}}, \bibinfo {author} {\bibfnamefont {R.}~\bibnamefont {Tibshirani}}, \emph {et~al.},\ }\href@noop {} {\emph {\bibinfo {title} {An introduction to statistical learning}}},\ Vol.\ \bibinfo {volume} {112}\ (\bibinfo  {publisher} {Springer},\ \bibinfo {year} {2013})\BibitemShut {NoStop}%
\bibitem [{\citenamefont {Hamann}\ \emph {et~al.}(2023)\citenamefont {Hamann}, \citenamefont {Kordts}, \citenamefont {Filinov}, \citenamefont {Bonitz}, \citenamefont {Dornheim},\ and\ \citenamefont {Vorberger}}]{Hamann2023rotonfeature}%
  \BibitemOpen
  \bibfield  {author} {\bibinfo {author} {\bibfnamefont {P.}~\bibnamefont {Hamann}}, \bibinfo {author} {\bibfnamefont {L.}~\bibnamefont {Kordts}}, \bibinfo {author} {\bibfnamefont {A.}~\bibnamefont {Filinov}}, \bibinfo {author} {\bibfnamefont {M.}~\bibnamefont {Bonitz}}, \bibinfo {author} {\bibfnamefont {T.}~\bibnamefont {Dornheim}},\ and\ \bibinfo {author} {\bibfnamefont {J.}~\bibnamefont {Vorberger}},\ }\bibfield  {title} {\bibinfo {title} {Prediction of a roton-type feature in warm dense hydrogen},\ }\href {https://doi.org/10.1103/PhysRevResearch.5.033039} {\bibfield  {journal} {\bibinfo  {journal} {Phys. Rev. Res.}\ }\textbf {\bibinfo {volume} {5}},\ \bibinfo {pages} {033039} (\bibinfo {year} {2023})}\BibitemShut {NoStop}%
\bibitem [{\citenamefont {Dornheim}\ \emph {et~al.}(2022{\natexlab{a}})\citenamefont {Dornheim}, \citenamefont {Moldabekov}, \citenamefont {Vorberger}, \citenamefont {K{\"a}hlert},\ and\ \citenamefont {Bonitz}}]{Dornheim_Nature_2022}%
  \BibitemOpen
  \bibfield  {author} {\bibinfo {author} {\bibfnamefont {T.}~\bibnamefont {Dornheim}}, \bibinfo {author} {\bibfnamefont {Z.}~\bibnamefont {Moldabekov}}, \bibinfo {author} {\bibfnamefont {J.}~\bibnamefont {Vorberger}}, \bibinfo {author} {\bibfnamefont {H.}~\bibnamefont {K{\"a}hlert}},\ and\ \bibinfo {author} {\bibfnamefont {M.}~\bibnamefont {Bonitz}},\ }\bibfield  {title} {\bibinfo {title} {Electronic pair alignment and roton feature in the warm dense electron gas},\ }\href {https://doi.org/10.1038/s42005-022-01078-9} {\bibfield  {journal} {\bibinfo  {journal} {Communications Physics}\ }\textbf {\bibinfo {volume} {5}},\ \bibinfo {pages} {304} (\bibinfo {year} {2022}{\natexlab{a}})}\BibitemShut {NoStop}%
\bibitem [{\citenamefont {Koskelo}\ \emph {et~al.}(2025)\citenamefont {Koskelo}, \citenamefont {Reining},\ and\ \citenamefont {Gatti}}]{koskelo2025short}%
  \BibitemOpen
  \bibfield  {author} {\bibinfo {author} {\bibfnamefont {J.}~\bibnamefont {Koskelo}}, \bibinfo {author} {\bibfnamefont {L.}~\bibnamefont {Reining}},\ and\ \bibinfo {author} {\bibfnamefont {M.}~\bibnamefont {Gatti}},\ }\bibfield  {title} {\bibinfo {title} {Short-range excitonic phenomena in low-density metals},\ }\href@noop {} {\bibfield  {journal} {\bibinfo  {journal} {Physical Review Letters}\ }\textbf {\bibinfo {volume} {134}},\ \bibinfo {pages} {046402} (\bibinfo {year} {2025})}\BibitemShut {NoStop}%
\bibitem [{\citenamefont {Filinov}\ \emph {et~al.}(2023)\citenamefont {Filinov}, \citenamefont {Ara},\ and\ \citenamefont {Tkachenko}}]{Filinov_PRB_2023}%
  \BibitemOpen
  \bibfield  {author} {\bibinfo {author} {\bibfnamefont {A.~V.}\ \bibnamefont {Filinov}}, \bibinfo {author} {\bibfnamefont {J.}~\bibnamefont {Ara}},\ and\ \bibinfo {author} {\bibfnamefont {I.~M.}\ \bibnamefont {Tkachenko}},\ }\bibfield  {title} {\bibinfo {title} {Dynamical response in strongly coupled uniform electron liquids: Observation of plasmon-roton coexistence using nine sum rules, shannon information entropy, and path-integral monte carlo simulations},\ }\href {https://doi.org/10.1103/PhysRevB.107.195143} {\bibfield  {journal} {\bibinfo  {journal} {Phys. Rev. B}\ }\textbf {\bibinfo {volume} {107}},\ \bibinfo {pages} {195143} (\bibinfo {year} {2023})}\BibitemShut {NoStop}%
\bibitem [{\citenamefont {Takada}(2016)}]{Takada_PRB_2016}%
  \BibitemOpen
  \bibfield  {author} {\bibinfo {author} {\bibfnamefont {Y.}~\bibnamefont {Takada}},\ }\bibfield  {title} {\bibinfo {title} {Emergence of an excitonic collective mode in the dilute electron gas},\ }\href {https://doi.org/10.1103/PhysRevB.94.245106} {\bibfield  {journal} {\bibinfo  {journal} {Phys. Rev. B}\ }\textbf {\bibinfo {volume} {94}},\ \bibinfo {pages} {245106} (\bibinfo {year} {2016})}\BibitemShut {NoStop}%
\bibitem [{\citenamefont {Godfrin}\ \emph {et~al.}(2012)\citenamefont {Godfrin}, \citenamefont {Meschke}, \citenamefont {Lauter}, \citenamefont {Sultan}, \citenamefont {B{\"o}hm}, \citenamefont {Krotscheck},\ and\ \citenamefont {Panholzer}}]{Godfrin2012}%
  \BibitemOpen
  \bibfield  {author} {\bibinfo {author} {\bibfnamefont {H.}~\bibnamefont {Godfrin}}, \bibinfo {author} {\bibfnamefont {M.}~\bibnamefont {Meschke}}, \bibinfo {author} {\bibfnamefont {H.-J.}\ \bibnamefont {Lauter}}, \bibinfo {author} {\bibfnamefont {A.}~\bibnamefont {Sultan}}, \bibinfo {author} {\bibfnamefont {H.~M.}\ \bibnamefont {B{\"o}hm}}, \bibinfo {author} {\bibfnamefont {E.}~\bibnamefont {Krotscheck}},\ and\ \bibinfo {author} {\bibfnamefont {M.}~\bibnamefont {Panholzer}},\ }\bibfield  {title} {\bibinfo {title} {Observation of a roton collective mode in a two-dimensional fermi liquid},\ }\href {https://doi.org/10.1038/nature10919} {\bibfield  {journal} {\bibinfo  {journal} {Nature}\ }\textbf {\bibinfo {volume} {483}},\ \bibinfo {pages} {576} (\bibinfo {year} {2012})}\BibitemShut {NoStop}%
\bibitem [{\citenamefont {Dornheim}\ \emph {et~al.}(2022{\natexlab{b}})\citenamefont {Dornheim}, \citenamefont {Moldabekov}, \citenamefont {Vorberger},\ and\ \citenamefont {Militzer}}]{Dornheim_SciRep_2022}%
  \BibitemOpen
  \bibfield  {author} {\bibinfo {author} {\bibfnamefont {T.}~\bibnamefont {Dornheim}}, \bibinfo {author} {\bibfnamefont {Z.~A.}\ \bibnamefont {Moldabekov}}, \bibinfo {author} {\bibfnamefont {J.}~\bibnamefont {Vorberger}},\ and\ \bibinfo {author} {\bibfnamefont {B.}~\bibnamefont {Militzer}},\ }\bibfield  {title} {\bibinfo {title} {Path integral monte carlo approach to the structural properties and collective excitations of liquid $^3$he without fixed nodes},\ }\href {https://doi.org/10.1038/s41598-021-04355-9} {\bibfield  {journal} {\bibinfo  {journal} {Scientific Reports}\ }\textbf {\bibinfo {volume} {12}},\ \bibinfo {pages} {708} (\bibinfo {year} {2022}{\natexlab{b}})}\BibitemShut {NoStop}%
\bibitem [{\citenamefont {Bobrov}\ \emph {et~al.}(2016)\citenamefont {Bobrov}, \citenamefont {Trigger},\ and\ \citenamefont {Litinski}}]{Trigger}%
  \BibitemOpen
  \bibfield  {author} {\bibinfo {author} {\bibfnamefont {V.}~\bibnamefont {Bobrov}}, \bibinfo {author} {\bibfnamefont {S.}~\bibnamefont {Trigger}},\ and\ \bibinfo {author} {\bibfnamefont {D.}~\bibnamefont {Litinski}},\ }\bibfield  {title} {\bibinfo {title} {Universality of the phonon–roton spectrum in liquids and superfluidity of 4he},\ }\href {https://doi.org/doi:10.1515/zna-2015-0397} {\bibfield  {journal} {\bibinfo  {journal} {Zeitschrift für Naturforschung A}\ }\textbf {\bibinfo {volume} {71}},\ \bibinfo {pages} {565} (\bibinfo {year} {2016})}\BibitemShut {NoStop}%
\bibitem [{\citenamefont {Ferr\'e}\ and\ \citenamefont {Boronat}(2016)}]{Ferre_PRB_2016}%
  \BibitemOpen
  \bibfield  {author} {\bibinfo {author} {\bibfnamefont {G.}~\bibnamefont {Ferr\'e}}\ and\ \bibinfo {author} {\bibfnamefont {J.}~\bibnamefont {Boronat}},\ }\bibfield  {title} {\bibinfo {title} {Dynamic structure factor of liquid $^{4}\mathrm{He}$ across the normal-superfluid transition},\ }\href {https://doi.org/10.1103/PhysRevB.93.104510} {\bibfield  {journal} {\bibinfo  {journal} {Phys. Rev. B}\ }\textbf {\bibinfo {volume} {93}},\ \bibinfo {pages} {104510} (\bibinfo {year} {2016})}\BibitemShut {NoStop}%
\bibitem [{\citenamefont {Murillo}(1998)}]{murillo1998SLFC}%
  \BibitemOpen
  \bibfield  {author} {\bibinfo {author} {\bibfnamefont {M.}~\bibnamefont {Murillo}},\ }\bibfield  {title} {\bibinfo {title} {Static local field correction description of acoustic waves in strongly coupling dusty plasmas},\ }\href@noop {} {\bibfield  {journal} {\bibinfo  {journal} {Physics of Plasmas}\ }\textbf {\bibinfo {volume} {5}},\ \bibinfo {pages} {3116} (\bibinfo {year} {1998})}\BibitemShut {NoStop}%
\bibitem [{\citenamefont {Murillo}(2000)}]{murillo2000DLFC}%
  \BibitemOpen
  \bibfield  {author} {\bibinfo {author} {\bibfnamefont {M.}~\bibnamefont {Murillo}},\ }\bibfield  {title} {\bibinfo {title} {Longitudinal collective modes of strongly coupled dusty plasmas at finite frequencies and wavevectors},\ }\href@noop {} {\bibfield  {journal} {\bibinfo  {journal} {Physics of Plasmas}\ }\textbf {\bibinfo {volume} {7}},\ \bibinfo {pages} {33} (\bibinfo {year} {2000})}\BibitemShut {NoStop}%
\bibitem [{\citenamefont {Korolov}\ \emph {et~al.}(2015)\citenamefont {Korolov}, \citenamefont {Kalman}, \citenamefont {Silvestri},\ and\ \citenamefont {Donkó}}]{Korolov_CPP_2015}%
  \BibitemOpen
  \bibfield  {author} {\bibinfo {author} {\bibfnamefont {I.}~\bibnamefont {Korolov}}, \bibinfo {author} {\bibfnamefont {G.~J.}\ \bibnamefont {Kalman}}, \bibinfo {author} {\bibfnamefont {L.}~\bibnamefont {Silvestri}},\ and\ \bibinfo {author} {\bibfnamefont {Z.}~\bibnamefont {Donkó}},\ }\bibfield  {title} {\bibinfo {title} {The dynamical structure function of the one-component plasma revisited},\ }\href {https://doi.org/https://doi.org/10.1002/ctpp.201400098} {\bibfield  {journal} {\bibinfo  {journal} {Contributions to Plasma Physics}\ }\textbf {\bibinfo {volume} {55}},\ \bibinfo {pages} {421} (\bibinfo {year} {2015})}\BibitemShut {NoStop}%
\bibitem [{\citenamefont {Choi}\ \emph {et~al.}(2019)\citenamefont {Choi}, \citenamefont {Dharuman},\ and\ \citenamefont {Murillo}}]{Choi_PRE_2019}%
  \BibitemOpen
  \bibfield  {author} {\bibinfo {author} {\bibfnamefont {Y.}~\bibnamefont {Choi}}, \bibinfo {author} {\bibfnamefont {G.}~\bibnamefont {Dharuman}},\ and\ \bibinfo {author} {\bibfnamefont {M.~S.}\ \bibnamefont {Murillo}},\ }\bibfield  {title} {\bibinfo {title} {High-frequency response of classical strongly coupled plasmas},\ }\href {https://doi.org/10.1103/PhysRevE.100.013206} {\bibfield  {journal} {\bibinfo  {journal} {Phys. Rev. E}\ }\textbf {\bibinfo {volume} {100}},\ \bibinfo {pages} {013206} (\bibinfo {year} {2019})}\BibitemShut {NoStop}%
\bibitem [{\citenamefont {Takada}\ and\ \citenamefont {Yasuhara}(2002)}]{Takada_PRL_2002}%
  \BibitemOpen
  \bibfield  {author} {\bibinfo {author} {\bibfnamefont {Y.}~\bibnamefont {Takada}}\ and\ \bibinfo {author} {\bibfnamefont {H.}~\bibnamefont {Yasuhara}},\ }\bibfield  {title} {\bibinfo {title} {Dynamical structure factor of the homogeneous electron liquid: Its accurate shape and the interpretation of experiments on aluminum},\ }\href {https://doi.org/10.1103/PhysRevLett.89.216402} {\bibfield  {journal} {\bibinfo  {journal} {Phys. Rev. Lett.}\ }\textbf {\bibinfo {volume} {89}},\ \bibinfo {pages} {216402} (\bibinfo {year} {2002})}\BibitemShut {NoStop}%
\bibitem [{\citenamefont {Ullrich}(2011)}]{ullrich2011time}%
  \BibitemOpen
  \bibfield  {author} {\bibinfo {author} {\bibfnamefont {C.~A.}\ \bibnamefont {Ullrich}},\ }\href@noop {} {\emph {\bibinfo {title} {Time-dependent density-functional theory: concepts and applications}}}\ (\bibinfo  {publisher} {OUP Oxford},\ \bibinfo {year} {2011})\BibitemShut {NoStop}%
\bibitem [{\citenamefont {Panholzer}\ \emph {et~al.}(2018)\citenamefont {Panholzer}, \citenamefont {Gatti},\ and\ \citenamefont {Reining}}]{Panholzer_PRL_2018}%
  \BibitemOpen
  \bibfield  {author} {\bibinfo {author} {\bibfnamefont {M.}~\bibnamefont {Panholzer}}, \bibinfo {author} {\bibfnamefont {M.}~\bibnamefont {Gatti}},\ and\ \bibinfo {author} {\bibfnamefont {L.}~\bibnamefont {Reining}},\ }\bibfield  {title} {\bibinfo {title} {Nonlocal and nonadiabatic effects in the charge-density response of solids: A time-dependent density-functional approach},\ }\href {https://doi.org/10.1103/PhysRevLett.120.166402} {\bibfield  {journal} {\bibinfo  {journal} {Phys. Rev. Lett.}\ }\textbf {\bibinfo {volume} {120}},\ \bibinfo {pages} {166402} (\bibinfo {year} {2018})}\BibitemShut {NoStop}%
\bibitem [{\citenamefont {Ruzsinszky}\ \emph {et~al.}(2020)\citenamefont {Ruzsinszky}, \citenamefont {Nepal}, \citenamefont {Pitarke},\ and\ \citenamefont {Perdew}}]{Ruzsinszky_PRB_2020}%
  \BibitemOpen
  \bibfield  {author} {\bibinfo {author} {\bibfnamefont {A.}~\bibnamefont {Ruzsinszky}}, \bibinfo {author} {\bibfnamefont {N.~K.}\ \bibnamefont {Nepal}}, \bibinfo {author} {\bibfnamefont {J.~M.}\ \bibnamefont {Pitarke}},\ and\ \bibinfo {author} {\bibfnamefont {J.~P.}\ \bibnamefont {Perdew}},\ }\bibfield  {title} {\bibinfo {title} {Constraint-based wave vector and frequency dependent exchange-correlation kernel of the uniform electron gas},\ }\href {https://doi.org/10.1103/PhysRevB.101.245135} {\bibfield  {journal} {\bibinfo  {journal} {Phys. Rev. B}\ }\textbf {\bibinfo {volume} {101}},\ \bibinfo {pages} {245135} (\bibinfo {year} {2020})}\BibitemShut {NoStop}%
\bibitem [{\citenamefont {Boon}\ and\ \citenamefont {Yip}(1991)}]{boon1991molecular}%
  \BibitemOpen
  \bibfield  {author} {\bibinfo {author} {\bibfnamefont {J.~P.}\ \bibnamefont {Boon}}\ and\ \bibinfo {author} {\bibfnamefont {S.}~\bibnamefont {Yip}},\ }\href@noop {} {\emph {\bibinfo {title} {Molecular hydrodynamics}}}\ (\bibinfo  {publisher} {Courier Corporation},\ \bibinfo {year} {1991})\BibitemShut {NoStop}%
\bibitem [{\citenamefont {Ichimaru}(2018)}]{ichimaru2018plasmavol1}%
  \BibitemOpen
  \bibfield  {author} {\bibinfo {author} {\bibfnamefont {S.}~\bibnamefont {Ichimaru}},\ }\href@noop {} {\emph {\bibinfo {title} {Statistical plasma physics, volume I: basic principles}}}\ (\bibinfo  {publisher} {CRC Press},\ \bibinfo {year} {2018})\BibitemShut {NoStop}%
\bibitem [{\citenamefont {Böhme}\ \emph {et~al.}(2023)\citenamefont {Böhme}, \citenamefont {Fletcher}, \citenamefont {Döppner}, \citenamefont {Kraus}, \citenamefont {Baczewski}, \citenamefont {Preston}, \citenamefont {MacDonald}, \citenamefont {Graziani}, \citenamefont {Moldabekov}, \citenamefont {Vorberger},\ and\ \citenamefont {Dornheim}}]{boehme2023evidence}%
  \BibitemOpen
  \bibfield  {author} {\bibinfo {author} {\bibfnamefont {M.~P.}\ \bibnamefont {Böhme}}, \bibinfo {author} {\bibfnamefont {L.~B.}\ \bibnamefont {Fletcher}}, \bibinfo {author} {\bibfnamefont {T.}~\bibnamefont {Döppner}}, \bibinfo {author} {\bibfnamefont {D.}~\bibnamefont {Kraus}}, \bibinfo {author} {\bibfnamefont {A.~D.}\ \bibnamefont {Baczewski}}, \bibinfo {author} {\bibfnamefont {T.~R.}\ \bibnamefont {Preston}}, \bibinfo {author} {\bibfnamefont {M.~J.}\ \bibnamefont {MacDonald}}, \bibinfo {author} {\bibfnamefont {F.~R.}\ \bibnamefont {Graziani}}, \bibinfo {author} {\bibfnamefont {Z.~A.}\ \bibnamefont {Moldabekov}}, \bibinfo {author} {\bibfnamefont {J.}~\bibnamefont {Vorberger}},\ and\ \bibinfo {author} {\bibfnamefont {T.}~\bibnamefont {Dornheim}},\ }\href@noop {} {\bibinfo {title} {Evidence of free-bound transitions in warm dense matter and their impact on equation-of-state measurements}} (\bibinfo {year} {2023}),\ \Eprint {https://arxiv.org/abs/2306.17653} {arXiv:2306.17653 [physics.plasm-ph]} \BibitemShut
  {NoStop}%
\bibitem [{\citenamefont {Kwong}\ and\ \citenamefont {Bonitz}(2000)}]{kwong_prl-00}%
  \BibitemOpen
  \bibfield  {author} {\bibinfo {author} {\bibfnamefont {N.-H.}\ \bibnamefont {Kwong}}\ and\ \bibinfo {author} {\bibfnamefont {M.}~\bibnamefont {Bonitz}},\ }\bibfield  {title} {\bibinfo {title} {Real-time kadanoff-baym approach to plasma oscillations in a correlated electron gas},\ }\href {https://journals.aps.org/prl/abstract/10.1103/PhysRevLett.84.1768} {\bibfield  {journal} {\bibinfo  {journal} {Phys. Rev. Lett}\ }\textbf {\bibinfo {volume} {84}},\ \bibinfo {pages} {1768} (\bibinfo {year} {2000})}\BibitemShut {NoStop}%
\bibitem [{\citenamefont {Kas}\ and\ \citenamefont {Rehr}(2017)}]{Kas_PRL_2017}%
  \BibitemOpen
  \bibfield  {author} {\bibinfo {author} {\bibfnamefont {J.~J.}\ \bibnamefont {Kas}}\ and\ \bibinfo {author} {\bibfnamefont {J.~J.}\ \bibnamefont {Rehr}},\ }\bibfield  {title} {\bibinfo {title} {Finite temperature green's function approach for excited state and thermodynamic properties of cool to warm dense matter},\ }\href {https://doi.org/10.1103/PhysRevLett.119.176403} {\bibfield  {journal} {\bibinfo  {journal} {Phys. Rev. Lett.}\ }\textbf {\bibinfo {volume} {119}},\ \bibinfo {pages} {176403} (\bibinfo {year} {2017})}\BibitemShut {NoStop}%
\bibitem [{\citenamefont {Schl\"unzen}\ \emph {et~al.}(2020)\citenamefont {Schl\"unzen}, \citenamefont {Joost},\ and\ \citenamefont {Bonitz}}]{Schluenzen_PRL_2020}%
  \BibitemOpen
  \bibfield  {author} {\bibinfo {author} {\bibfnamefont {N.}~\bibnamefont {Schl\"unzen}}, \bibinfo {author} {\bibfnamefont {J.-P.}\ \bibnamefont {Joost}},\ and\ \bibinfo {author} {\bibfnamefont {M.}~\bibnamefont {Bonitz}},\ }\bibfield  {title} {\bibinfo {title} {Achieving the scaling limit for nonequilibrium green functions simulations},\ }\href {https://doi.org/10.1103/PhysRevLett.124.076601} {\bibfield  {journal} {\bibinfo  {journal} {Phys. Rev. Lett.}\ }\textbf {\bibinfo {volume} {124}},\ \bibinfo {pages} {076601} (\bibinfo {year} {2020})}\BibitemShut {NoStop}%
\bibitem [{\citenamefont {Holas}\ and\ \citenamefont {Rahman}(1987)}]{Holas_PRB_1987}%
  \BibitemOpen
  \bibfield  {author} {\bibinfo {author} {\bibfnamefont {A.}~\bibnamefont {Holas}}\ and\ \bibinfo {author} {\bibfnamefont {S.}~\bibnamefont {Rahman}},\ }\bibfield  {title} {\bibinfo {title} {Dynamic local-field factor of an electron liquid in the quantum versions of the singwi-tosi-land-sj\"olander and vashishta-singwi theories},\ }\href {https://doi.org/10.1103/PhysRevB.35.2720} {\bibfield  {journal} {\bibinfo  {journal} {Phys. Rev. B}\ }\textbf {\bibinfo {volume} {35}},\ \bibinfo {pages} {2720} (\bibinfo {year} {1987})}\BibitemShut {NoStop}%
\bibitem [{\citenamefont {Tolias}\ \emph {et~al.}(2024)\citenamefont {Tolias}, \citenamefont {Dornheim},\ and\ \citenamefont {Vorberger}}]{tolias2024density}%
  \BibitemOpen
  \bibfield  {author} {\bibinfo {author} {\bibfnamefont {P.}~\bibnamefont {Tolias}}, \bibinfo {author} {\bibfnamefont {T.}~\bibnamefont {Dornheim}},\ and\ \bibinfo {author} {\bibfnamefont {J.}~\bibnamefont {Vorberger}},\ }\bibfield  {title} {\bibinfo {title} {On the density-density correlations of the non-interacting finite temperature electron gas},\ }\href@noop {} {\bibfield  {journal} {\bibinfo  {journal} {arXiv preprint arXiv:2410.22942}\ } (\bibinfo {year} {2024})}\BibitemShut {NoStop}%
\bibitem [{\citenamefont {Kim}\ \emph {et~al.}(2015)\citenamefont {Kim}, \citenamefont {Petreczky},\ and\ \citenamefont {Rothkopf}}]{KimPetreczkyRothkopf2015ComparisonBRMandMEM}%
  \BibitemOpen
  \bibfield  {author} {\bibinfo {author} {\bibfnamefont {S.}~\bibnamefont {Kim}}, \bibinfo {author} {\bibfnamefont {P.}~\bibnamefont {Petreczky}},\ and\ \bibinfo {author} {\bibfnamefont {A.}~\bibnamefont {Rothkopf}},\ }\bibfield  {title} {\bibinfo {title} {Lattice nrqcd study of s-and p-wave bottomonium states in a thermal medium with n f= 2+ 1 light flavors},\ }\href@noop {} {\bibfield  {journal} {\bibinfo  {journal} {Physical Review D}\ }\textbf {\bibinfo {volume} {91}},\ \bibinfo {pages} {054511} (\bibinfo {year} {2015})}\BibitemShut {NoStop}%
\bibitem [{\citenamefont {Asakawa}\ \emph {et~al.}(2001)\citenamefont {Asakawa}, \citenamefont {Nakahara},\ and\ \citenamefont {Hatsuda}}]{Asakawa2001MEM}%
  \BibitemOpen
  \bibfield  {author} {\bibinfo {author} {\bibfnamefont {M.}~\bibnamefont {Asakawa}}, \bibinfo {author} {\bibfnamefont {Y.}~\bibnamefont {Nakahara}},\ and\ \bibinfo {author} {\bibfnamefont {T.}~\bibnamefont {Hatsuda}},\ }\bibfield  {title} {\bibinfo {title} {Maximum entropy analysis of the spectral functions in lattice qcd},\ }\href {https://doi.org/https://doi.org/10.1016/S0146-6410(01)00150-8} {\bibfield  {journal} {\bibinfo  {journal} {Progress in Particle and Nuclear Physics}\ }\textbf {\bibinfo {volume} {46}},\ \bibinfo {pages} {459} (\bibinfo {year} {2001})}\BibitemShut {NoStop}%
\bibitem [{\citenamefont {Gull}(1989)}]{gull1989MEMBayesianWeighting}%
  \BibitemOpen
  \bibfield  {author} {\bibinfo {author} {\bibfnamefont {S.~F.}\ \bibnamefont {Gull}},\ }\bibfield  {title} {\bibinfo {title} {Developments in maximum entropy data analysis},\ }in\ \href@noop {} {\emph {\bibinfo {booktitle} {Maximum Entropy and Bayesian Methods: Cambridge, England, 1988}}}\ (\bibinfo  {publisher} {Springer},\ \bibinfo {year} {1989})\ pp.\ \bibinfo {pages} {53--71}\BibitemShut {NoStop}%
\bibitem [{\citenamefont {Bryan}(1990)}]{bryan1990algorithm}%
  \BibitemOpen
  \bibfield  {author} {\bibinfo {author} {\bibfnamefont {R.}~\bibnamefont {Bryan}},\ }\bibfield  {title} {\bibinfo {title} {Maximum entropy analysis of oversampled data problems},\ }\href@noop {} {\bibfield  {journal} {\bibinfo  {journal} {European Biophysics Journal}\ }\textbf {\bibinfo {volume} {18}},\ \bibinfo {pages} {165} (\bibinfo {year} {1990})}\BibitemShut {NoStop}%
\bibitem [{\citenamefont {Rothkopf}(2020)}]{rothkopf2020bryan}%
  \BibitemOpen
  \bibfield  {author} {\bibinfo {author} {\bibfnamefont {A.}~\bibnamefont {Rothkopf}},\ }\bibfield  {title} {\bibinfo {title} {Bryan’s maximum entropy method—diagnosis of a flawed argument and its remedy},\ }\href@noop {} {\bibfield  {journal} {\bibinfo  {journal} {Data}\ }\textbf {\bibinfo {volume} {5}},\ \bibinfo {pages} {85} (\bibinfo {year} {2020})}\BibitemShut {NoStop}%
\bibitem [{\citenamefont {Boninsegni}\ and\ \citenamefont {Ceperley}(1996)}]{Boninsegni1996MEM}%
  \BibitemOpen
  \bibfield  {author} {\bibinfo {author} {\bibfnamefont {M.}~\bibnamefont {Boninsegni}}\ and\ \bibinfo {author} {\bibfnamefont {D.~M.}\ \bibnamefont {Ceperley}},\ }\bibfield  {title} {\bibinfo {title} {Density fluctuations in liquid4he. path integrals and maximum entropy},\ }\href {https://doi.org/10.1007/BF00751861} {\bibfield  {journal} {\bibinfo  {journal} {Journal of Low Temperature Physics}\ }\textbf {\bibinfo {volume} {104}},\ \bibinfo {pages} {339} (\bibinfo {year} {1996})}\BibitemShut {NoStop}%
\bibitem [{\citenamefont {Kugler}(1975)}]{kugler1975LFCs}%
  \BibitemOpen
  \bibfield  {author} {\bibinfo {author} {\bibfnamefont {A.~A.}\ \bibnamefont {Kugler}},\ }\bibfield  {title} {\bibinfo {title} {Theory of the local field correction in an electron gas},\ }\href {http://link.springer.com/article/10.1007/BF01024183} {\bibfield  {journal} {\bibinfo  {journal} {J. Stat. Phys}\ }\textbf {\bibinfo {volume} {12}},\ \bibinfo {pages} {35} (\bibinfo {year} {1975})}\BibitemShut {NoStop}%
\bibitem [{\citenamefont {Vashishta}\ and\ \citenamefont {Singwi}(1972)}]{VashishtaSingwi1972VS-LFC}%
  \BibitemOpen
  \bibfield  {author} {\bibinfo {author} {\bibfnamefont {P.}~\bibnamefont {Vashishta}}\ and\ \bibinfo {author} {\bibfnamefont {K.~S.}\ \bibnamefont {Singwi}},\ }\bibfield  {title} {\bibinfo {title} {Electron correlations at metallic densities v},\ }\href {http://link.aps.org/doi/10.1103/PhysRevB.6.875} {\bibfield  {journal} {\bibinfo  {journal} {Phys. Rev. B}\ }\textbf {\bibinfo {volume} {6}},\ \bibinfo {pages} {875} (\bibinfo {year} {1972})}\BibitemShut {NoStop}%
\bibitem [{\citenamefont {Singwi}\ \emph {et~al.}(1968)\citenamefont {Singwi}, \citenamefont {Tosi}, \citenamefont {Land},\ and\ \citenamefont {Sj\"olander}}]{Singwi1968STLS-LFC}%
  \BibitemOpen
  \bibfield  {author} {\bibinfo {author} {\bibfnamefont {K.~S.}\ \bibnamefont {Singwi}}, \bibinfo {author} {\bibfnamefont {M.~P.}\ \bibnamefont {Tosi}}, \bibinfo {author} {\bibfnamefont {R.~H.}\ \bibnamefont {Land}},\ and\ \bibinfo {author} {\bibfnamefont {A.}~\bibnamefont {Sj\"olander}},\ }\bibfield  {title} {\bibinfo {title} {Electron correlations at metallic densities},\ }\href {http://link.aps.org/doi/10.1103/PhysRev.176.589} {\bibfield  {journal} {\bibinfo  {journal} {Phys. Rev}\ }\textbf {\bibinfo {volume} {176}},\ \bibinfo {pages} {589} (\bibinfo {year} {1968})}\BibitemShut {NoStop}%
\bibitem [{\citenamefont {Sjostrom}\ and\ \citenamefont {Dufty}(2013)}]{Sjostrom2013STLS2-LFC}%
  \BibitemOpen
  \bibfield  {author} {\bibinfo {author} {\bibfnamefont {T.}~\bibnamefont {Sjostrom}}\ and\ \bibinfo {author} {\bibfnamefont {J.}~\bibnamefont {Dufty}},\ }\bibfield  {title} {\bibinfo {title} {Uniform electron gas at finite temperatures},\ }\href {http://link.aps.org/doi/10.1103/PhysRevB.88.115123} {\bibfield  {journal} {\bibinfo  {journal} {Phys. Rev. B}\ }\textbf {\bibinfo {volume} {88}},\ \bibinfo {pages} {115123} (\bibinfo {year} {2013})}\BibitemShut {NoStop}%
\bibitem [{\citenamefont {Atwal}\ and\ \citenamefont {Ashcroft}(2002)}]{atwal2002fullyconserving}%
  \BibitemOpen
  \bibfield  {author} {\bibinfo {author} {\bibfnamefont {G.}~\bibnamefont {Atwal}}\ and\ \bibinfo {author} {\bibfnamefont {N.}~\bibnamefont {Ashcroft}},\ }\bibfield  {title} {\bibinfo {title} {Relaxation of an electron system: Conserving approximation},\ }\href@noop {} {\bibfield  {journal} {\bibinfo  {journal} {Physical Review B}\ }\textbf {\bibinfo {volume} {65}},\ \bibinfo {pages} {115109} (\bibinfo {year} {2002})}\BibitemShut {NoStop}%
\bibitem [{\citenamefont {Morawetz}\ and\ \citenamefont {Fuhrmann}(2000)}]{Morawetz2000quantumliquids}%
  \BibitemOpen
  \bibfield  {author} {\bibinfo {author} {\bibfnamefont {K.}~\bibnamefont {Morawetz}}\ and\ \bibinfo {author} {\bibfnamefont {U.}~\bibnamefont {Fuhrmann}},\ }\bibfield  {title} {\bibinfo {title} {General response function for interacting quantum liquids},\ }\href {https://doi.org/10.1103/PhysRevE.61.2272} {\bibfield  {journal} {\bibinfo  {journal} {Phys. Rev. E}\ }\textbf {\bibinfo {volume} {61}},\ \bibinfo {pages} {2272} (\bibinfo {year} {2000})}\BibitemShut {NoStop}%
\bibitem [{\citenamefont {Murillo}(2004)}]{murillo2004strongly}%
  \BibitemOpen
  \bibfield  {author} {\bibinfo {author} {\bibfnamefont {M.~S.}\ \bibnamefont {Murillo}},\ }\bibfield  {title} {\bibinfo {title} {Strongly coupled plasma physics and high energy-density matter},\ }\href@noop {} {\bibfield  {journal} {\bibinfo  {journal} {Physics of Plasmas}\ }\textbf {\bibinfo {volume} {11}},\ \bibinfo {pages} {2964} (\bibinfo {year} {2004})}\BibitemShut {NoStop}%
\bibitem [{\citenamefont {Ichimaru}(1982)}]{ichimaru1982stronglycoupledplasma}%
  \BibitemOpen
  \bibfield  {author} {\bibinfo {author} {\bibfnamefont {S.}~\bibnamefont {Ichimaru}},\ }\bibfield  {title} {\bibinfo {title} {Strongly coupled plasmas: high-density classical plasmas and degenerate electron liquids},\ }\href {https://journals.aps.org/rmp/abstract/10.1103/RevModPhys.54.1017} {\bibfield  {journal} {\bibinfo  {journal} {Rev. Mod. Phys}\ }\textbf {\bibinfo {volume} {54}},\ \bibinfo {pages} {1017} (\bibinfo {year} {1982})}\BibitemShut {NoStop}%
\bibitem [{\citenamefont {Kugler}(1970)}]{kugler1970bounds}%
  \BibitemOpen
  \bibfield  {author} {\bibinfo {author} {\bibfnamefont {A.~A.}\ \bibnamefont {Kugler}},\ }\bibfield  {title} {\bibinfo {title} {Bounds for some equilibrium properties of an electron gas},\ }\href@noop {} {\bibfield  {journal} {\bibinfo  {journal} {Physical Review A}\ }\textbf {\bibinfo {volume} {1}},\ \bibinfo {pages} {1688} (\bibinfo {year} {1970})}\BibitemShut {NoStop}%
\bibitem [{\citenamefont {Olsen}\ \emph {et~al.}(2019)\citenamefont {Olsen}, \citenamefont {Patrick}, \citenamefont {Bates}, \citenamefont {Ruzsinszky},\ and\ \citenamefont {Thygesen}}]{olsen2019beyond}%
  \BibitemOpen
  \bibfield  {author} {\bibinfo {author} {\bibfnamefont {T.}~\bibnamefont {Olsen}}, \bibinfo {author} {\bibfnamefont {C.~E.}\ \bibnamefont {Patrick}}, \bibinfo {author} {\bibfnamefont {J.~E.}\ \bibnamefont {Bates}}, \bibinfo {author} {\bibfnamefont {A.}~\bibnamefont {Ruzsinszky}},\ and\ \bibinfo {author} {\bibfnamefont {K.~S.}\ \bibnamefont {Thygesen}},\ }\bibfield  {title} {\bibinfo {title} {Beyond the rpa and gw methods with adiabatic xc-kernels for accurate ground state and quasiparticle energies},\ }\href@noop {} {\bibfield  {journal} {\bibinfo  {journal} {npj Computational Materials}\ }\textbf {\bibinfo {volume} {5}},\ \bibinfo {pages} {106} (\bibinfo {year} {2019})}\BibitemShut {NoStop}%
\bibitem [{\citenamefont {Moldabekov}\ \emph {et~al.}(2023)\citenamefont {Moldabekov}, \citenamefont {Pavanello}, \citenamefont {B\"ohme}, \citenamefont {Vorberger},\ and\ \citenamefont {Dornheim}}]{Moldabekov_PRR_2023}%
  \BibitemOpen
  \bibfield  {author} {\bibinfo {author} {\bibfnamefont {Z.~A.}\ \bibnamefont {Moldabekov}}, \bibinfo {author} {\bibfnamefont {M.}~\bibnamefont {Pavanello}}, \bibinfo {author} {\bibfnamefont {M.~P.}\ \bibnamefont {B\"ohme}}, \bibinfo {author} {\bibfnamefont {J.}~\bibnamefont {Vorberger}},\ and\ \bibinfo {author} {\bibfnamefont {T.}~\bibnamefont {Dornheim}},\ }\bibfield  {title} {\bibinfo {title} {Linear-response time-dependent density functional theory approach to warm dense matter with adiabatic exchange-correlation kernels},\ }\href {https://doi.org/10.1103/PhysRevResearch.5.023089} {\bibfield  {journal} {\bibinfo  {journal} {Phys. Rev. Res.}\ }\textbf {\bibinfo {volume} {5}},\ \bibinfo {pages} {023089} (\bibinfo {year} {2023})}\BibitemShut {NoStop}%
\bibitem [{\citenamefont {Runge}\ and\ \citenamefont {Gross}(1984)}]{RG84}%
  \BibitemOpen
  \bibfield  {author} {\bibinfo {author} {\bibfnamefont {E.}~\bibnamefont {Runge}}\ and\ \bibinfo {author} {\bibfnamefont {E.~K.~U.}\ \bibnamefont {Gross}},\ }\bibfield  {title} {\bibinfo {title} {Density-functional theory for time-dependent systems},\ }\href {https://doi.org/10.1103/PhysRevLett.52.997} {\bibfield  {journal} {\bibinfo  {journal} {Phys. Rev. Lett.}\ }\textbf {\bibinfo {volume} {52}},\ \bibinfo {pages} {997} (\bibinfo {year} {1984})}\BibitemShut {NoStop}%
\bibitem [{\citenamefont {Dornheim}\ \emph {et~al.}(2023{\natexlab{b}})\citenamefont {Dornheim}, \citenamefont {Moldabekov}, \citenamefont {Tolias}, \citenamefont {Böhme},\ and\ \citenamefont {Vorberger}}]{Dornheim_MRE_2023}%
  \BibitemOpen
  \bibfield  {author} {\bibinfo {author} {\bibfnamefont {T.}~\bibnamefont {Dornheim}}, \bibinfo {author} {\bibfnamefont {Z.}~\bibnamefont {Moldabekov}}, \bibinfo {author} {\bibfnamefont {P.}~\bibnamefont {Tolias}}, \bibinfo {author} {\bibfnamefont {M.}~\bibnamefont {Böhme}},\ and\ \bibinfo {author} {\bibfnamefont {J.}~\bibnamefont {Vorberger}},\ }\bibfield  {title} {\bibinfo {title} {Physical insights from imaginary-time density--density correlation functions},\ }\href {https://doi.org/10.1063/5.0149638} {\bibfield  {journal} {\bibinfo  {journal} {Matter and Radiation at Extremes}\ }\textbf {\bibinfo {volume} {8}},\ \bibinfo {pages} {056601} (\bibinfo {year} {2023}{\natexlab{b}})}\BibitemShut {NoStop}%
\bibitem [{\citenamefont {Dornheim}\ \emph {et~al.}(2021{\natexlab{b}})\citenamefont {Dornheim}, \citenamefont {Moldabekov},\ and\ \citenamefont {Tolias}}]{Dornheim_PRB_ESA_2021}%
  \BibitemOpen
  \bibfield  {author} {\bibinfo {author} {\bibfnamefont {T.}~\bibnamefont {Dornheim}}, \bibinfo {author} {\bibfnamefont {Z.~A.}\ \bibnamefont {Moldabekov}},\ and\ \bibinfo {author} {\bibfnamefont {P.}~\bibnamefont {Tolias}},\ }\bibfield  {title} {\bibinfo {title} {Analytical representation of the local field correction of the uniform electron gas within the effective static approximation},\ }\href {https://doi.org/10.1103/PhysRevB.103.165102} {\bibfield  {journal} {\bibinfo  {journal} {Phys. Rev. B}\ }\textbf {\bibinfo {volume} {103}},\ \bibinfo {pages} {165102} (\bibinfo {year} {2021}{\natexlab{b}})}\BibitemShut {NoStop}%
\bibitem [{\citenamefont {Hou}\ \emph {et~al.}(2022)\citenamefont {Hou}, \citenamefont {Wang}, \citenamefont {Haule}, \citenamefont {Deng},\ and\ \citenamefont {Chen}}]{Hou_PRB_2022}%
  \BibitemOpen
  \bibfield  {author} {\bibinfo {author} {\bibfnamefont {P.-C.}\ \bibnamefont {Hou}}, \bibinfo {author} {\bibfnamefont {B.-Z.}\ \bibnamefont {Wang}}, \bibinfo {author} {\bibfnamefont {K.}~\bibnamefont {Haule}}, \bibinfo {author} {\bibfnamefont {Y.}~\bibnamefont {Deng}},\ and\ \bibinfo {author} {\bibfnamefont {K.}~\bibnamefont {Chen}},\ }\bibfield  {title} {\bibinfo {title} {Exchange-correlation effect in the charge response of a warm dense electron gas},\ }\href {https://doi.org/10.1103/PhysRevB.106.L081126} {\bibfield  {journal} {\bibinfo  {journal} {Phys. Rev. B}\ }\textbf {\bibinfo {volume} {106}},\ \bibinfo {pages} {L081126} (\bibinfo {year} {2022})}\BibitemShut {NoStop}%
\bibitem [{\citenamefont {Dornheim}\ \emph {et~al.}(2024{\natexlab{a}})\citenamefont {Dornheim}, \citenamefont {Tolias}, \citenamefont {Vorberger},\ and\ \citenamefont {Moldabekov}}]{dornheim2024QuantumDelocalization}%
  \BibitemOpen
  \bibfield  {author} {\bibinfo {author} {\bibfnamefont {T.}~\bibnamefont {Dornheim}}, \bibinfo {author} {\bibfnamefont {P.}~\bibnamefont {Tolias}}, \bibinfo {author} {\bibfnamefont {J.}~\bibnamefont {Vorberger}},\ and\ \bibinfo {author} {\bibfnamefont {Z.~A.}\ \bibnamefont {Moldabekov}},\ }\bibfield  {title} {\bibinfo {title} {Quantum delocalization, structural order, and density response of the strongly coupled electron liquid},\ }\href {https://doi.org/10.1209/0295-5075/ad5d88} {\bibfield  {journal} {\bibinfo  {journal} {Europhysics Letters}\ }\textbf {\bibinfo {volume} {147}},\ \bibinfo {pages} {36001} (\bibinfo {year} {2024}{\natexlab{a}})}\BibitemShut {NoStop}%
\bibitem [{\citenamefont {Dornheim}\ \emph {et~al.}(2020{\natexlab{b}})\citenamefont {Dornheim}, \citenamefont {Cangi}, \citenamefont {Ramakrishna}, \citenamefont {B\"ohme}, \citenamefont {Tanaka},\ and\ \citenamefont {Vorberger}}]{Dornheim_PRL_2020_ESA}%
  \BibitemOpen
  \bibfield  {author} {\bibinfo {author} {\bibfnamefont {T.}~\bibnamefont {Dornheim}}, \bibinfo {author} {\bibfnamefont {A.}~\bibnamefont {Cangi}}, \bibinfo {author} {\bibfnamefont {K.}~\bibnamefont {Ramakrishna}}, \bibinfo {author} {\bibfnamefont {M.}~\bibnamefont {B\"ohme}}, \bibinfo {author} {\bibfnamefont {S.}~\bibnamefont {Tanaka}},\ and\ \bibinfo {author} {\bibfnamefont {J.}~\bibnamefont {Vorberger}},\ }\bibfield  {title} {\bibinfo {title} {Effective static approximation: A fast and reliable tool for warm-dense matter theory},\ }\href {https://doi.org/10.1103/PhysRevLett.125.235001} {\bibfield  {journal} {\bibinfo  {journal} {Phys. Rev. Lett.}\ }\textbf {\bibinfo {volume} {125}},\ \bibinfo {pages} {235001} (\bibinfo {year} {2020}{\natexlab{b}})}\BibitemShut {NoStop}%
\bibitem [{\citenamefont {Dornheim}\ \emph {et~al.}(2021{\natexlab{c}})\citenamefont {Dornheim}, \citenamefont {Moldabekov},\ and\ \citenamefont {Tolias}}]{Dornheim_PRB_2021}%
  \BibitemOpen
  \bibfield  {author} {\bibinfo {author} {\bibfnamefont {T.}~\bibnamefont {Dornheim}}, \bibinfo {author} {\bibfnamefont {Z.~A.}\ \bibnamefont {Moldabekov}},\ and\ \bibinfo {author} {\bibfnamefont {P.}~\bibnamefont {Tolias}},\ }\bibfield  {title} {\bibinfo {title} {Analytical representation of the local field correction of the uniform electron gas within the effective static approximation},\ }\href {https://doi.org/10.1103/PhysRevB.103.165102} {\bibfield  {journal} {\bibinfo  {journal} {Phys. Rev. B}\ }\textbf {\bibinfo {volume} {103}},\ \bibinfo {pages} {165102} (\bibinfo {year} {2021}{\natexlab{c}})}\BibitemShut {NoStop}%
\bibitem [{\citenamefont {Dornheim}\ \emph {et~al.}(2024{\natexlab{b}})\citenamefont {Dornheim}, \citenamefont {Tolias}, \citenamefont {Kalkavouras}, \citenamefont {Moldabekov},\ and\ \citenamefont {Vorberger}}]{Dornheim_PRB_2024}%
  \BibitemOpen
  \bibfield  {author} {\bibinfo {author} {\bibfnamefont {T.}~\bibnamefont {Dornheim}}, \bibinfo {author} {\bibfnamefont {P.}~\bibnamefont {Tolias}}, \bibinfo {author} {\bibfnamefont {F.}~\bibnamefont {Kalkavouras}}, \bibinfo {author} {\bibfnamefont {Z.~A.}\ \bibnamefont {Moldabekov}},\ and\ \bibinfo {author} {\bibfnamefont {J.}~\bibnamefont {Vorberger}},\ }\bibfield  {title} {\bibinfo {title} {Dynamic exchange correlation effects in the strongly coupled electron liquid},\ }\href {https://doi.org/10.1103/PhysRevB.110.075137} {\bibfield  {journal} {\bibinfo  {journal} {Phys. Rev. B}\ }\textbf {\bibinfo {volume} {110}},\ \bibinfo {pages} {075137} (\bibinfo {year} {2024}{\natexlab{b}})}\BibitemShut {NoStop}%
\bibitem [{\citenamefont {Dornheim}\ \emph {et~al.}(2023{\natexlab{c}})\citenamefont {Dornheim}, \citenamefont {Wicaksono}, \citenamefont {Suarez-Cardona}, \citenamefont {Tolias}, \citenamefont {B\"ohme}, \citenamefont {Moldabekov}, \citenamefont {Hecht},\ and\ \citenamefont {Vorberger}}]{Dornheim_moments_2023}%
  \BibitemOpen
  \bibfield  {author} {\bibinfo {author} {\bibfnamefont {T.}~\bibnamefont {Dornheim}}, \bibinfo {author} {\bibfnamefont {D.~C.}\ \bibnamefont {Wicaksono}}, \bibinfo {author} {\bibfnamefont {J.~E.}\ \bibnamefont {Suarez-Cardona}}, \bibinfo {author} {\bibfnamefont {P.}~\bibnamefont {Tolias}}, \bibinfo {author} {\bibfnamefont {M.~P.}\ \bibnamefont {B\"ohme}}, \bibinfo {author} {\bibfnamefont {Z.~A.}\ \bibnamefont {Moldabekov}}, \bibinfo {author} {\bibfnamefont {M.}~\bibnamefont {Hecht}},\ and\ \bibinfo {author} {\bibfnamefont {J.}~\bibnamefont {Vorberger}},\ }\bibfield  {title} {\bibinfo {title} {Extraction of the frequency moments of spectral densities from imaginary-time correlation function data},\ }\href {https://doi.org/10.1103/PhysRevB.107.155148} {\bibfield  {journal} {\bibinfo  {journal} {Phys. Rev. B}\ }\textbf {\bibinfo {volume} {107}},\ \bibinfo {pages} {155148} (\bibinfo {year} {2023}{\natexlab{c}})}\BibitemShut {NoStop}%
\bibitem [{\citenamefont {Dornheim}\ \emph {et~al.}(2023{\natexlab{d}})\citenamefont {Dornheim}, \citenamefont {Vorberger}, \citenamefont {Moldabekov},\ and\ \citenamefont {Böhme}}]{Dornheim_PTR_2023}%
  \BibitemOpen
  \bibfield  {author} {\bibinfo {author} {\bibfnamefont {T.}~\bibnamefont {Dornheim}}, \bibinfo {author} {\bibfnamefont {J.}~\bibnamefont {Vorberger}}, \bibinfo {author} {\bibfnamefont {Z.~A.}\ \bibnamefont {Moldabekov}},\ and\ \bibinfo {author} {\bibfnamefont {M.}~\bibnamefont {Böhme}},\ }\bibfield  {title} {\bibinfo {title} {Analysing the dynamic structure of warm dense matter in the imaginary-time domain: theoretical models and simulations},\ }\href {https://doi.org/10.1098/rsta.2022.0217} {\bibfield  {journal} {\bibinfo  {journal} {Philosophical Transactions of the Royal Society A: Mathematical, Physical and Engineering Sciences}\ }\textbf {\bibinfo {volume} {381}},\ \bibinfo {pages} {20220217} (\bibinfo {year} {2023}{\natexlab{d}})}\BibitemShut {NoStop}%
\bibitem [{\citenamefont {Wu}\ \emph {et~al.}(2007)\citenamefont {Wu}, \citenamefont {Huang}, \citenamefont {Long},\ and\ \citenamefont {Peng}}]{wu2007trend}%
  \BibitemOpen
  \bibfield  {author} {\bibinfo {author} {\bibfnamefont {Z.}~\bibnamefont {Wu}}, \bibinfo {author} {\bibfnamefont {N.~E.}\ \bibnamefont {Huang}}, \bibinfo {author} {\bibfnamefont {S.~R.}\ \bibnamefont {Long}},\ and\ \bibinfo {author} {\bibfnamefont {C.-K.}\ \bibnamefont {Peng}},\ }\bibfield  {title} {\bibinfo {title} {On the trend, detrending, and variability of nonlinear and nonstationary time series},\ }\href@noop {} {\bibfield  {journal} {\bibinfo  {journal} {Proceedings of the National Academy of Sciences}\ }\textbf {\bibinfo {volume} {104}},\ \bibinfo {pages} {14889} (\bibinfo {year} {2007})}\BibitemShut {NoStop}%
\bibitem [{\citenamefont {Brockwell}\ and\ \citenamefont {Davis}(2002)}]{brockwell2002TSA}%
  \BibitemOpen
  \bibfield  {author} {\bibinfo {author} {\bibfnamefont {P.~J.}\ \bibnamefont {Brockwell}}\ and\ \bibinfo {author} {\bibfnamefont {R.~A.}\ \bibnamefont {Davis}},\ }\href@noop {} {\emph {\bibinfo {title} {Introduction to time series and forecasting}}}\ (\bibinfo  {publisher} {Springer},\ \bibinfo {year} {2002})\BibitemShut {NoStop}%
\bibitem [{\citenamefont {Adhikari}\ and\ \citenamefont {Agrawal}(2013)}]{adhikari2013TSA}%
  \BibitemOpen
  \bibfield  {author} {\bibinfo {author} {\bibfnamefont {R.}~\bibnamefont {Adhikari}}\ and\ \bibinfo {author} {\bibfnamefont {R.~K.}\ \bibnamefont {Agrawal}},\ }\bibfield  {title} {\bibinfo {title} {An introductory study on time series modeling and forecasting},\ }\href@noop {} {\bibfield  {journal} {\bibinfo  {journal} {arXiv preprint arXiv:1302.6613}\ } (\bibinfo {year} {2013})}\BibitemShut {NoStop}%
\bibitem [{\citenamefont {Virtanen}\ \emph {et~al.}(2020)\citenamefont {Virtanen}, \citenamefont {Gommers}, \citenamefont {Oliphant}, \citenamefont {Haberland}, \citenamefont {Reddy}, \citenamefont {Cournapeau}, \citenamefont {Burovski}, \citenamefont {Peterson}, \citenamefont {Weckesser}, \citenamefont {Bright}, \citenamefont {{van der Walt}}, \citenamefont {Brett}, \citenamefont {Wilson}, \citenamefont {Millman}, \citenamefont {Mayorov}, \citenamefont {Nelson}, \citenamefont {Jones}, \citenamefont {Kern}, \citenamefont {Larson}, \citenamefont {Carey}, \citenamefont {Polat}, \citenamefont {Feng}, \citenamefont {Moore}, \citenamefont {{VanderPlas}}, \citenamefont {Laxalde}, \citenamefont {Perktold}, \citenamefont {Cimrman}, \citenamefont {Henriksen}, \citenamefont {Quintero}, \citenamefont {Harris}, \citenamefont {Archibald}, \citenamefont {Ribeiro}, \citenamefont {Pedregosa}, \citenamefont {{van Mulbregt}},\ and\ \citenamefont {{SciPy 1.0 Contributors}}}]{SciPy2020NMeth}%
  \BibitemOpen
  \bibfield  {author} {\bibinfo {author} {\bibfnamefont {P.}~\bibnamefont {Virtanen}}, \bibinfo {author} {\bibfnamefont {R.}~\bibnamefont {Gommers}}, \bibinfo {author} {\bibfnamefont {T.~E.}\ \bibnamefont {Oliphant}}, \bibinfo {author} {\bibfnamefont {M.}~\bibnamefont {Haberland}}, \bibinfo {author} {\bibfnamefont {T.}~\bibnamefont {Reddy}}, \bibinfo {author} {\bibfnamefont {D.}~\bibnamefont {Cournapeau}}, \bibinfo {author} {\bibfnamefont {E.}~\bibnamefont {Burovski}}, \bibinfo {author} {\bibfnamefont {P.}~\bibnamefont {Peterson}}, \bibinfo {author} {\bibfnamefont {W.}~\bibnamefont {Weckesser}}, \bibinfo {author} {\bibfnamefont {J.}~\bibnamefont {Bright}}, \bibinfo {author} {\bibfnamefont {S.~J.}\ \bibnamefont {{van der Walt}}}, \bibinfo {author} {\bibfnamefont {M.}~\bibnamefont {Brett}}, \bibinfo {author} {\bibfnamefont {J.}~\bibnamefont {Wilson}}, \bibinfo {author} {\bibfnamefont {K.~J.}\ \bibnamefont {Millman}}, \bibinfo {author} {\bibfnamefont {N.}~\bibnamefont {Mayorov}}, \bibinfo {author} {\bibfnamefont
  {A.~R.~J.}\ \bibnamefont {Nelson}}, \bibinfo {author} {\bibfnamefont {E.}~\bibnamefont {Jones}}, \bibinfo {author} {\bibfnamefont {R.}~\bibnamefont {Kern}}, \bibinfo {author} {\bibfnamefont {E.}~\bibnamefont {Larson}}, \bibinfo {author} {\bibfnamefont {C.~J.}\ \bibnamefont {Carey}}, \bibinfo {author} {\bibfnamefont {{\.I}.}~\bibnamefont {Polat}}, \bibinfo {author} {\bibfnamefont {Y.}~\bibnamefont {Feng}}, \bibinfo {author} {\bibfnamefont {E.~W.}\ \bibnamefont {Moore}}, \bibinfo {author} {\bibfnamefont {J.}~\bibnamefont {{VanderPlas}}}, \bibinfo {author} {\bibfnamefont {D.}~\bibnamefont {Laxalde}}, \bibinfo {author} {\bibfnamefont {J.}~\bibnamefont {Perktold}}, \bibinfo {author} {\bibfnamefont {R.}~\bibnamefont {Cimrman}}, \bibinfo {author} {\bibfnamefont {I.}~\bibnamefont {Henriksen}}, \bibinfo {author} {\bibfnamefont {E.~A.}\ \bibnamefont {Quintero}}, \bibinfo {author} {\bibfnamefont {C.~R.}\ \bibnamefont {Harris}}, \bibinfo {author} {\bibfnamefont {A.~M.}\ \bibnamefont {Archibald}}, \bibinfo {author}
  {\bibfnamefont {A.~H.}\ \bibnamefont {Ribeiro}}, \bibinfo {author} {\bibfnamefont {F.}~\bibnamefont {Pedregosa}}, \bibinfo {author} {\bibfnamefont {P.}~\bibnamefont {{van Mulbregt}}},\ and\ \bibinfo {author} {\bibnamefont {{SciPy 1.0 Contributors}}},\ }\bibfield  {title} {\bibinfo {title} {{{SciPy} 1.0: Fundamental Algorithms for Scientific Computing in Python}},\ }\href {https://doi.org/10.1038/s41592-019-0686-2} {\bibfield  {journal} {\bibinfo  {journal} {Nature Methods}\ }\textbf {\bibinfo {volume} {17}},\ \bibinfo {pages} {261} (\bibinfo {year} {2020})}\BibitemShut {NoStop}%
\bibitem [{\citenamefont {Samworth}\ and\ \citenamefont {Cambridge}(2012)}]{samworth2012stein}%
  \BibitemOpen
  \bibfield  {author} {\bibinfo {author} {\bibfnamefont {R.~J.}\ \bibnamefont {Samworth}}\ and\ \bibinfo {author} {\bibfnamefont {S.}~\bibnamefont {Cambridge}},\ }\bibfield  {title} {\bibinfo {title} {Stein’s paradox},\ }\href@noop {} {\bibfield  {journal} {\bibinfo  {journal} {eureka}\ }\textbf {\bibinfo {volume} {62}},\ \bibinfo {pages} {38} (\bibinfo {year} {2012})}\BibitemShut {NoStop}%
\bibitem [{\citenamefont {Hatano}(1994)}]{hatano1994data}%
  \BibitemOpen
  \bibfield  {author} {\bibinfo {author} {\bibfnamefont {N.}~\bibnamefont {Hatano}},\ }\bibfield  {title} {\bibinfo {title} {Data analysis for quantum monte carlo simulations with the negative-sign problem},\ }\href@noop {} {\bibfield  {journal} {\bibinfo  {journal} {Journal of the Physical Society of Japan}\ }\textbf {\bibinfo {volume} {63}},\ \bibinfo {pages} {1691} (\bibinfo {year} {1994})}\BibitemShut {NoStop}%
\bibitem [{\citenamefont {Dornheim}\ \emph {et~al.}(2024{\natexlab{c}})\citenamefont {Dornheim}, \citenamefont {Böhme},\ and\ \citenamefont {Schwalbe}}]{ISHTAR}%
  \BibitemOpen
  \bibfield  {author} {\bibinfo {author} {\bibfnamefont {T.}~\bibnamefont {Dornheim}}, \bibinfo {author} {\bibfnamefont {M.}~\bibnamefont {Böhme}},\ and\ \bibinfo {author} {\bibfnamefont {S.}~\bibnamefont {Schwalbe}},\ }\href {https://doi.org/10.5281/zenodo.10497098} {\bibinfo {title} {{ISHTAR - Imaginary-time Stochastic High- performance Tool for Ab initio Research}}} (\bibinfo {year} {2024}{\natexlab{c}})\BibitemShut {NoStop}%
\bibitem [{\citenamefont {Dornheim}\ \emph {et~al.}(2021{\natexlab{d}})\citenamefont {Dornheim}, \citenamefont {B\"ohme}, \citenamefont {Militzer},\ and\ \citenamefont {Vorberger}}]{Dornheim_PRB_nk_2021}%
  \BibitemOpen
  \bibfield  {author} {\bibinfo {author} {\bibfnamefont {T.}~\bibnamefont {Dornheim}}, \bibinfo {author} {\bibfnamefont {M.}~\bibnamefont {B\"ohme}}, \bibinfo {author} {\bibfnamefont {B.}~\bibnamefont {Militzer}},\ and\ \bibinfo {author} {\bibfnamefont {J.}~\bibnamefont {Vorberger}},\ }\bibfield  {title} {\bibinfo {title} {Ab initio path integral monte carlo approach to the momentum distribution of the uniform electron gas at finite temperature without fixed nodes},\ }\href {https://doi.org/10.1103/PhysRevB.103.205142} {\bibfield  {journal} {\bibinfo  {journal} {Phys. Rev. B}\ }\textbf {\bibinfo {volume} {103}},\ \bibinfo {pages} {205142} (\bibinfo {year} {2021}{\natexlab{d}})}\BibitemShut {NoStop}%
\bibitem [{\citenamefont {Boninsegni}\ \emph {et~al.}(2006{\natexlab{a}})\citenamefont {Boninsegni}, \citenamefont {Prokofev},\ and\ \citenamefont {Svistunov}}]{boninsegni1}%
  \BibitemOpen
  \bibfield  {author} {\bibinfo {author} {\bibfnamefont {M.}~\bibnamefont {Boninsegni}}, \bibinfo {author} {\bibfnamefont {N.~V.}\ \bibnamefont {Prokofev}},\ and\ \bibinfo {author} {\bibfnamefont {B.~V.}\ \bibnamefont {Svistunov}},\ }\bibfield  {title} {\bibinfo {title} {Worm algorithm and diagrammatic {M}onte {C}arlo: A new approach to continuous-space path integral {M}onte {C}arlo simulations},\ }\href {https://journals.aps.org/pre/abstract/10.1103/PhysRevE.74.036701} {\bibfield  {journal} {\bibinfo  {journal} {Phys. Rev. E}\ }\textbf {\bibinfo {volume} {74}},\ \bibinfo {pages} {036701} (\bibinfo {year} {2006}{\natexlab{a}})}\BibitemShut {NoStop}%
\bibitem [{\citenamefont {Boninsegni}\ \emph {et~al.}(2006{\natexlab{b}})\citenamefont {Boninsegni}, \citenamefont {Prokofev},\ and\ \citenamefont {Svistunov}}]{boninsegni2}%
  \BibitemOpen
  \bibfield  {author} {\bibinfo {author} {\bibfnamefont {M.}~\bibnamefont {Boninsegni}}, \bibinfo {author} {\bibfnamefont {N.~V.}\ \bibnamefont {Prokofev}},\ and\ \bibinfo {author} {\bibfnamefont {B.~V.}\ \bibnamefont {Svistunov}},\ }\bibfield  {title} {\bibinfo {title} {Worm algorithm for continuous-space path integral {M}onte {C}arlo simulations},\ }\href {https://journals.aps.org/prl/abstract/10.1103/PhysRevLett.96.070601} {\bibfield  {journal} {\bibinfo  {journal} {Phys. Rev. Lett}\ }\textbf {\bibinfo {volume} {96}},\ \bibinfo {pages} {070601} (\bibinfo {year} {2006}{\natexlab{b}})}\BibitemShut {NoStop}%
\bibitem [{\citenamefont {Chuna}\ and\ \citenamefont {Murillo}(2025)}]{chuna2024conservative}%
  \BibitemOpen
  \bibfield  {author} {\bibinfo {author} {\bibfnamefont {T.}~\bibnamefont {Chuna}}\ and\ \bibinfo {author} {\bibfnamefont {M.~S.}\ \bibnamefont {Murillo}},\ }\bibfield  {title} {\bibinfo {title} {Conservative dielectric functions and electrical conductivities from the multicomponent bhatnagar-gross-krook equation},\ }\href {https://doi.org/10.1103/PhysRevE.111.035206} {\bibfield  {journal} {\bibinfo  {journal} {Phys. Rev. E}\ }\textbf {\bibinfo {volume} {111}},\ \bibinfo {pages} {035206} (\bibinfo {year} {2025})}\BibitemShut {NoStop}%
\bibitem [{\citenamefont {Selchow}\ and\ \citenamefont {Morawetz}(1999)}]{selchow1999dielectric}%
  \BibitemOpen
  \bibfield  {author} {\bibinfo {author} {\bibfnamefont {A.}~\bibnamefont {Selchow}}\ and\ \bibinfo {author} {\bibfnamefont {K.}~\bibnamefont {Morawetz}},\ }\bibfield  {title} {\bibinfo {title} {Dielectric properties of interacting storage ring plasmas},\ }\href@noop {} {\bibfield  {journal} {\bibinfo  {journal} {Physical Review E}\ }\textbf {\bibinfo {volume} {59}},\ \bibinfo {pages} {1015} (\bibinfo {year} {1999})}\BibitemShut {NoStop}%
\bibitem [{\citenamefont {Vitali}\ \emph {et~al.}(2010)\citenamefont {Vitali}, \citenamefont {Rossi}, \citenamefont {Reatto},\ and\ \citenamefont {Galli}}]{Vitali_PRB_2010}%
  \BibitemOpen
  \bibfield  {author} {\bibinfo {author} {\bibfnamefont {E.}~\bibnamefont {Vitali}}, \bibinfo {author} {\bibfnamefont {M.}~\bibnamefont {Rossi}}, \bibinfo {author} {\bibfnamefont {L.}~\bibnamefont {Reatto}},\ and\ \bibinfo {author} {\bibfnamefont {D.~E.}\ \bibnamefont {Galli}},\ }\bibfield  {title} {\bibinfo {title} {Ab initio low-energy dynamics of superfluid and solid $^{4}\textnormal{H}\textnormal{e}$},\ }\href {https://doi.org/10.1103/PhysRevB.82.174510} {\bibfield  {journal} {\bibinfo  {journal} {Phys. Rev. B}\ }\textbf {\bibinfo {volume} {82}},\ \bibinfo {pages} {174510} (\bibinfo {year} {2010})}\BibitemShut {NoStop}%
\bibitem [{\citenamefont {Zhang}(2016)}]{zhang2016Sq}%
  \BibitemOpen
  \bibfield  {author} {\bibinfo {author} {\bibfnamefont {K.}~\bibnamefont {Zhang}},\ }\bibfield  {title} {\bibinfo {title} {On the concept of static structure factor},\ }\href@noop {} {\bibfield  {journal} {\bibinfo  {journal} {arXiv preprint arXiv:1606.03610}\ } (\bibinfo {year} {2016})}\BibitemShut {NoStop}%
\bibitem [{\citenamefont {Sandvik}\ \emph {et~al.}(1998)\citenamefont {Sandvik}, \citenamefont {Capponi}, \citenamefont {Poilblanc},\ and\ \citenamefont {Dagotto}}]{sandvik1998numerical}%
  \BibitemOpen
  \bibfield  {author} {\bibinfo {author} {\bibfnamefont {A.~W.}\ \bibnamefont {Sandvik}}, \bibinfo {author} {\bibfnamefont {S.}~\bibnamefont {Capponi}}, \bibinfo {author} {\bibfnamefont {D.}~\bibnamefont {Poilblanc}},\ and\ \bibinfo {author} {\bibfnamefont {E.}~\bibnamefont {Dagotto}},\ }\bibfield  {title} {\bibinfo {title} {Numerical calculations of the b 1 g raman spectrum of the two-dimensional heisenberg model},\ }\href@noop {} {\bibfield  {journal} {\bibinfo  {journal} {Physical Review B}\ }\textbf {\bibinfo {volume} {57}},\ \bibinfo {pages} {8478} (\bibinfo {year} {1998})}\BibitemShut {NoStop}%
\bibitem [{\citenamefont {Chiesa}\ \emph {et~al.}(2006)\citenamefont {Chiesa}, \citenamefont {Ceperley}, \citenamefont {Martin},\ and\ \citenamefont {Holzmann}}]{Chiesa_PRL_2006}%
  \BibitemOpen
  \bibfield  {author} {\bibinfo {author} {\bibfnamefont {S.}~\bibnamefont {Chiesa}}, \bibinfo {author} {\bibfnamefont {D.~M.}\ \bibnamefont {Ceperley}}, \bibinfo {author} {\bibfnamefont {R.~M.}\ \bibnamefont {Martin}},\ and\ \bibinfo {author} {\bibfnamefont {M.}~\bibnamefont {Holzmann}},\ }\bibfield  {title} {\bibinfo {title} {Finite-size error in many-body simulations with long-range interactions},\ }\href {https://doi.org/10.1103/PhysRevLett.97.076404} {\bibfield  {journal} {\bibinfo  {journal} {Phys. Rev. Lett.}\ }\textbf {\bibinfo {volume} {97}},\ \bibinfo {pages} {076404} (\bibinfo {year} {2006})}\BibitemShut {NoStop}%
\bibitem [{\citenamefont {Dornheim}\ \emph {et~al.}(2016)\citenamefont {Dornheim}, \citenamefont {Groth}, \citenamefont {Sjostrom}, \citenamefont {Malone}, \citenamefont {Foulkes},\ and\ \citenamefont {Bonitz}}]{dornheim_prl}%
  \BibitemOpen
  \bibfield  {author} {\bibinfo {author} {\bibfnamefont {T.}~\bibnamefont {Dornheim}}, \bibinfo {author} {\bibfnamefont {S.}~\bibnamefont {Groth}}, \bibinfo {author} {\bibfnamefont {T.}~\bibnamefont {Sjostrom}}, \bibinfo {author} {\bibfnamefont {F.~D.}\ \bibnamefont {Malone}}, \bibinfo {author} {\bibfnamefont {W.~M.~C.}\ \bibnamefont {Foulkes}},\ and\ \bibinfo {author} {\bibfnamefont {M.}~\bibnamefont {Bonitz}},\ }\bibfield  {title} {\bibinfo {title} {Ab initio quantum {M}onte {C}arlo simulation of the warm dense electron gas in the thermodynamic limit},\ }\href {http://link.aps.org/doi/10.1103/PhysRevLett.117.156403} {\bibfield  {journal} {\bibinfo  {journal} {Phys. Rev. Lett.}\ }\textbf {\bibinfo {volume} {117}},\ \bibinfo {pages} {156403} (\bibinfo {year} {2016})}\BibitemShut {NoStop}%
\bibitem [{\citenamefont {Dornheim}\ \emph {et~al.}(2024{\natexlab{d}})\citenamefont {Dornheim}, \citenamefont {Schwalbe}, \citenamefont {Tolias}, \citenamefont {B{\"o}hme}, \citenamefont {Moldabekov},\ and\ \citenamefont {Vorberger}}]{dornheim_MRE_2024}%
  \BibitemOpen
  \bibfield  {author} {\bibinfo {author} {\bibfnamefont {T.}~\bibnamefont {Dornheim}}, \bibinfo {author} {\bibfnamefont {S.}~\bibnamefont {Schwalbe}}, \bibinfo {author} {\bibfnamefont {P.}~\bibnamefont {Tolias}}, \bibinfo {author} {\bibfnamefont {M.~P.}\ \bibnamefont {B{\"o}hme}}, \bibinfo {author} {\bibfnamefont {Z.~A.}\ \bibnamefont {Moldabekov}},\ and\ \bibinfo {author} {\bibfnamefont {J.}~\bibnamefont {Vorberger}},\ }\bibfield  {title} {\bibinfo {title} {Ab initio density response and local field factor of warm dense hydrogen},\ }\href@noop {} {\bibfield  {journal} {\bibinfo  {journal} {Matter and Radiation at Extremes}\ }\textbf {\bibinfo {volume} {9}} (\bibinfo {year} {2024}{\natexlab{d}})}\BibitemShut {NoStop}%
\bibitem [{\citenamefont {Dornheim}\ \emph {et~al.}(2024{\natexlab{e}})\citenamefont {Dornheim}, \citenamefont {Schwalbe}, \citenamefont {Böhme}, \citenamefont {Moldabekov}, \citenamefont {Vorberger},\ and\ \citenamefont {Tolias}}]{Dornheim_JCP_2024}%
  \BibitemOpen
  \bibfield  {author} {\bibinfo {author} {\bibfnamefont {T.}~\bibnamefont {Dornheim}}, \bibinfo {author} {\bibfnamefont {S.}~\bibnamefont {Schwalbe}}, \bibinfo {author} {\bibfnamefont {M.~P.}\ \bibnamefont {Böhme}}, \bibinfo {author} {\bibfnamefont {Z.~A.}\ \bibnamefont {Moldabekov}}, \bibinfo {author} {\bibfnamefont {J.}~\bibnamefont {Vorberger}},\ and\ \bibinfo {author} {\bibfnamefont {P.}~\bibnamefont {Tolias}},\ }\bibfield  {title} {\bibinfo {title} {{Ab initio path integral Monte Carlo simulations of warm dense two-component systems without fixed nodes: Structural properties}},\ }\href {https://doi.org/10.1063/5.0206787} {\bibfield  {journal} {\bibinfo  {journal} {The Journal of Chemical Physics}\ }\textbf {\bibinfo {volume} {160}},\ \bibinfo {pages} {164111} (\bibinfo {year} {2024}{\natexlab{e}})}\BibitemShut {NoStop}%
\bibitem [{\citenamefont {Dornheim}\ \emph {et~al.}(2024{\natexlab{f}})\citenamefont {Dornheim}, \citenamefont {Döppner}, \citenamefont {Tolias}, \citenamefont {Böhme}, \citenamefont {Fletcher}, \citenamefont {Gawne}, \citenamefont {Graziani}, \citenamefont {Kraus}, \citenamefont {MacDonald}, \citenamefont {Moldabekov}, \citenamefont {Schwalbe}, \citenamefont {Gericke},\ and\ \citenamefont {Vorberger}}]{Dornheim_NatComm_2024}%
  \BibitemOpen
  \bibfield  {author} {\bibinfo {author} {\bibfnamefont {T.}~\bibnamefont {Dornheim}}, \bibinfo {author} {\bibfnamefont {T.}~\bibnamefont {Döppner}}, \bibinfo {author} {\bibfnamefont {P.}~\bibnamefont {Tolias}}, \bibinfo {author} {\bibfnamefont {M.}~\bibnamefont {Böhme}}, \bibinfo {author} {\bibfnamefont {L.}~\bibnamefont {Fletcher}}, \bibinfo {author} {\bibfnamefont {T.}~\bibnamefont {Gawne}}, \bibinfo {author} {\bibfnamefont {F.}~\bibnamefont {Graziani}}, \bibinfo {author} {\bibfnamefont {D.}~\bibnamefont {Kraus}}, \bibinfo {author} {\bibfnamefont {M.}~\bibnamefont {MacDonald}}, \bibinfo {author} {\bibfnamefont {Z.}~\bibnamefont {Moldabekov}}, \bibinfo {author} {\bibfnamefont {S.}~\bibnamefont {Schwalbe}}, \bibinfo {author} {\bibfnamefont {D.}~\bibnamefont {Gericke}},\ and\ \bibinfo {author} {\bibfnamefont {J.}~\bibnamefont {Vorberger}},\ }\href@noop {} {\bibinfo {title} {Unraveling electronic correlations in warm dense quantum plasmas}} (\bibinfo {year} {2024}{\natexlab{f}}),\ \Eprint
  {https://arxiv.org/abs/2402.19113} {arXiv:2402.19113 [physics.plasm-ph]} \BibitemShut {NoStop}%
\bibitem [{\citenamefont {D{\"o}ppner}\ \emph {et~al.}(2023)\citenamefont {D{\"o}ppner}, \citenamefont {Bethkenhagen}, \citenamefont {Kraus}, \citenamefont {Neumayer}, \citenamefont {Chapman}, \citenamefont {Bachmann}, \citenamefont {Baggott}, \citenamefont {B{\"o}hme}, \citenamefont {Divol}, \citenamefont {Falcone}, \citenamefont {Fletcher}, \citenamefont {Landen}, \citenamefont {MacDonald}, \citenamefont {Saunders}, \citenamefont {Sch{\"o}rner}, \citenamefont {Sterne}, \citenamefont {Vorberger}, \citenamefont {Witte}, \citenamefont {Yi}, \citenamefont {Redmer}, \citenamefont {Glenzer},\ and\ \citenamefont {Gericke}}]{Tilo_Nature_2023}%
  \BibitemOpen
  \bibfield  {author} {\bibinfo {author} {\bibfnamefont {T.}~\bibnamefont {D{\"o}ppner}}, \bibinfo {author} {\bibfnamefont {M.}~\bibnamefont {Bethkenhagen}}, \bibinfo {author} {\bibfnamefont {D.}~\bibnamefont {Kraus}}, \bibinfo {author} {\bibfnamefont {P.}~\bibnamefont {Neumayer}}, \bibinfo {author} {\bibfnamefont {D.~A.}\ \bibnamefont {Chapman}}, \bibinfo {author} {\bibfnamefont {B.}~\bibnamefont {Bachmann}}, \bibinfo {author} {\bibfnamefont {R.~A.}\ \bibnamefont {Baggott}}, \bibinfo {author} {\bibfnamefont {M.~P.}\ \bibnamefont {B{\"o}hme}}, \bibinfo {author} {\bibfnamefont {L.}~\bibnamefont {Divol}}, \bibinfo {author} {\bibfnamefont {R.~W.}\ \bibnamefont {Falcone}}, \bibinfo {author} {\bibfnamefont {L.~B.}\ \bibnamefont {Fletcher}}, \bibinfo {author} {\bibfnamefont {O.~L.}\ \bibnamefont {Landen}}, \bibinfo {author} {\bibfnamefont {M.~J.}\ \bibnamefont {MacDonald}}, \bibinfo {author} {\bibfnamefont {A.~M.}\ \bibnamefont {Saunders}}, \bibinfo {author} {\bibfnamefont {M.}~\bibnamefont {Sch{\"o}rner}},
  \bibinfo {author} {\bibfnamefont {P.~A.}\ \bibnamefont {Sterne}}, \bibinfo {author} {\bibfnamefont {J.}~\bibnamefont {Vorberger}}, \bibinfo {author} {\bibfnamefont {B.~B.~L.}\ \bibnamefont {Witte}}, \bibinfo {author} {\bibfnamefont {A.}~\bibnamefont {Yi}}, \bibinfo {author} {\bibfnamefont {R.}~\bibnamefont {Redmer}}, \bibinfo {author} {\bibfnamefont {S.~H.}\ \bibnamefont {Glenzer}},\ and\ \bibinfo {author} {\bibfnamefont {D.~O.}\ \bibnamefont {Gericke}},\ }\bibfield  {title} {\bibinfo {title} {Observing the onset of pressure-driven k-shell delocalization},\ }\href {https://doi.org/10.1038/s41586-023-05996-8} {\bibfield  {journal} {\bibinfo  {journal} {Nature}\ }\textbf {\bibinfo {volume} {618}},\ \bibinfo {pages} {270–275} (\bibinfo {year} {2023})}\BibitemShut {NoStop}%
\bibitem [{\citenamefont {Gawne}\ \emph {et~al.}(2024)\citenamefont {Gawne}, \citenamefont {Moldabekov}, \citenamefont {Humphries}, \citenamefont {Appel}, \citenamefont {Baehtz}, \citenamefont {Bouffetier}, \citenamefont {Brambrink}, \citenamefont {Cangi}, \citenamefont {G\"ode}, \citenamefont {Kon\^opkov\'a}, \citenamefont {Makita}, \citenamefont {Mishchenko}, \citenamefont {Nakatsutsumi}, \citenamefont {Ramakrishna}, \citenamefont {Randolph}, \citenamefont {Schwalbe}, \citenamefont {Vorberger}, \citenamefont {Wollenweber}, \citenamefont {Zastrau}, \citenamefont {Dornheim},\ and\ \citenamefont {Preston}}]{Gawne_PRB_2024}%
  \BibitemOpen
  \bibfield  {author} {\bibinfo {author} {\bibfnamefont {T.}~\bibnamefont {Gawne}}, \bibinfo {author} {\bibfnamefont {Z.~A.}\ \bibnamefont {Moldabekov}}, \bibinfo {author} {\bibfnamefont {O.~S.}\ \bibnamefont {Humphries}}, \bibinfo {author} {\bibfnamefont {K.}~\bibnamefont {Appel}}, \bibinfo {author} {\bibfnamefont {C.}~\bibnamefont {Baehtz}}, \bibinfo {author} {\bibfnamefont {V.}~\bibnamefont {Bouffetier}}, \bibinfo {author} {\bibfnamefont {E.}~\bibnamefont {Brambrink}}, \bibinfo {author} {\bibfnamefont {A.}~\bibnamefont {Cangi}}, \bibinfo {author} {\bibfnamefont {S.}~\bibnamefont {G\"ode}}, \bibinfo {author} {\bibfnamefont {Z.}~\bibnamefont {Kon\^opkov\'a}}, \bibinfo {author} {\bibfnamefont {M.}~\bibnamefont {Makita}}, \bibinfo {author} {\bibfnamefont {M.}~\bibnamefont {Mishchenko}}, \bibinfo {author} {\bibfnamefont {M.}~\bibnamefont {Nakatsutsumi}}, \bibinfo {author} {\bibfnamefont {K.}~\bibnamefont {Ramakrishna}}, \bibinfo {author} {\bibfnamefont {L.}~\bibnamefont {Randolph}}, \bibinfo {author}
  {\bibfnamefont {S.}~\bibnamefont {Schwalbe}}, \bibinfo {author} {\bibfnamefont {J.}~\bibnamefont {Vorberger}}, \bibinfo {author} {\bibfnamefont {L.}~\bibnamefont {Wollenweber}}, \bibinfo {author} {\bibfnamefont {U.}~\bibnamefont {Zastrau}}, \bibinfo {author} {\bibfnamefont {T.}~\bibnamefont {Dornheim}},\ and\ \bibinfo {author} {\bibfnamefont {T.~R.}\ \bibnamefont {Preston}},\ }\bibfield  {title} {\bibinfo {title} {Ultrahigh resolution x-ray thomson scattering measurements at the european x-ray free electron laser},\ }\href {https://doi.org/10.1103/PhysRevB.109.L241112} {\bibfield  {journal} {\bibinfo  {journal} {Phys. Rev. B}\ }\textbf {\bibinfo {volume} {109}},\ \bibinfo {pages} {L241112} (\bibinfo {year} {2024})}\BibitemShut {NoStop}%
\bibitem [{\citenamefont {Bellenbaum}\ \emph {et~al.}(2025)\citenamefont {Bellenbaum}, \citenamefont {Böhme}, \citenamefont {Bonitz}, \citenamefont {Döppner}, \citenamefont {Fletcher}, \citenamefont {Gawne}, \citenamefont {Kraus}, \citenamefont {Moldabekov}, \citenamefont {Schwalbe}, \citenamefont {Vorberger},\ and\ \citenamefont {Dornheim}}]{bellenbaum2025estimatingionizationstatescontinuum}%
  \BibitemOpen
  \bibfield  {author} {\bibinfo {author} {\bibfnamefont {H.~M.}\ \bibnamefont {Bellenbaum}}, \bibinfo {author} {\bibfnamefont {M.~P.}\ \bibnamefont {Böhme}}, \bibinfo {author} {\bibfnamefont {M.}~\bibnamefont {Bonitz}}, \bibinfo {author} {\bibfnamefont {T.}~\bibnamefont {Döppner}}, \bibinfo {author} {\bibfnamefont {L.~B.}\ \bibnamefont {Fletcher}}, \bibinfo {author} {\bibfnamefont {T.}~\bibnamefont {Gawne}}, \bibinfo {author} {\bibfnamefont {D.}~\bibnamefont {Kraus}}, \bibinfo {author} {\bibfnamefont {Z.~A.}\ \bibnamefont {Moldabekov}}, \bibinfo {author} {\bibfnamefont {S.}~\bibnamefont {Schwalbe}}, \bibinfo {author} {\bibfnamefont {J.}~\bibnamefont {Vorberger}},\ and\ \bibinfo {author} {\bibfnamefont {T.}~\bibnamefont {Dornheim}},\ }\href {https://arxiv.org/abs/2503.14014} {\bibinfo {title} {Estimating ionization states and continuum lowering from ab initio path integral monte carlo simulations for warm dense hydrogen}} (\bibinfo {year} {2025}),\ \Eprint {https://arxiv.org/abs/2503.14014} {arXiv:2503.14014
  [physics.chem-ph]} \BibitemShut {NoStop}%
\bibitem [{\citenamefont {Baczewski}\ \emph {et~al.}(2016)\citenamefont {Baczewski}, \citenamefont {Shulenburger}, \citenamefont {Desjarlais}, \citenamefont {Hansen},\ and\ \citenamefont {Magyar}}]{Baczewski_PRL_2016}%
  \BibitemOpen
  \bibfield  {author} {\bibinfo {author} {\bibfnamefont {A.~D.}\ \bibnamefont {Baczewski}}, \bibinfo {author} {\bibfnamefont {L.}~\bibnamefont {Shulenburger}}, \bibinfo {author} {\bibfnamefont {M.~P.}\ \bibnamefont {Desjarlais}}, \bibinfo {author} {\bibfnamefont {S.~B.}\ \bibnamefont {Hansen}},\ and\ \bibinfo {author} {\bibfnamefont {R.~J.}\ \bibnamefont {Magyar}},\ }\bibfield  {title} {\bibinfo {title} {X-ray thomson scattering in warm dense matter without the chihara decomposition},\ }\href {https://journals.aps.org/prl/abstract/10.1103/PhysRevLett.116.115004} {\bibfield  {journal} {\bibinfo  {journal} {Phys. Rev. Lett}\ }\textbf {\bibinfo {volume} {116}},\ \bibinfo {pages} {115004} (\bibinfo {year} {2016})}\BibitemShut {NoStop}%
\bibitem [{\citenamefont {Sch\"orner}\ \emph {et~al.}(2023)\citenamefont {Sch\"orner}, \citenamefont {Bethkenhagen}, \citenamefont {D\"oppner}, \citenamefont {Kraus}, \citenamefont {Fletcher}, \citenamefont {Glenzer},\ and\ \citenamefont {Redmer}}]{Schoerner_PRE_2023}%
  \BibitemOpen
  \bibfield  {author} {\bibinfo {author} {\bibfnamefont {M.}~\bibnamefont {Sch\"orner}}, \bibinfo {author} {\bibfnamefont {M.}~\bibnamefont {Bethkenhagen}}, \bibinfo {author} {\bibfnamefont {T.}~\bibnamefont {D\"oppner}}, \bibinfo {author} {\bibfnamefont {D.}~\bibnamefont {Kraus}}, \bibinfo {author} {\bibfnamefont {L.~B.}\ \bibnamefont {Fletcher}}, \bibinfo {author} {\bibfnamefont {S.~H.}\ \bibnamefont {Glenzer}},\ and\ \bibinfo {author} {\bibfnamefont {R.}~\bibnamefont {Redmer}},\ }\bibfield  {title} {\bibinfo {title} {X-ray thomson scattering spectra from density functional theory molecular dynamics simulations based on a modified chihara formula},\ }\href {https://doi.org/10.1103/PhysRevE.107.065207} {\bibfield  {journal} {\bibinfo  {journal} {Phys. Rev. E}\ }\textbf {\bibinfo {volume} {107}},\ \bibinfo {pages} {065207} (\bibinfo {year} {2023})}\BibitemShut {NoStop}%
\bibitem [{\citenamefont {Moldabekov}\ \emph {et~al.}(2025)\citenamefont {Moldabekov}, \citenamefont {Schwalbe}, \citenamefont {Gawne}, \citenamefont {Preston}, \citenamefont {Vorberger},\ and\ \citenamefont {Dornheim}}]{moldabekov2025applyingliouvillelanczosmethodtimedependent}%
  \BibitemOpen
  \bibfield  {author} {\bibinfo {author} {\bibfnamefont {Z.~A.}\ \bibnamefont {Moldabekov}}, \bibinfo {author} {\bibfnamefont {S.}~\bibnamefont {Schwalbe}}, \bibinfo {author} {\bibfnamefont {T.}~\bibnamefont {Gawne}}, \bibinfo {author} {\bibfnamefont {T.~R.}\ \bibnamefont {Preston}}, \bibinfo {author} {\bibfnamefont {J.}~\bibnamefont {Vorberger}},\ and\ \bibinfo {author} {\bibfnamefont {T.}~\bibnamefont {Dornheim}},\ }\href {https://arxiv.org/abs/2502.04921} {\bibinfo {title} {Applying the liouville-lanczos method of time-dependent density-functional theory to warm dense matter}} (\bibinfo {year} {2025}),\ \Eprint {https://arxiv.org/abs/2502.04921} {arXiv:2502.04921 [physics.plasm-ph]} \BibitemShut {NoStop}%
\bibitem [{\citenamefont {White}(2025)}]{White_2025}%
  \BibitemOpen
  \bibfield  {author} {\bibinfo {author} {\bibfnamefont {A.~J.}\ \bibnamefont {White}},\ }\bibfield  {title} {\bibinfo {title} {Dynamical structure factors of warm dense matter from time-dependent orbital-free and mixed-stochastic-deterministic density functional theory},\ }\href {https://doi.org/10.1088/2516-1075/adad24} {\bibfield  {journal} {\bibinfo  {journal} {Electronic Structure}\ }\textbf {\bibinfo {volume} {7}},\ \bibinfo {pages} {014001} (\bibinfo {year} {2025})}\BibitemShut {NoStop}%
\end{thebibliography}%

\appendix

\end{document}